\documentclass[final,3p,twocolumn]{elsarticle}
\usepackage{graphicx}
\usepackage{epsfig}
\usepackage{dcolumn}
\usepackage{bm}
\usepackage{amsmath}
\usepackage{color}
\usepackage[colorlinks,linkcolor=red,anchorcolor=green,citecolor=blue]{hyperref}
\usepackage{multirow}
\usepackage{relsize}
\usepackage{color,xcolor}
\usepackage{lineno}
\journal{Frontier of Physics}
\usepackage{threeparttable}
\usepackage{rotating}

\newcommand{\BR}{{\cal B}}

\newcommand{\ones}{\Upsilon(1S)}
\newcommand{\twos}{\Upsilon(2S)}

\newcommand{\allu}{\Upsilon(1S,2S,3S)}

\newcommand{\LL}{\ell^+\ell^-}

\newcommand{\beq}{\begin{equation}}
\newcommand{\eeq}{\end{equation}}
\newcommand{\bitm}{\begin{itemize}}
\newcommand{\eitm}{\end{itemize}}

\begin{document}

\begin{frontmatter}

\title{Experimental review of the $\allu$ physics at $e^+e^-$ colliders and the LHC}

\author[FUDAN]{Sen Jia}
\author[BUAA]{Xingyu Zhou}
\author[FUDAN]{Chengping Shen\corref{cor1}}
\cortext[cor1]{Corresponding author}
\ead{shencp@fudan.edu.cn}

\address[FUDAN]{Key Laboratory of Nuclear Physics and Ion-beam Application (MOE) and Institute of Modern Physics, Fudan University, Shanghai 200443, China}
\address[BUAA]{School of Physics and Information Engineering, Shanxi Normal University, Linfen 041004, China}

\begin{abstract}

The three lowest-lying $\Upsilon$ states, i.e. $\Upsilon(1S)$, $\Upsilon(2S)$, and $\Upsilon(3S)$, composed of $b\bar b$ pairs and below the $B\bar B$ threshold, provide a good platform for the researches of hadronic physics and physics beyond the Standard Model. They can be produced directly in $e^+e^-$ colliding  experiments, such as CLEO, Babar, and Belle, with low continuum backgrounds. In these experiments, many measurements of the exclusive $\ones$ and $\twos$ decays into light hadrons, which shed light on the ``80\% rule" for the Okubo-Zweig-Iizuka suppressed decays in the bottomonium sector, were carried out. Meanwhile, many studies of the charmonium and bottomonium productions in $\Upsilon(1S,2S,3S)$ decays were performed, to distinguish different Quantum Chromodynamics (QCD) models. Besides, exotic states and new physics were also extensively explored in $\Upsilon(1S,2S,3S)$ decays at CLEO, BaBar, and Belle.
The $\Upsilon(1S,2S,3S)$ states can also be produced in $pp$ collisions and in collisions involving heavy ions. The precision measurements of their cross sections and polarizations at the large hadron collider (LHC), especially in the CMS, ATLAS, and LHCb experiments, help to understand $\Upsilon$ production mechanisms in $pp$ collisions. The observation of the sequential $\Upsilon$ suppression in heavy ion collisions at CMS, LHCb, and ALICE is of great importance for verifying the quark-gluon plasma predicted by QCD. In this article, we review the experimental results on $\Upsilon(1S,2S,3S)$ at $e^+e^-$ colliders and the LHC, and summarize their prospects at Belle II and the LHC.

\end{abstract}


\end{frontmatter}

\bigskip
\centerline{Outline}
\bigskip

\tableofcontents
\clearpage
\section{Introduction}

The first $\Upsilon$ resonance, namely $\Upsilon(1S)$, was discovered in 1977 in the bombardment of a beam of high energy protons to a stationary nuclear target~\cite{Herb252,Innes1240}.
Soon, in the $e^+e^-$ collisions to the $\mu^+\mu^-$ final state, $\Upsilon(1S)$ was confirmed, and $\Upsilon(2S)$ and $\Upsilon(3S)$ were also observed~\cite{Berger243,Darden246,Bienlein360}.
The discovery of $\Upsilon(1S,2S,3S)$ opens a door to the study of the bottomonium spectrum. In the following decades many different experiments, at lepton, hadron, and ion colliders, reported the fruitful results on these $\Upsilon$ states.

Because $b$ quark is almost 3 times heavier than $c$ quark, the $b\bar b$ system is more non-relativistic than the $c\bar{c}$ system. Since non-relativistic  systems are easier to describe theoretically, the $\Upsilon$ system as a representative of bottomonium family plays an important role in the studies of strong interactions. Although the $t\bar t$ system is completely non-relativistic, experimental studies of strong interaction phenomena with $t\bar t$ are very difficult because the $t$ quark decays by weak interactions.

Unlike the $\Upsilon$ states above the $B\bar B$ threshold, the intrinsic widths of $\Upsilon(1S,2S,3S)$ are quite narrow~\cite{PDG}, merely 20 $\sim$ 50 keV. This can be explained by the Okubo-Zweig-Iizuka (OZI) rules~\cite{OZI1,OZI2,OZI3,OZI4}. For $\Upsilon(4S)$ and other $\Upsilon$ states with higher masses, they are allowed to decay via a typical strong interaction into a pair of heavy quark mesons: $(b\bar b)\to (b\bar q)(q\bar b)$ $(q=u,d,c)$, e.g., $\Upsilon(4S)\to B\bar B$, which results in a large natural width. Below the $B\bar B$ threshold, $\Upsilon(1S,2S,3S)$ can not decay via this OZI-favored way, thus leading to a narrow natural width. However, they can still decay by strong interactions with the $b\bar b$ pair annihilation via three gluons into hadrons.

In the process of $e^+e^-\to \gamma^* \to \Upsilon(1S,2S,3S)$, the entire center-of-mass (C.M.) energy of the initial $e^+e^-$ turns into the rest mass of the $\Upsilon$ state. The C.M. energy was matched to the resonance mass; thus only one $\Upsilon$ resonance was produced at a time.
Accordingly, a particularly clean environment for studies of the properties of the $\Upsilon$ states is provided by $e^+e^-$ colliders at $B$-factories (Belle and BaBar) and CLEO. The hadronic and radiative decays of the $\allu$ resonances have been extensively studied at $B$-factories and CLEO to test various theoretical predictions by potential models~\cite{Eichten34,Buchmuller24,Gupta34,Liu46,Barnes72,Radford75}, effective field approaches~\cite{Brambilla77,Bodwin51}, lattice gauge calculations~\cite{Okamoto65,Serenone}, etc. The bottomonium spectroscopy has also been greatly enriched in the past decades.

In the hadronic or radiative decays of $\Upsilon(1S,2S,3S)$, the searches for exotic states and new physics (NP) are very promising at $B$-factories and CLEO.  The ample gluons in the hadronic decays of $\Upsilon(1S,2S,3S)$ provide an entry to many potential exotic states, including the glueballs, charmoniumlike states (so-called $XYZ$ states), and stable six-quark states. The $\Upsilon(1S,2S,3S)$ decays also provide a good platform to explore the dark sector physics. In the   Standard Model (SM), invisible decays of $\Upsilon(1S)$ involving neutrinos in the final states are produced with $\BR[\Upsilon(1S)\to \nu\bar \nu]$ $\sim$ $10^{-5}$~\cite{Chang441}. Low mass dark matter (DM) particle ($\chi$), if it exists, should enhance the invisible decays of $\Upsilon(1S)$~\cite{McElrath72}. The $\chi$ can also be produced in interactions of SM particles through the exchange of a $CP$-odd Higgs boson $A^0$, which is part of the Next-to-Minimal Super-symmetric Model (NMSSM)~\cite{Ellwanger491}. Such $A^0$ states in the final states of $\chi\chi$ (invisible), $\mu^+\mu^-$, $\tau^+\tau^-$, and hadrons have been searched for in the radiative decays of $\Upsilon(1S,2S,3S)$ at Belle, BaBar, and CLEO.

Historically, the $\allu$ productions in $pp$ collisions were poorly understood. Several effective QCD models, such as the color-singlet model~\cite{Artoisenet101,Lansberg38}, the color-octet mechanism~\cite{Wang85}, and the color-evaporation model~\cite{Frawley462}, predict different cross sections and polarizations. A precise measurement of $\allu$ production cross sections is thus crucial for distinguishing these models. Thanks to the good momentum resolutions of the detectors, the experiments at the LHC, especially the CMS, ATLAS, and LHCb experiments, have the ability to distinguish the three $\allu$ structures in the dimuon decay channel. The $\allu$ cross sections in $pp$ collisions have been provided by CMS, ATLAS, and LHCb at different C.M. energies, within the complementary rapidity and momentum coverages. The $\allu$ polarizations in $pp$ collisions have also been explored at the LHC.

QCD predicts that strongly interacting matter at a critical temperature undergoes a phase transition to a deconfined state, often referred to as quark-gluon plasma (QGP), in which quarks and gluons are no longer bounded within hadrons in the medium. If QGP is formed in heavy-ion collisions, it is expected to screen the confining potential of heavy quark-antiquark pairs, leading to the melting of charmonia and bottomonia~\cite{Matsui416}. The dissociation of the quarkonium states depends on the temperature of the medium, and is expected to occur sequentially, along the increasing values of their binding energies~\cite{Digal094015}. For example, the loosely bound states $\Upsilon(2S)$ and $\Upsilon(3S)$ are more likely to unbind than the tightly bound state $\Upsilon(1S)$. Up to now, the pattern of QGP in $\allu$ states has been established at CMS, LHCb, and ALICE by observing a sequential suppression of their yields in $PbPb$ and $pPb$ collisions.

In this review, studies on $\allu$ at $e^+e^-$ colliders and the LHC are summarized. With regard to $e^+e^-$ colliders, we discuss the hadronic and radiative decays of $\allu$, especially those for exotic states and NP, and the transitions between the $\allu$ and other $\Upsilon$ states with higher masses. Several effective models of QCD were compared with and challenged by these measurements. The $\allu$ cross sections and polarizations in $pp$ collisions were studied at the LHC. The QGP formation predicted by QCD was probed by comparing the $\allu$ yields from different types of collisions.

\section{$\allu$ studies at $e^+e^-$ colliders}

\subsection{The $\allu$ datasets at $e^+e^-$ colliders}

Table~\ref{t1} summaries the data samples collected at the $\Upsilon(1S,2S,3S)$ resonances in the CLEO, BaBar, and Belle experiments~\cite{Kou1808}. From this table, Belle owns the largest $\Upsilon(1S)$ and $\Upsilon(2S)$ samples, and Babar has the biggest $\Upsilon(3S)$ sample. These datasets provide a solid platform for investigating the $\allu$ decays.

\linespread{1.2}
\begin{table*}[htbp]
\caption{$\allu$ datasets at $e^+e^-$ colliders~\cite{Kou1808}.}
\vspace{0.2cm}
\label{t1}
\centering
\begin{tabular}{c  c  c  c  c  c  c}
\hline
\multirow{2}*{Experiment} &\multicolumn{2}{c}{$\Upsilon(1S)$} & \multicolumn{2}{c}{$\Upsilon(2S)$} & \multicolumn{2}{c}{$\Upsilon(3S)$} \\
& ~~~fb$^{-1}$~~~ & ~~~$10^6$~~~ & ~~~fb$^{-1}$~~~ & ~~~$10^6$~~~ & ~~~fb$^{-1}$~~~ & ~~~$10^6$~~~ \\\hline
CLEO & 1.2 & 21 & 1.2 & 10 & 1.2 & 5 \\
BaBar & - & - & 14 & 99 & 30 & 122 \\
Belle & 6 & 102 & 25 & 158 & 3 & 12 \\\hline
\end{tabular}
\end{table*}

\subsection{Hadronic and radiative decays of $\allu$}

\subsubsection{Study of gluon fragmentation in $\Upsilon(1S,2S,3S)\to ggg$ and $\Upsilon(1S,2S,3S)\to \gamma gg$}

Below the $B\bar B$ threshold, the $\allu$ states decay in the OZI-suppressed manner. To be specific, they can decay via three gluons ($ggg$) or two gluons plus a photon ($gg\gamma$). The ratio of these two decay rates was predicted by perturbative quantum chromodynamics (pQCD)~\cite{Brodsky28}:
$$
R_{\gamma} = \frac{\Gamma_{gg\gamma}}{\Gamma_{ggg}} = \frac{38}{5} q^2_b \frac{\alpha_{em}}{\alpha_{s}} [1+(2.2\pm0.8)\alpha_{s}/\pi], \label{eq:1}
$$
where $q_b=-1/3$ is the electric charge of $b$-quark. Therefore, one can estimate the strong coupling constant $\alpha_s$ according to the formula with $R_{\gamma}$ from experimental measurements.

The exclusive productions of $ggg$ and $gg\gamma$ in the $\allu$ decays were studied by CLEO~\cite{Csorna56,Nemati55,Besson74}, ARGUS~\cite{Albrecht199}, and Crystal Ball~\cite{Bizzeti267}. The recent results of $\allu\to gg\gamma$ were given by CLEO~\cite{Besson74} in the measurements of the direct photon momentum spectrum. The signal yield of $\Upsilon(1S,2S,3S)\to gg\gamma$ is determined in the $x_{\gamma}$ = $p_{\gamma}/E_{\rm beam}$ spectrum after excluding all possible backgrounds, where $p_{\gamma}$ is the momentum of the isolated photon in the $e^+e^-$ C.M. frame, and $E_{\rm beam}$ is the beam energy.
Two main backgrounds are photons that come from initial state radiation (ISR), which are the dominant background at the highest photon energy ($x_\gamma$ $>$ 0.65), while at lower energies ($x_\gamma$ $<$ 0.65) the dominant background comes from photons resulting from $\pi^0$ decays. The ISR background is well simulated by the \textsc{jetset} \oldstylenums{7}$_.$\oldstylenums{4}~\cite{jetset} event generator. A data-driven method is used to estimate the background contribution from $\pi^0$ decays. Continuum processes have been subtracted using an off-resonance data sample. By importing the branching fractions of $\Upsilon(1S,2S,3S)\to ggg$ from PDG~\cite{PDG}, the $R_{\gamma}$ ratios are obtained to be $R_{\gamma}(1S)=(2.70\pm0.01\pm0.13\pm0.24)\%$, $R_{\gamma}(2S)=(3.18\pm0.04\pm0.22\pm0.41)\%$, and $R_{\gamma}(3S)=(2.72\pm0.06\pm0.32\pm0.37)\%$, where the first, second, and third uncertainties are statistical, systematic, and theoretical model dependent~\cite{GS1,GS2,GS3,GS4}, respectively. Hereinafter if there are two and more uncertainties in the formulae, the first one is statistical and the second is systematic.
The above values of $R_{\gamma}(1S)$, $R_{\gamma}(2S)$, and $R_{\gamma}(3S)$ imply the strong coupling constant $\alpha_s$ = $0.1114\pm0.0002\pm0.0029\pm0.0053$, $0.1026\pm0.0007\pm0.0041\pm0.0077$, and $0.113\pm0.001\pm0.0007\pm0.008$ at the $\Upsilon(1S)$, $\Upsilon(2S)$, and $\Upsilon(3S)$ resonances, respectively~\cite{Besson74}.
Considering the strong coupling constant $\alpha_s$ can be written as a function of the QCD scale parameter $\Lambda_{\overline {MS}}$, defined in the modified minimal subtraction scheme~\cite{PDG}, we can determine the value of $\Lambda_{\overline {MS}}$ further.

\subsubsection{Exclusive $\Upsilon(1S,2S)$ decays into light hadrons}

The OZI-suppressed decays of $J/\psi$ and $\psi^{\prime}$ to hadrons proceed via the annihilation of the charm-anticharm pair into three gluons, or two gluons together with a photon. For both cases, pQCD predicts~\cite{Appelquist34,Rujula34}
$$
Q_{\psi}=\frac{\BR_{\psi^{\prime}\to {\rm hadrons}}}{\BR_{J/\psi\to {\rm hadrons}}}=\frac{\BR_{\psi^{\prime}\to e^+e^-}}{\BR_{J/\psi\to e^+e^-}} \approx 12\%,
$$
which is referred to as the ``12\% rule'' and is expected to apply with reasonable accuracy to both inclusive and exclusive decays.
However, the rule was found to be severely violated for $\rho\pi$ and other Vector-Pseudoscalar (VP) and Vector-Tensor (VT) final states~\cite{Franklin51,Bai69}. This is the so-called ``$\rho\pi$ puzzle".
None of the many existing theoretical explanations that have been proposed is able to accommodate all of the measurements reported to date~\cite{Brambilla1534,Brambilla2981,Gu63}. A similar rule can be derived for OZI-suppressed bottomonium decays, in which case we expect
$$
Q_{\Upsilon}=\frac{\BR_{\Upsilon(2S)\to {\rm hadrons}}}{\BR_{\Upsilon(1S)\to {\rm hadrons}}}=\frac{\BR_{\Upsilon(2S)\to e^+e^-}}{\BR_{\Upsilon(1S)\to e^+e^-}} = 0.80\pm0.08.
$$
This rule should hold better than the ``12\% rule" for charmonium decays, since the bottomonium states have higher mass, pQCD and the potential models should be more applicable, as demonstrated in the calculations of the $b\bar b$ meson spectrum.

\begin{figure*}[htbp]
\centering
\includegraphics[width=5.5cm]{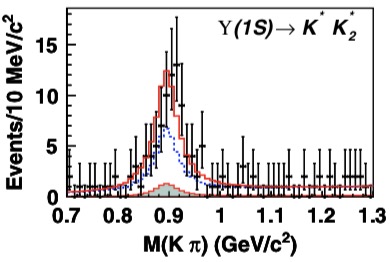}
\hspace{0.15cm}
\includegraphics[width=5.5cm]{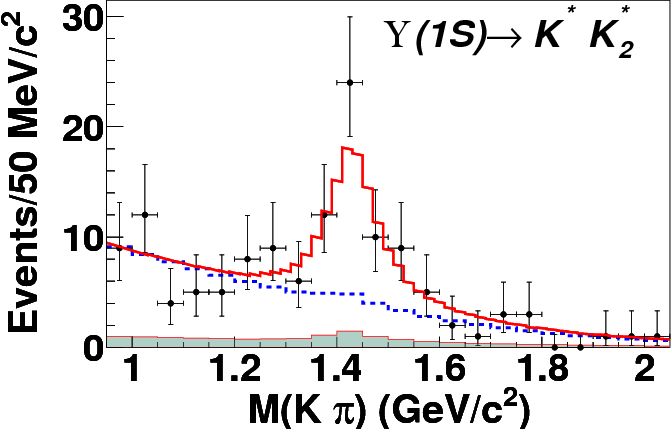}

\includegraphics[width=5.5cm]{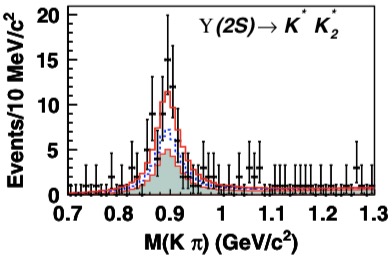}
\hspace{0.15cm}
\includegraphics[width=5.5cm]{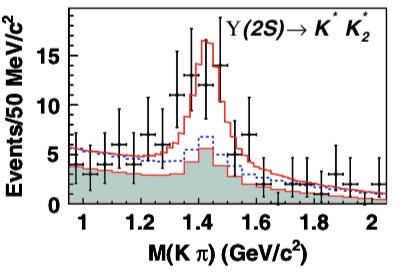}
\caption{The invariant mass distributions for $K^*(892)^0$ candidates and ${\bar K}^*_2(1430)^0$ candidates from $\Upsilon(1S)$ and $\Upsilon(2S)$ decays from Belle~\cite{Shen86}. The open histograms show the results of the two-dimensional fits, the dotted curves show the total background estimates, and the grey histograms are the normalized continuum contributions, which are determined from the data at $\sqrt{s}=10.52$ GeV, and extrapolated down to the $\Upsilon(1S)$ and $\Upsilon(2S)$ data.}\label{SIIA21}
\end{figure*}

\begin{figure*}[htbp]
\centering
\includegraphics[width=5.5cm]{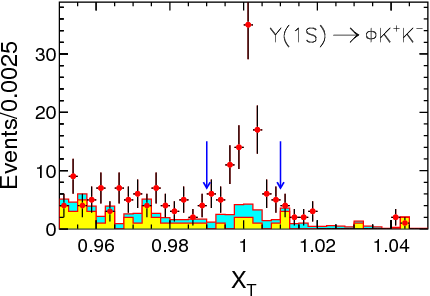}
\hspace{0.15cm}
\includegraphics[width=5.5cm]{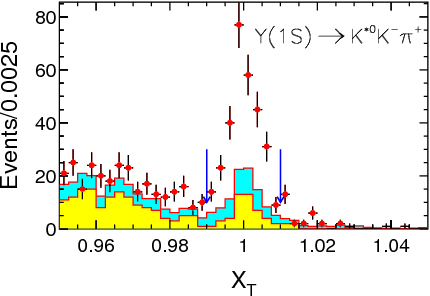}
\hspace{0.15cm}

\includegraphics[width=5.5cm]{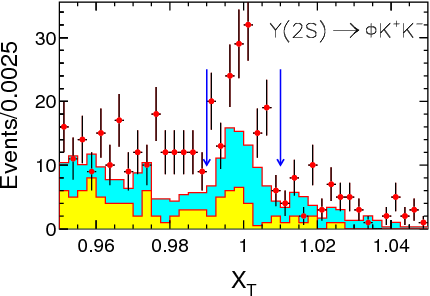}
\hspace{0.15cm}
\includegraphics[width=5.5cm]{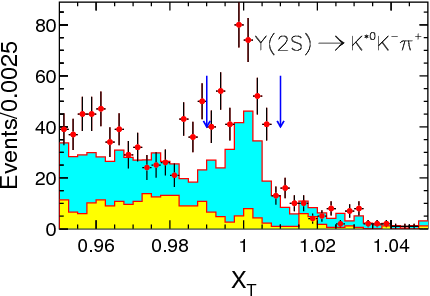}
\hspace{0.15cm}
\caption{Scaled total energy, $X_T$, distributions from $\Upsilon(1S)$ and $\Upsilon(2S)$ decays to $\phi K^+ K^-$ and $K^*(892)^0K^-\pi^+$ from Belle~\cite{Shen86}. The red dots with error bars are from resonance data, the yellow-shaded histograms are from the normalized continuum contributions described in the text, and the cyan-shaded histograms are from the normalized inclusive $\Upsilon(1S)$ and $\Upsilon(2S)$ MC events. The blue arrows show the required signal regions.}\label{SIIA22}
\end{figure*}

\begin{figure*}[htbp]
\centering
\includegraphics[width=5.5cm]{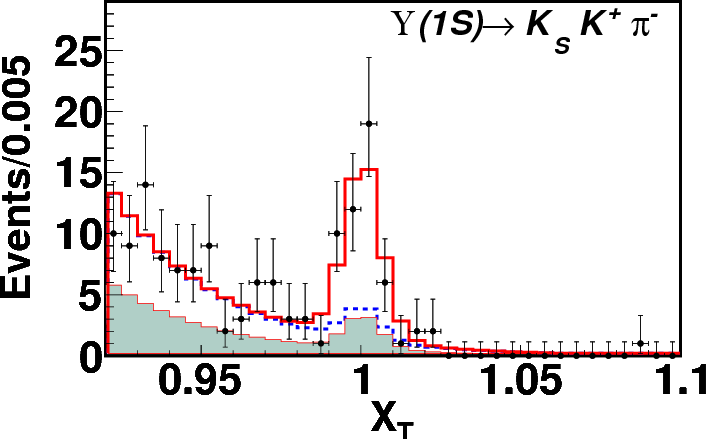}
\hspace{0.15cm}
\includegraphics[width=5.5cm]{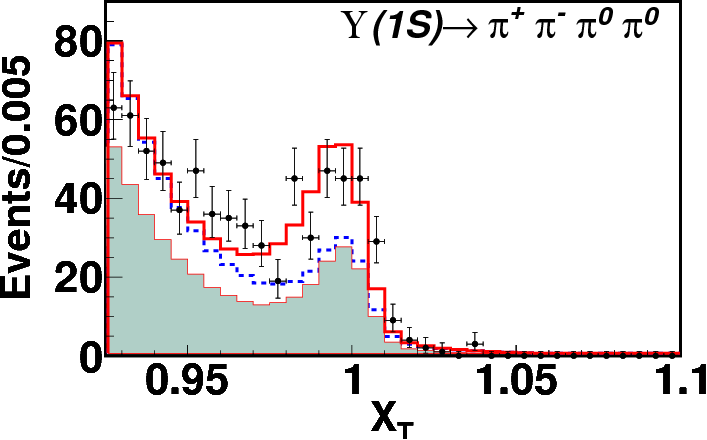}

\includegraphics[width=5.5cm]{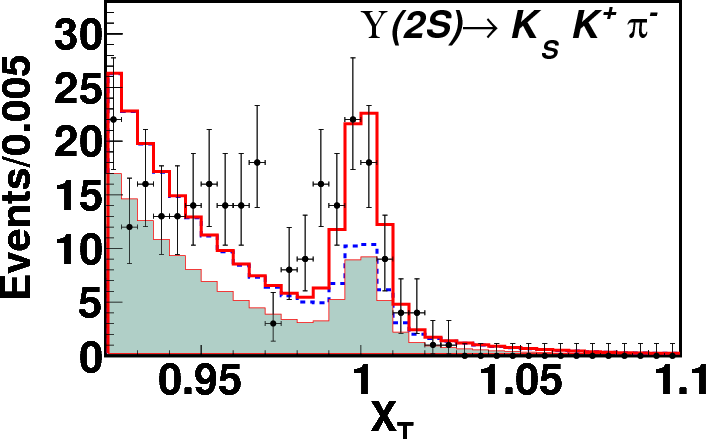}
\hspace{0.15cm}
\includegraphics[width=5.5cm]{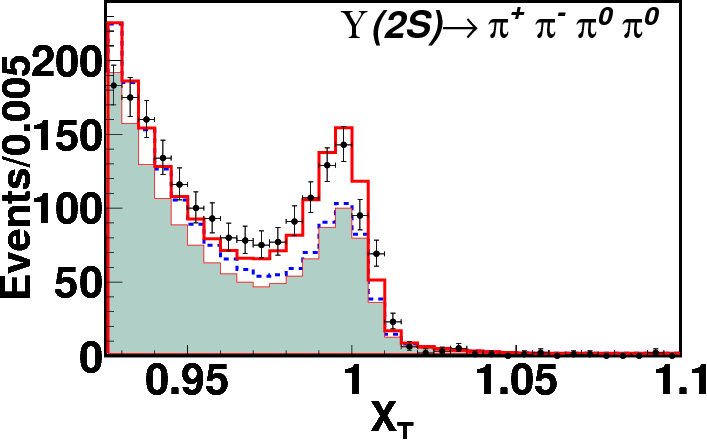}
\caption{The fits to the scaled total energy $X_T$ distributions from $\Upsilon(1S)$ and $\Upsilon(2S)$ decays to $K^0_SK^+\pi^-$ and $\pi^+\pi^-\pi^0\pi^0$ from Belle~\cite{Shen88}. Solid dots with error bars are from resonance data. The open histograms show the best fits, the dashed curves are the total background estimates, and the grey histograms are the normalized continuum background contributions.}\label{SIIA23}
\end{figure*}

To verify the ``80\% rule", Belle measured the exclusive $\Upsilon(1S)$ and $\Upsilon(2S)$ decays into light hadrons, including two-body VT, VP and Axial-vector-Pseudoscalar (AP) final states, three-body final states, and four-body final states~\cite{Shen86,Shen88}. Among these decay modes, the~evidences of both $\Upsilon(1S)$ and $\Upsilon(2S)$ are found in the $K^*(892)^0{\bar K}^*_2(1430)^0$, $\phi K^+ K^-$, $K^*(892)^0K^-\pi^+$, $\pi^+\pi^-\pi^0\pi^0$, and $K^0_SK^+\pi^-$ final states, as demonstrated in Figs.~\ref{SIIA21},~\ref{SIIA22}, and~\ref{SIIA23}. Note that charge-conjugate modes are implied throughout this review.

In the $K^*(892)^0{\bar K}^*_2(1430)^0$ mode, the numbers of signal events are extracted by performing unbinned two-dimensional maximum likelihood fits to the invariant mass distributions for $K^*(892)^0$ candidates and ${\bar K}^*_2(1430)^0$ candidates; they are $42.2\pm9.5$ and $32\pm11$ with statistical significances of 5.4$\sigma$ and 3.3$\sigma$ for $\Upsilon(1S)$ and $\Upsilon(2S)$ decays, respectively.

For three-body and four-body final states, a requirement on the energy conservation variable $X_T=\Sigma_h E_h/\sqrt{s}$ is imposed to extract the signals, where $E_h$ is the energy of the final-state particle $h$ in the $e^+e^-$ C.M. frame, and $\sqrt{s}$ is the C.M. energy. The $X_T$ distributions for $\Upsilon(1S,2S)\to \phi K^+ K^-$, $K^*(892)^0K^-\pi^+$, $K^0_SK^+\pi^-$, and $\pi^+\pi^-\pi^0\pi^0$ are shown in Figs.~\ref{SIIA22} and~\ref{SIIA23}. For $\Upsilon(1S,2S)\to K^0_SK^+\pi^-$ and $\pi^+\pi^-\pi^0\pi^0$, unbinned simultaneous maximum likelihood fits to the $X_T$ distributions are performed to extract the signal and background yields in the $\Upsilon(1S,2S)$ and continuum data samples, as shown by the open histograms in Fig.~\ref{SIIA23}. For $\Upsilon(1S,2S)\to \phi K^+ K^-$ and $K^*(892)^0K^-\pi^+$, unbinned simultaneous maximum likelihood fits to the $K^+K^-$ and $K^+\pi^-$ invariant mass spectra for $\phi$ and $K^*(892)^0$ candidates are applied to extract the signal and background yields after requiring events within the $X_T$ signal range of [0.985, 1.015]. The statistical signal significances for $\Upsilon(1S)$ ($\Upsilon(2S)$) $\to \phi K^+ K^-$, $K^*(892)^0K^-\pi^+$, $K^0_SK^+\pi^-$, and $\pi^+\pi^-\pi^0\pi^0$ are 8.6$\sigma$ (6.5$\sigma$), 11$\sigma$ (6.4$\sigma$), 6.2$\sigma$ (4.0$\sigma$), and 7.1$\sigma$ (7.4$\sigma$), respectively. The branching fractions for these decays are obtained for the first time.

With 1.09 fb$^{-1}$ $\Upsilon(1S)$ and 1.28 fb$^{-1}$ $\Upsilon(2S)$ events of CLEO, the authors in Ref.~\cite{Dobbs86} measured the branching fractions for a number of exclusive decays into different final states consisting of 4 -- 10 light hadrons, pions, kaons, and protons~\cite{Dobbs86}. Significance strength (significance $>$ 2$\sigma$) is found in 17 decay modes for both $\Upsilon(1S)$ and $\Upsilon(2S)$, with branching fractions ranging from 1.3 $\times$ $10^{-5}$ to 109.5 $\times$ $10^{-5}$.

\begin{figure*}[htbp]
\centering
\includegraphics[width=8cm,height=10cm]{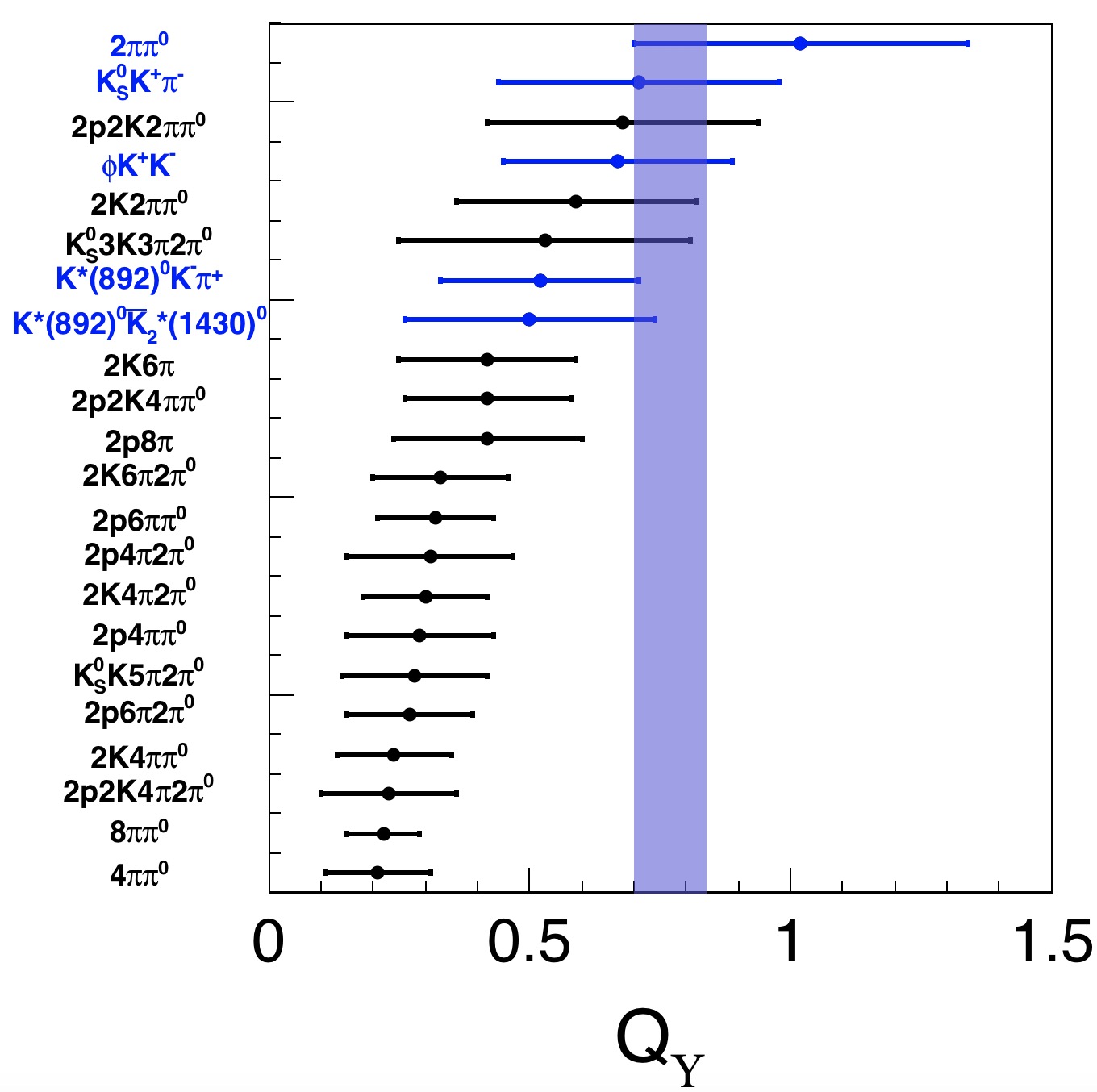}
\caption{The $Q_{\Upsilon}$ values for 22 exclusive light hadron decay modes of $\Upsilon(1S)$ and $\Upsilon(2S)$ from Belle~\cite{Shen86,Shen88} (blue dots with error bars) and measurements using CLEO data samples~\cite{Dobbs86} (black dots with error bars). The band shows the theoretical prediction and its uncertainty.}\label{77rulef}
\end{figure*}

The detailed branching fractions and the ratios $Q_{\Upsilon}$ corresponding to them~\cite{Shen86,Shen88,Dobbs86} are listed in Table~\ref{77rulet}, and displayed in Fig.~\ref{77rulef}. From Fig.~\ref{77rulef}, the $Q_{\Upsilon}$ values measured by Belle~\cite{Shen86,Shen88} are close to the ``80\% rule" line. However, most of the $Q_{\Upsilon}$ values from the measurements using CLEO data~\cite{Dobbs86} are below the ``80\% rule" line. More precise measurements are needed to understand the large discrepancy.

\linespread{1.2}
\begin{table*}[htbp]
\caption{The branching fractions for 22 exclusive light hadron decay modes of the $\Upsilon(1S)$ and $\Upsilon(2S)$ and the corresponding $Q_{\Upsilon}$ values from Belle~\cite{Shen86,Shen88} and measurements using CLEO data~\cite{Dobbs86}. Here, the uncertainties include statistical and systematic uncertainties.}
\vspace{0.2cm}
\label{77rulet}
\centering
\begin{tabular}{c | c | c | c | c}
\hline\hline
\multirow{2}*{Measurements} & \multirow{2}*{Mode} & $\BR(\Upsilon(1S)\to hadrons)$ & $\BR(\Upsilon(2S)\to hadrons)$ & \multirow{2}*{$Q_{\Upsilon}$}  \\
&&($\times$$10^{-5}$) & ($\times$$10^{-5}$)  & \\\hline\hline
\multirow{5}*{from Belle~\cite{Shen86,Shen88}} & $K^*(892)^0{\bar K}^*_2(1430)^0$	&	0.30 	$\pm$	0.08 	&	0.15 	$\pm$	0.06 	&	0.50 	$\pm$	0.24 	\\
& $K^*(892)^0K^-\pi^+$	&	0.44 	$\pm$	0.08 	&	0.23 	$\pm$	0.07 	&	0.52 	$\pm$	0.19 	\\
& $\phi K^+K^-$	&	0.24 	$\pm$	0.05 	&	0.16 	$\pm$	0.04 	&	0.67 	$\pm$	0.22 	\\
& $K^0_SK^+\pi^-$	&	0.16 	$\pm$	0.04 	&	0.11 	$\pm$	0.03 	&	0.71 	$\pm$	0.27 	\\
& $2\pi2\pi^0$	&	1.28 	$\pm$	0.30 	&	1.30 	$\pm$	0.28 	&	1.02 	$\pm$	0.32 	\\\hline\hline
& $4\pi\pi^0$	&	6.10 	$\pm$	0.88 	&	1.29 	$\pm$	0.56 	&	0.21 	$\pm$	0.10 	\\
& $8\pi\pi^0$	&	55.49 	$\pm$	9.16 	&	12.45 	$\pm$	3.36 	&	0.22 	$\pm$	0.07 	\\
& $2p2K4\pi2\pi^0$	&	22.58 	$\pm$	5.22 	&	5.23 	$\pm$	2.68 	&	0.23 	$\pm$	0.13 	\\
& $2K4\pi\pi^0$	&	30.81 	$\pm$	4.97 	&	7.42 	$\pm$	3.19 	&	0.24 	$\pm$	0.11 	\\
& $2p6\pi2\pi^0$	&	49.55 	$\pm$	10.08 	&	13.34 	$\pm$	5.14 	&	0.27 	$\pm$	0.12 	\\
& $K^0_SK5\pi2\pi^0$	&	101.43 	$\pm$	19.63 	&	28.26 	$\pm$	13.19 	&	0.28 	$\pm$	0.14 	\\
& $2p4\pi\pi^0$	&	13.54 	$\pm$	2.27 	&	3.90 	$\pm$	1.76 	&	0.29 	$\pm$	0.14 	\\
& $2K4\pi2\pi^0$	&	61.67 	$\pm$	11.20 	&	18.80 	$\pm$	6.46 	&	0.30 	$\pm$	0.12 	\\
using CLEO data~\cite{Dobbs86}& $2p4\pi2\pi^0$	&	22.68 	$\pm$	4.32 	&	7.12 	$\pm$	3.34 	&	0.31 	$\pm$	0.16 	\\
& $2p6\pi\pi^0$	&	32.82 	$\pm$	5.91 	&	10.58 	$\pm$	3.18 	&	0.32 	$\pm$	0.11 	\\
& $2K6\pi2\pi^0$	&	109.53 	$\pm$	21.84 	&	36.18 	$\pm$	12.36 	&	0.33 	$\pm$	0.13 	\\
& $2p8\pi$	&	7.69 	$\pm$	1.68 	&	3.21 	$\pm$	1.17 	&	0.42 	$\pm$	0.18 	\\
& $2p2K4\pi\pi^0$	&	15.03 	$\pm$	3.07 	&	6.29 	$\pm$	2.10 	&	0.42 	$\pm$	0.16 	\\
& $2K6\pi$	&	13.93 	$\pm$	2.52 	&	5.92 	$\pm$	2.12 	&	0.42 	$\pm$	0.17 	\\
& $K^0_S3K3\pi2\pi^0$	&	29.36 	$\pm$	6.78 	&	15.68 	$\pm$	7.47 	&	0.53 	$\pm$	0.28 	\\
& $2K2\pi\pi^0$	&	5.43 	$\pm$	0.91 	&	3.19 	$\pm$	1.15 	&	0.59 	$\pm$	0.23 	\\
& $2p2K2\pi\pi^0$	&	5.18 	$\pm$	1.04 	&	3.54 	$\pm$	1.12 	&	0.68 	$\pm$	0.26 	\\\hline\hline
\end{tabular}
\end{table*}

\subsubsection{Charmonium and bottomonium productions in $\allu$ decays}

\vspace{0.5cm}
$\bullet$ $\Upsilon(1S,2S)\to charmonium+anything$
\vspace{0.1cm}

Although the $c\bar c$ systems have been studied for decades, their production mechanisms, especially in gluon-rich environments, have not yet been fully understood. Several theoretical papers suggested that $J/\psi$ can be produced abundantly in the $\Upsilon(1S)$ decay via the color-singlet~\cite{Li482} or color-octet mechanism~\cite{Cheung54,Napsuciale57,Trottier320}. In both mechanisms, the branching fraction of $\Upsilon(1S) \to J/\psi + anything$ is predicted to be about a few times $10^{-4}$. However, the momentum distributions given by the two scenarios are significantly different. In the color-octet mechanism, the $J/\psi$ momentum is expected to be accumulated near the kinematic end point. In contrast, the process with color-singlet mechanism inherently results in a soft $J/\psi$ momentum spectrum because of the two additional charm quarks in the final states.

\begin{figure*}[htbp]
\centering
\includegraphics[width=12cm]{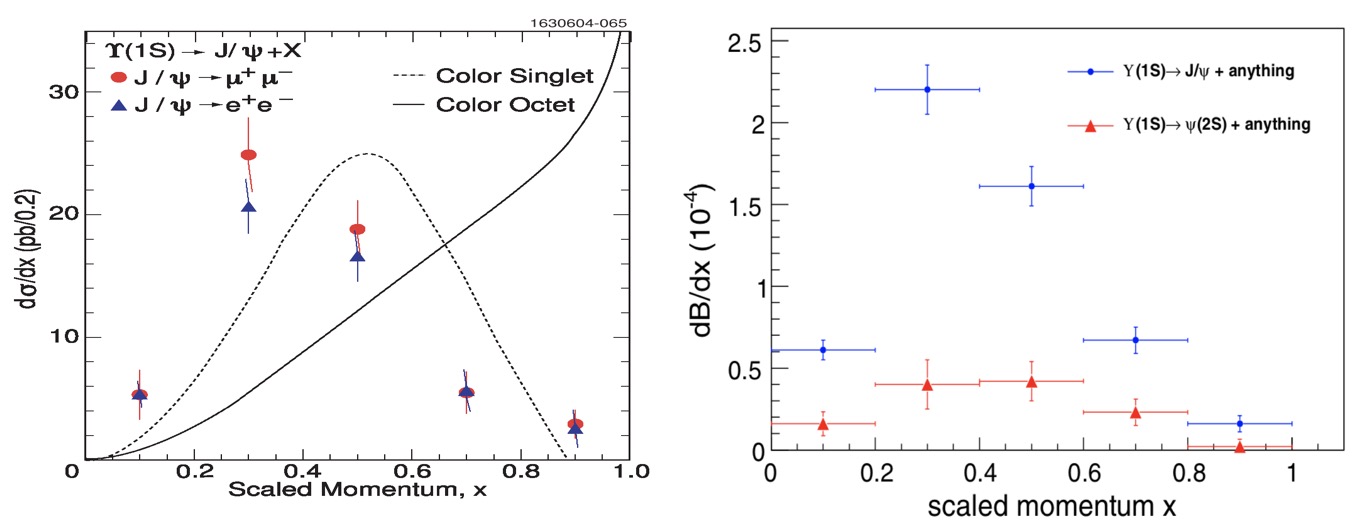}
\put(-315, 90){\large \bf (a)}
\put(-140, 90){\large \bf (b)}
\caption{(a) The differential cross sections in $x$ for $\Upsilon(1S)\to J/\psi+anything$ from CLEO~\cite{Briere70}. (b) The differential branching fractions of $\Upsilon(1S)$ inclusive decays into the $J/\psi$ and $\psi(2S)$ from Belle~\cite{Shen93}.}\label{SIIA31}
\end{figure*}

CLEO analyzed $\Upsilon(1S)\to J/\psi + anything$ using its $\Upsilon(1S)$ data sample~\cite{Briere70}. To determine the differential branching fractions, MC and data samples are partitioned according to the scaled momentum $x=p^*_{J/\psi}/(\frac{1}{2\sqrt{s}}\times(s-m^2_{J/\psi}))$, where $p^*_{J/\psi}$ is the momentum of the $J/\psi$ candidate in the $e^+e^-$ C.M. frame, and $m_{J/\psi}$ is the $J/\psi$ nominal mass~\cite{PDG}. The value of $\frac{1}{2\sqrt{s}}\times(s-m^2_{J/\psi})$ is the value of $p^*_{J/\psi}$ for the case where the $J/\psi$ candidate recoils against a massless particle. The branching fraction for $\Upsilon(1S)\to J/\psi + anything$ in the whole momentum region is measured to be $(6.4\pm0.4\pm0.6)\times10^{-4}$, which is consistent with the predictions of both the color-octet and color-singlet mechanisms~\cite{Li482,Cheung54,Napsuciale57,Trottier320}.
The feed-down contributions of $\psi^{\prime}$, $\chi_{c1}$ and $\chi_{c2}$ to $\Upsilon(1S)\to J/\psi+anything$ are $\BR(\Upsilon(1S)\to \psi^{\prime}/\chi_{c1}/\chi_{c2}+anything)\BR(\psi^{\prime}/\chi_{c1}/\chi_{c2}\to J/\psi+anything)/\BR(\Upsilon(1S)\to J/\psi+anything)$ = $(24\pm6\pm5)\%/(11\pm3\pm2)\%/(10\pm2\pm2)\%$, which are larger than the predictions in either the color-singlet~\cite{Li482} or color-octet model~\cite{Cheung54}.
The differential cross sections in $x$ for $\Upsilon(1S)\to J/\psi+anything$ from CLEO is shown in Fig.~\ref{SIIA31}(a)~\cite{Briere70}. With a larger $\Upsilon(1S)$ data sample, Belle reported $\BR(\Upsilon(1S)\to J/\psi+anything)=(5.25\pm0.13\pm0.25)\times10^{-4}$ and $\BR(\Upsilon(1S)\to \psi^{\prime}+anything)=(1.23\pm0.17\pm0.11)\times10^{-4}$, with substantially improved precisions~\cite{Shen93}. The differential branching fractions of $\Upsilon(1S)$ inclusive decays into $J/\psi$ and $\psi^{\prime}$ are shown in Fig.~\ref{SIIA31}(b). From Fig.~\ref{SIIA31}, the scaled momentum spectra for $\Upsilon(1S)\to J/\psi+anything$ from both experiments are relatively soft, peaking around $x\sim0.3$, which favor the expectation of the color-singlet process~\cite{Li482}, and are in sharp contrast to the prediction of the color-octet model~\cite{Cheung54,Napsuciale57}. The study of $\Upsilon(1S)\to charmonium+anything$ therefore provides a powerful platform for distinguishing the roles of color-singlet versus color-octet mechanisms.

Besides these charmonium productions in $\Upsilon(1S)$ decays, Belle also studied the $\chi_{c1}$ and $\chi_{c2}$ productions in $\Upsilon(2S)$ decays, and corresponding branching fractions $\BR(\Upsilon(2S)\to \chi_{c1}+anything)$ and $\BR(\Upsilon(2S)\to \chi_{c2}+anything)$ were measured to be $(2.24\pm0.44\pm0.20)\times10^{-4}$ and $(2.28\pm0.73\pm0.34)\times10^{-4}$~\cite{Jia95}, respectively. In addition, it is worth mentioning that, similar to $\Upsilon(1S)\to \chi_{c1}+anything$, the scaled momentum spectrum for $\Upsilon(2S)\to \chi_{c1}+anything$ also peaks around $x\sim0.3$~\cite{Jia95}.

\vspace{0.5cm}
$\bullet$ $\Upsilon(1S) \to$ {\it double charmonia}
\vspace{0.1cm}

The double charmonium production at $B$-factories is still a matter of debate. The cross sections of the processes $e^+e^-\to J/\psi\eta_c$, $J/\psi\eta_c(2S)$, $\psi^{\prime}\eta_c$, $\psi^{\prime}\eta_c(2S)$, $J/\psi\chi_{c0}$, and $\psi^{\prime}\chi_{c0}$ measured by Belle~\cite{Abe89,Abe70} and BaBar~\cite{Aubert72} exceeded the nonrelativistic quantum chromodynamics (NRQCD) calculations by approximately an order of magnitude~\cite{Braaten67,Liu557,Ma70,Liu69,Bondar612,Liu77}.
Taking into account the next-to-leading order (NLO) correction to NRQCD, the discrepancy between theory and experiment can be largely removed~\cite{Zhang092001,Zhang092003}.~Inspired by the unexpectedly high double-charmonium production in $e^+e^-$ annihilation, interest has turned to the double charmonium states produced in bottomonium decays. Comprehensive studies of the exclusive decay of $\Upsilon(1S,2S,3S)$ into double charmonia have been performed using the NRQCD factorization approach~\cite{Jia76,Xu87}. The branching fractions are predicted to be of order $10^{-6}$ for $\Upsilon(1S,2S,3S)\to J/\psi(\psi^{\prime})+\eta_c(\eta_c(2S))$~\cite{Jia76}. For the $J/\psi+\chi_{c0}/\chi_{c1}/\chi_{c2}$ decay modes, the branching fractions are calculated at the lowest order; that of $\Upsilon(1S,2S,3S)\to J/\psi+\chi_{c1}$ is the largest, about a few times $10^{-6}$, while that of $J/\psi+\chi_{c2}$ is only of order $10^{-7}$~\cite{Xu87}.

\begin{figure}[htbp]
\centering
\includegraphics[width=6cm]{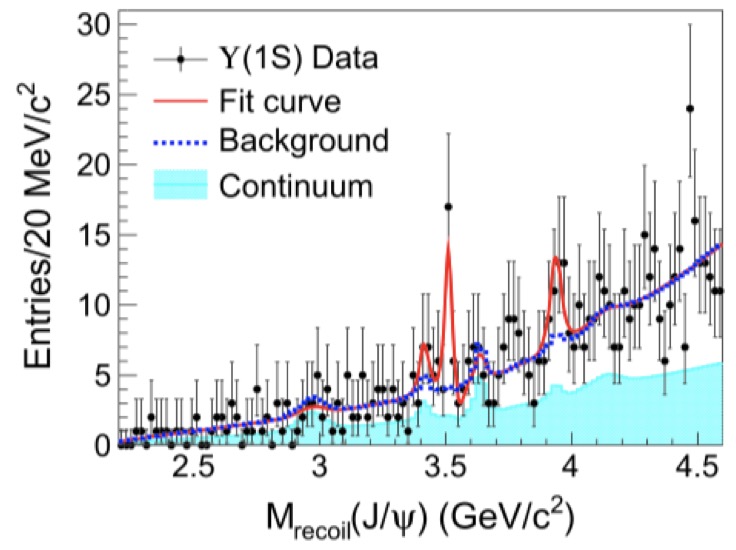}
\caption{The recoil mass spectrum against $J/\psi$ in $\Upsilon(1S)$ decays from Belle~\cite{Yang90}. The red solid curve is the nominal fit and the blue-dashed curve shows the total background. The fitted normalized continuum contribution is represented by the cyan-shaded histogram.}\label{SIIA32}
\end{figure}

Belle searched for double charmonia $J/\psi(\psi^{\prime})+X$, where $X$ is the $\eta_c$, $\chi_{cJ}$ ($J$ = 0, 1, 2), or $\eta_c(2S)$, in $\Upsilon(1S)$ and $\Upsilon(2S)$ decays~\cite{Yang90}. To increase the signal detection efficiencies, only the $J/\psi$ or $\psi^{\prime}$ candidate is fully reconstructed, and the other charmonium state $X$ is searched for in the recoil mass distribution of the fully reconstructed $J/\psi$ or $\psi^{\prime}$. The evidence for $\Upsilon(1S)\to J/\psi + \chi_{c1}$ was found with a signal significance of 4.6$\sigma$, as indicated by the dots with error bars in Fig.~\ref{SIIA32}. The measured branching fraction is $(3.90\pm1.21\pm0.23)\times10^{-6}$. For other cases, considering the significances are less than $3\sigma$, the 90\% confidence level (C.L.) upper limits on the branching fractions are determined; they are mostly at the level of $10^{-6}$. The results are basically consistent with the theoretical calculations using the NRQCD factorization approach~\cite{Jia76,Xu87}.

$\bullet$ $\Upsilon(1S,2S)\to \gamma +  charmonium$
\vspace{0.1cm}

\begin{figure*}[htbp]
\centering
\includegraphics[width=12cm]{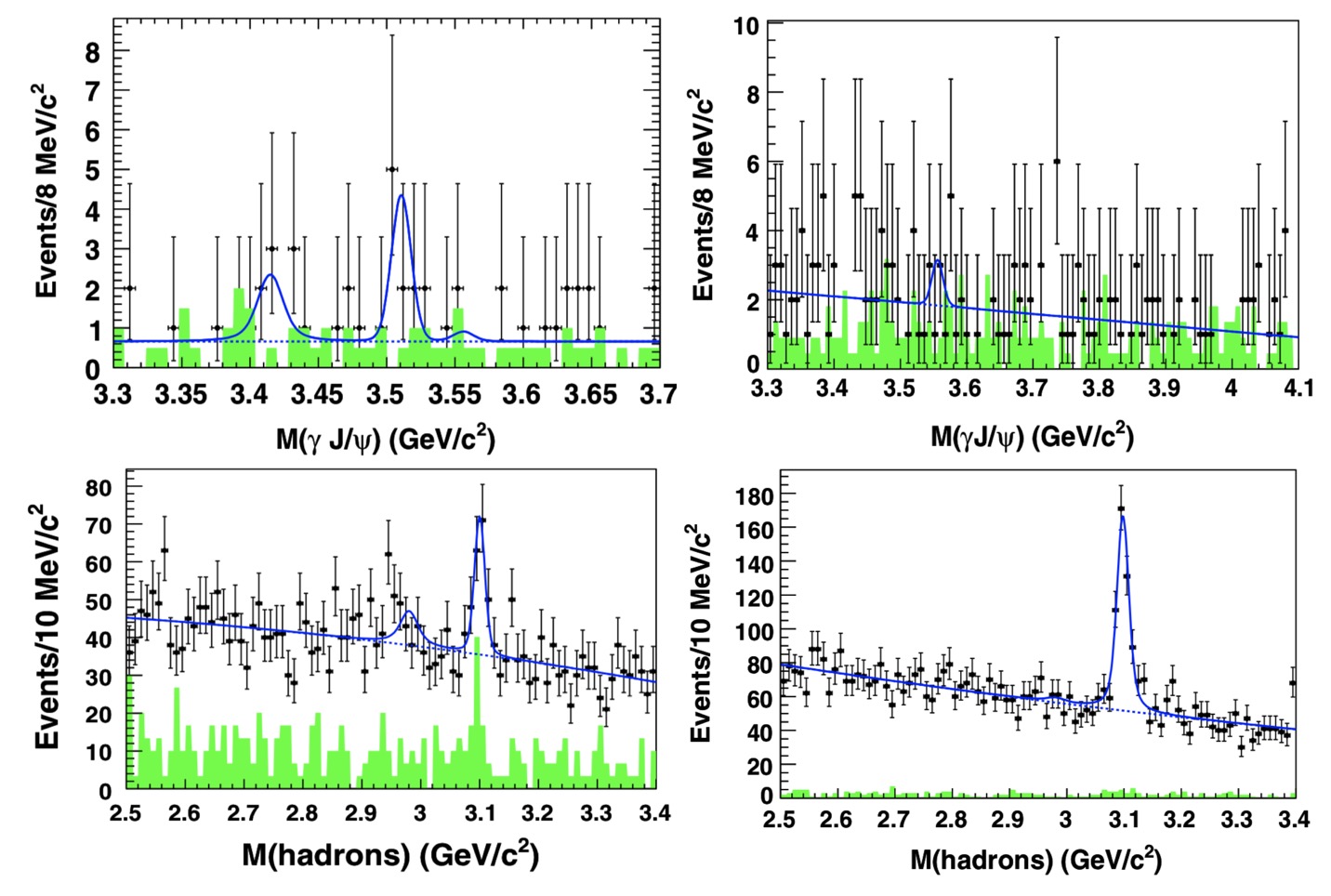}
\put(-195, 210){\bf (a)}
\put(-30, 210){\bf (b)}
\put(-195, 95){\bf (c)}
\put(-30, 95){\bf (d)}
\caption{The $\gamma J/\psi$ invariant mass distributions in the (a) $\Upsilon(1S)$ and (b) $\Upsilon(2S)$ decays, and invariant mass spectra for $\eta_c$ candidates in the (c) $\Upsilon(1S)$ and (d) $\Upsilon(2S)$ decays from Belle~\cite{Shen82,Wang84}. The shaded histograms in (a) and (b) are from the normalized $J/\psi$ sidebands, and the shaded histograms (not normalized) in (c) and (d) are from the continuum contributions.}\label{SIIA33}
\end{figure*}

The $\chi_{cJ}$ ($J$ = 0, 1, 2) and $\eta_c$ productions in $\Upsilon(1S,2S)$ radiative decays were studied by Belle~\cite{Shen82,Wang84}. The $\chi_{cJ}$ states are reconstructed via their $E1$ transitions to $J/\psi$. The $\eta_c$ is reconstructed using the following five hadronic decay modes: $K^0_SK^+\pi^-$, $\pi^+\pi^-K^+K^-$, $2(K^+K^-)$, $2(\pi^+\pi^-)$, and $3(\pi^+\pi^-)$. To suppress the ISR background $e^+e^-\to \gamma_{\rm ISR}\psi^{\prime}\to \gamma_{\rm ISR}\gamma_{\rm miss}\chi_{cJ}(\to \gamma J/\psi)$ in $\Upsilon(1S,2S)\to \gamma + \chi_{cJ}$, the square of the missing mass of the photon from $\chi_{cJ}$ and lepton pair is required to be between $-0.5$ GeV$^2$/$c^{4}$ and $0.5$ GeV$^2$/$c^{4}$ since this background has at least two missing photons ($\gamma_{\rm ISR}$ and $\gamma_{\rm miss}$) and the missing mass tends to be large. Figure~\ref{SIIA33} shows the $\gamma J/\psi$ invariant mass distributions and invariant mass spectra for $\eta_c$ candidates in the $\Upsilon(1S,2S)$ data samples. No significant $\chi_{cJ}$ or $\eta_c$ signal is observed in the analyses. The upper limits at 90\% C.L. on the branching fractions for $\Upsilon(1S,2S)\to \gamma \chi_{cJ}$ are within a range of (3.6-100)$\times10^{-6}$ and for $\Upsilon(1S,2S)\to \gamma \eta_c$ within (2.7-5.7)$\times10^{-6}$, which are consistent with the NRQCD predictions~\cite{Gao0701009}.

\begin{figure}[htbp]
\centering
\includegraphics[width=7cm]{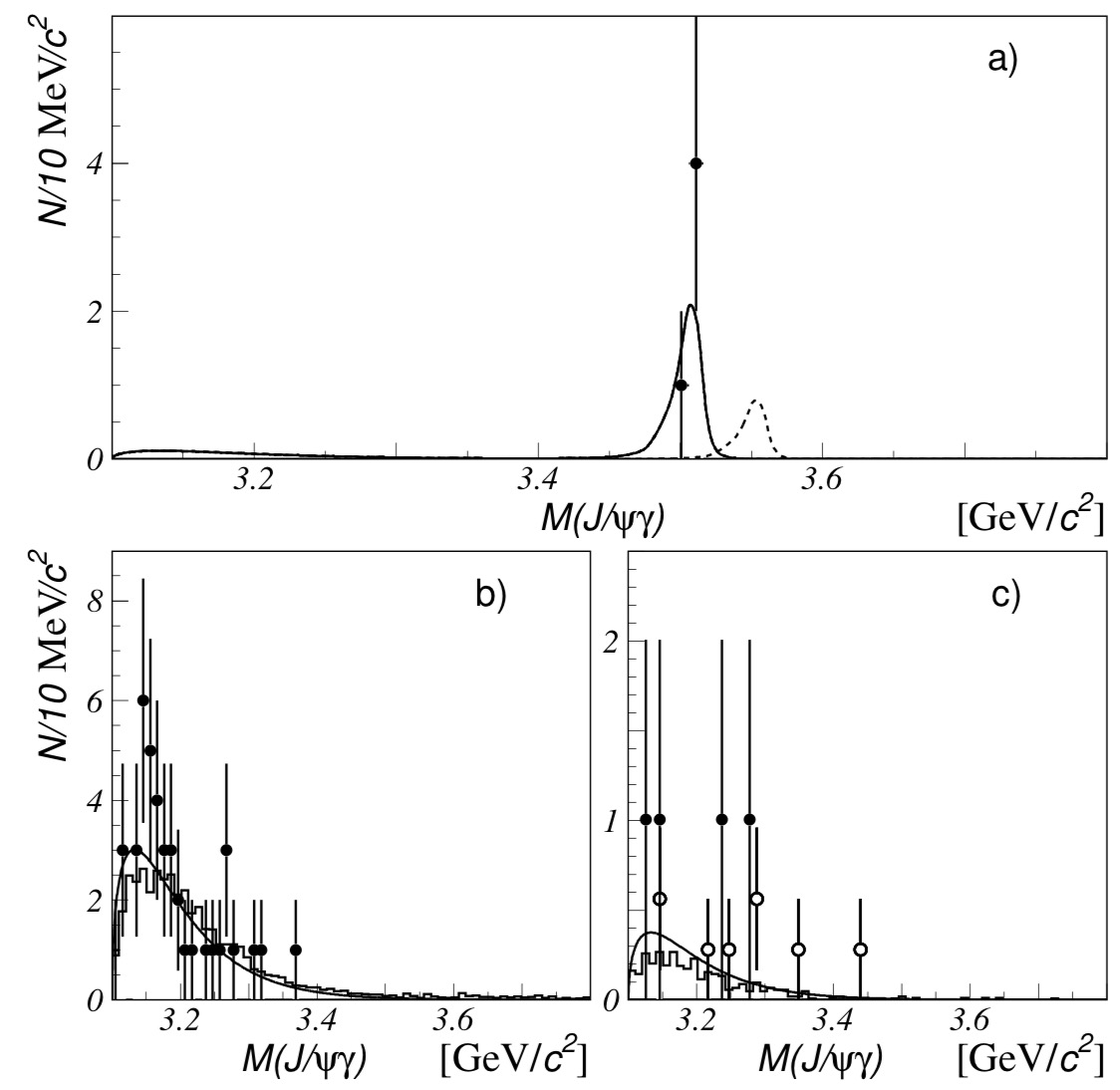}
\caption{The $J/\psi\gamma$ invariant mass spectrum in the $\Upsilon(1S)$ data from Belle~\cite{Katrenko}: a) $M_{\rm rec}$ (the mass recoiling against a pion pair) signal window, b) 20 times wider $J/\psi$ mass sidebands, c) 20 times wider $M_{\rm rec}$ mass sidebands. Histograms are the background expectation from the MC simulation from: b) $\Upsilon(1S)\to \gamma_{\rm FSR}\mu^+\mu^-$, c) $e^+e^-\to \gamma_{\rm ISR}\psi^{\prime}$. The solid lines show the result of the simultaneous fit to all these distributions. The dotted line in a) shows the upper limit on the $\chi_{c2}$ signal yield at 90\% C.L.}\label{SIIA3add1}
\end{figure}

Very recently, Belle reported the first observation of the radiative decay of $\Upsilon(1S)$ into a charmonium state, i.e., $\Upsilon(1S) \to \gamma \chi_{c1}$~\cite{Katrenko}. Unlike the previous Belle analysis using the $\Upsilon(1S)$ data sample~\cite{Shen82}, the authors use the $\Upsilon(2S)$ data sample and tag $\Upsilon(1S)$ via the $\Upsilon(2S)\to\Upsilon(1S)\pi^+\pi^-$ transition. Although the number of tagged $\Upsilon(1S)$ events is several times smaller than the number of directly produced $\Upsilon(1S)$ events used in the previous analysis, the tagging procedure drastically suppresses the backgrounds, especially those from the processes with ISR or final-state radiation (FSR), which have an event topology similar to that of the signal. Moreover, two extra pion tracks increase a trigger efficiency for low-multiplicity final states of the charmonium decay. In order to estimate the statistical significance of the observed signal, a simultaneous unbinned likelihood fit to $\gamma J/\psi$ mass spectra in $M_{\rm rec}$ (the mass recoiling against a pion pair) signal, and $J/\psi$ and $M_{\rm rec}$ sidebands regions is performed, as illustrated in Fig.~\ref{SIIA3add1}. The fit yields the number of signal events to be $5.0^{+2.5}_{-1.9}$, and the estimated background contribution in the signal region is less than 0.1. The significance of the $\chi_{c1}$ signal is 7.5$\sigma$. Finally, the branching fraction is calculated to be $\BR(\Upsilon(1S)\to\gamma\chi_{c1})$ = $(4.7^{+2.4}_{-1.8}$$^{+0.4}_{-0.5})$ $\times$ $10^{-5}$. The obtained result is slightly higher than the previous upper limit~\cite{Shen82} and much higher than the theoretical expectation~\cite{Gao0701009}. However, the recent observation of production in the process $e^+e^-\to \gamma\chi_{c1}$ with a large cross section~\cite{Jia092015} perhaps indicates a similarity of the mechanism of $\chi_{c1}$ formation from the initial vector state with emission of a photon.

\vspace{0.5cm}
$\bullet$ $\Upsilon(1S,2S,3S)\to \gamma + bottomonium$
\vspace{0.1cm}

\begin{figure*}[htbp]
\centering
\includegraphics[width=13cm]{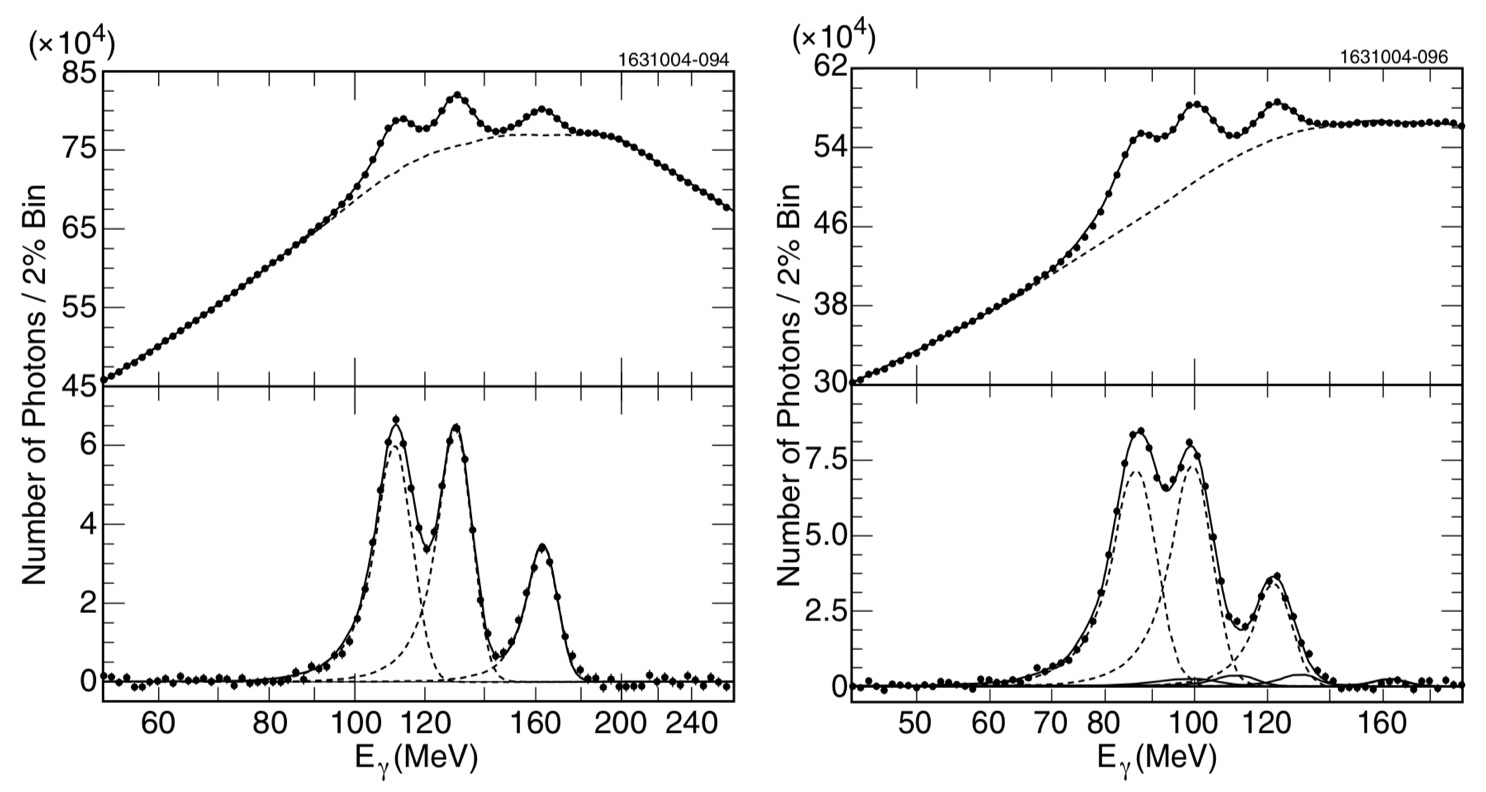}
\put(-335, 163){\large \bf (a)}
\put(-150, 163){\large \bf (b)}
\caption{The energy spectra of photons in (a) $\Upsilon(2S)$ and (b) $\Upsilon(3S)$ decays from CLEO~\cite{Artuso94}. The dots represent the data (top plot) or the data after subtracting the backgrounds (bottom plot).}\label{SIIA34}
\end{figure*}

Long-lived $b\bar b$ states are especially suited for testing lattice QCD calculations~\cite{Davies92}, potential models~\cite{Besson43,Eichten49,Mutuk2018}, and effective field theories~\cite{Brambilla094005,Pineda074024,Brambilla054005,Segovia074011}.
Electron-positron colliders can directly produce the narrow $S$-wave states $\Upsilon(2S)$ and $\Upsilon(3S)$, whose radiative decays provide access to the triplet $P$-wave states $\chi_{bJ}(1P)$ and $\chi_{bJ}(2P)$ with $J=$ 0, 1, 2. The precise measurements for $\Upsilon(2S)\to\gamma\chi_{bJ}(1P)$ and $\Upsilon(3S)\to\gamma\chi_{bJ}(2P)$ were performed by CLEO~\cite{Artuso94}. The signals were searched for in the inclusive photon spectra in decays of these narrow resonances. The energy spectra of photons in $\Upsilon(2S)$ and $\Upsilon(3S)$ decays are shown in Fig.~\ref{SIIA34}, where the bottom shows the distributions in data after subtracting the backgrounds. Three peaks are obvious, indicated as $\chi_{b2}(1P)/\chi_{b1}(1P)/\chi_{b0}(1P)$ and $\chi_{b2}(2P)/\chi_{b1}(2P)/\chi_{b0}(2P)$ (from left to right), in $\Upsilon(2S)$ and $\Upsilon(3S)$ decays. The measured branching fractions for $\Upsilon(2S)\to\gamma\chi_{bJ}(1P)$ and $\Upsilon(3S)\to\gamma\chi_{bJ}(2P)$ ($J$ = 0, 1, 2) transitions are listed in Table~\ref{YtogchibJ}. These branching fractions are basically consistent with previous measurements~\cite{Han49,Klopfenstein51,Heintz46,Nernst54,Albrecht160,Haas52,Morrison67,Edwards59}.
It is worth mentioning that the transition $\Upsilon(3S)\to \gamma\chi_{bJ}(1P)$ with $J=$ 0, 1, 2 was also studied in Refs.~\cite{Artuso94,Heintz46,Kornicer83,Lees072002}. In comparison with the $\Upsilon(nS)\to \gamma\chi_{bJ}$ $[(n-1)P]$ for $n$ = 2 and 3, the measured branching fraction of $\Upsilon(3S)\to\gamma\chi_{bJ}(1P)$ decreases by one order of magnitude.

\linespread{1.2}
\begin{table}[htbp]
\caption{The branching fractions for $\Upsilon(2S)\to\chi_{bJ}(1P)$ and $\Upsilon(3S)\to\chi_{bJ}(2P)$ ($J$ = 0, 1, 2) transitions from CLEO~\cite{Artuso94}.}
\vspace{0.2cm}
\label{YtogchibJ}
\centering
\begin{tabular}{c c}
\hline\hline
$\BR(\Upsilon(2S)\to\chi_{b0}(1P))$ & $(3.75\pm0.12\pm0.47)\%$\\
$\BR(\Upsilon(2S)\to\chi_{b1}(1P))$ & $(6.93\pm0.12\pm0.41)\%$\\
$\BR(\Upsilon(2S)\to\chi_{b2}(1P))$ & $(7.24\pm0.11\pm0.40)\%$\\\hline
$\BR(\Upsilon(3S)\to\chi_{b0}(2P))$ & $(6.77\pm0.20\pm0.65)\%$\\
$\BR(\Upsilon(3S)\to\chi_{b1}(2P))$ & $(14.54\pm0.18\pm0.73)\%$\\
$\BR(\Upsilon(3S)\to\chi_{b2}(2P))$ & $(15.79\pm0.17\pm0.73)\%$\\\hline\hline
\end{tabular}
\end{table}

With the substantial productions of $\chi_{bJ}$ states in the $\Upsilon(2S,3S)$ radiative decays, the exclusive decays of the $\chi_{bJ}$ states into light hadron final states and double charmonia can be studied further.
CLEO presented the observations of 14 exclusive final states of light hadrons in both $\chi_{bJ}(1P)$ and $\chi_{bJ}(2P)$ decays~\cite{Asner78}. Later, Belle used 74 hadronic final states to reconstruct $\chi_{bJ}(1P)$ states, where 41 modes are observed with at least five standard deviation significance~\cite{Abdesselam}. These measurements enriched the understanding of the $\chi_{bJ}$ states. Belle searched  for the first time for double charmonium decays of $\chi_{bJ}$ states ($\Upsilon(2S)\to \gamma\chi_{bJ}, \chi_{bJ}\to J/\psi J/\psi, J/\psi\psi^{\prime}, \psi^{\prime}\psi^{\prime}$)~\cite{Shen85}. No significant $\chi_{bJ}$ signal is observed in the double charmonium mass spectra. The upper limits at 90\% C.L. on the branching fractions of $\chi_{bJ}$ decays were determined to be at the level of 10$^{-5}$, except for $\BR(\chi_{c0}\to J/\psi \psi^{\prime})$ $<$ $1.2\times10^{-4}$.
These upper limits are much lower than the central values predicted by the light cone formalism~\cite{Braguta80,Braguta73} and pQCD calculation~\cite{Braguta72}, but are consistent with calculations using the NRQCD factorization approach~\cite{Zhang84,Sang84}.

\begin{figure}[htbp]
\centering
\includegraphics[width=4.8cm]{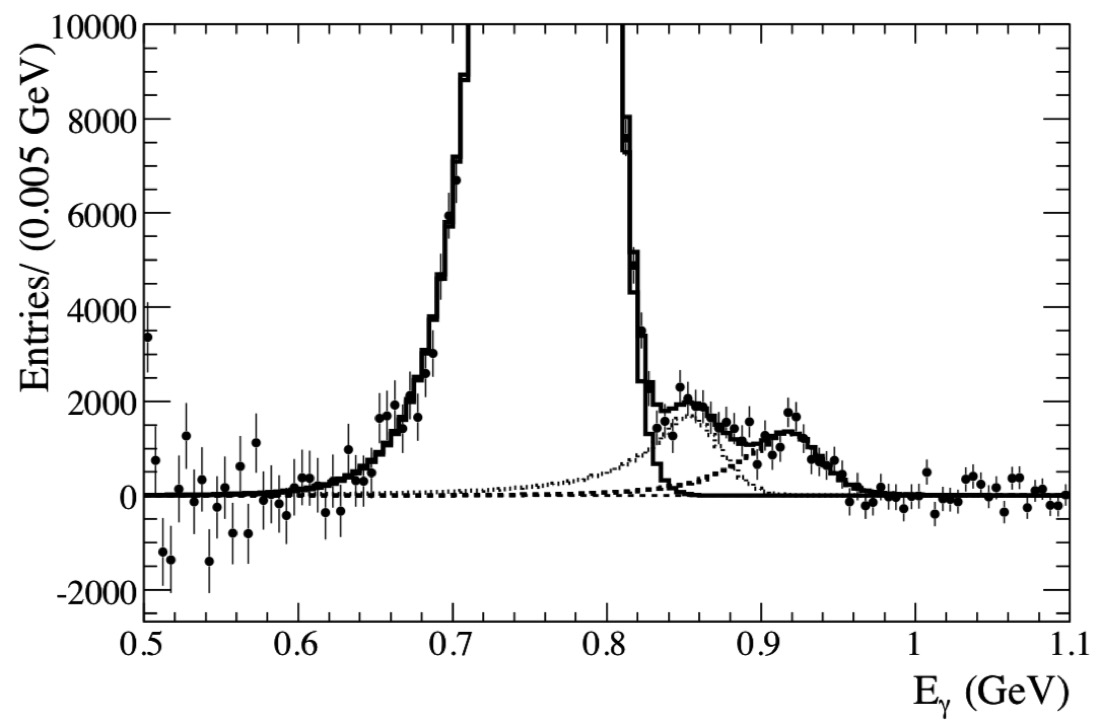}
\caption{$\Upsilon(3S)$ inclusive photon spectrum after subtracting the non-peaking background from BaBar~\cite{Aubert101}. The dots with error bars show the inclusive photon spectrum after subtracting the non-peaking background, and the solid, dotted, and dashed lines show the $\chi_{bJ}(2P)$ peak, ISR $\Upsilon(1S)$, and $\eta_b(1S)$ signal.}\label{SIIA35}
\end{figure}

Besides the triplet $P$-wave states $\chi_{bJ}(1P)$ and $\chi_{bJ}(2P)$, another interest in $\Upsilon$ radiative decays is to search for the spin-singlet pseudoscalar partner --- $\eta_b(nS)$ state. Measurement of the hyperfine mass splittings between the triplet and singlet quarkonium states is critical for understanding the role of spin-spin interactions in quarkonium models and for testing QCD calculations. BaBar reported the first observation of $\eta_b(1S)$ in the radiative decay of $\Upsilon(3S)$~\cite{Aubert101}. The signal of $\Upsilon(3S)\to \gamma\eta_b(1S)$ is extracted from a fit to the inclusive photon energy spectrum in the $e^+e^-$ C.M. frame, as shown in Fig.~\ref{SIIA35}. A peak is observed at $E_{\gamma}$ = ($921.2^{+2.1}_{-2.8}\pm2.4$) MeV with a significance of 10$\sigma$, which corresponds to an $\eta_b(1S)$ mass of ($9388.9^{+3.1}_{-2.3}\pm2.7$) MeV/$c^2$. The branching fraction for this radiative $\Upsilon(3S)$ decay is obtained to be $(4.8\pm0.5\pm1.2)\times 10^{-4}$.

In the $\Upsilon(2S)$ radiative decay, the searches for $\eta_b(1S)$ were also performed by BaBar~\cite{Aubert161801} and Belle~\cite{Fulsom121}.
The first observation of $\Upsilon(2S)\to \gamma\eta_b(1S)$ was reported by Belle~\cite{Fulsom121}. The branching fraction for $\Upsilon(2S)\to \gamma\eta_b(1S)$ is $(6.1^{+0.6}_{-0.7}$$^{+0.9}_{-0.6})\times 10^{-4}$~\cite{Fulsom121}. It is compatible with the BaBar result $(3.9\pm1.1^{+1.1}_{-0.9})\times 10^{-4}$~\cite{Aubert161801} and also in agreement with the recent lattice NRQCD calculation $(5.4\pm1.8)\times10^{-4}$~\cite{Hughes92}.

\begin{figure*}[htbp]
\centering
\includegraphics[width=12cm]{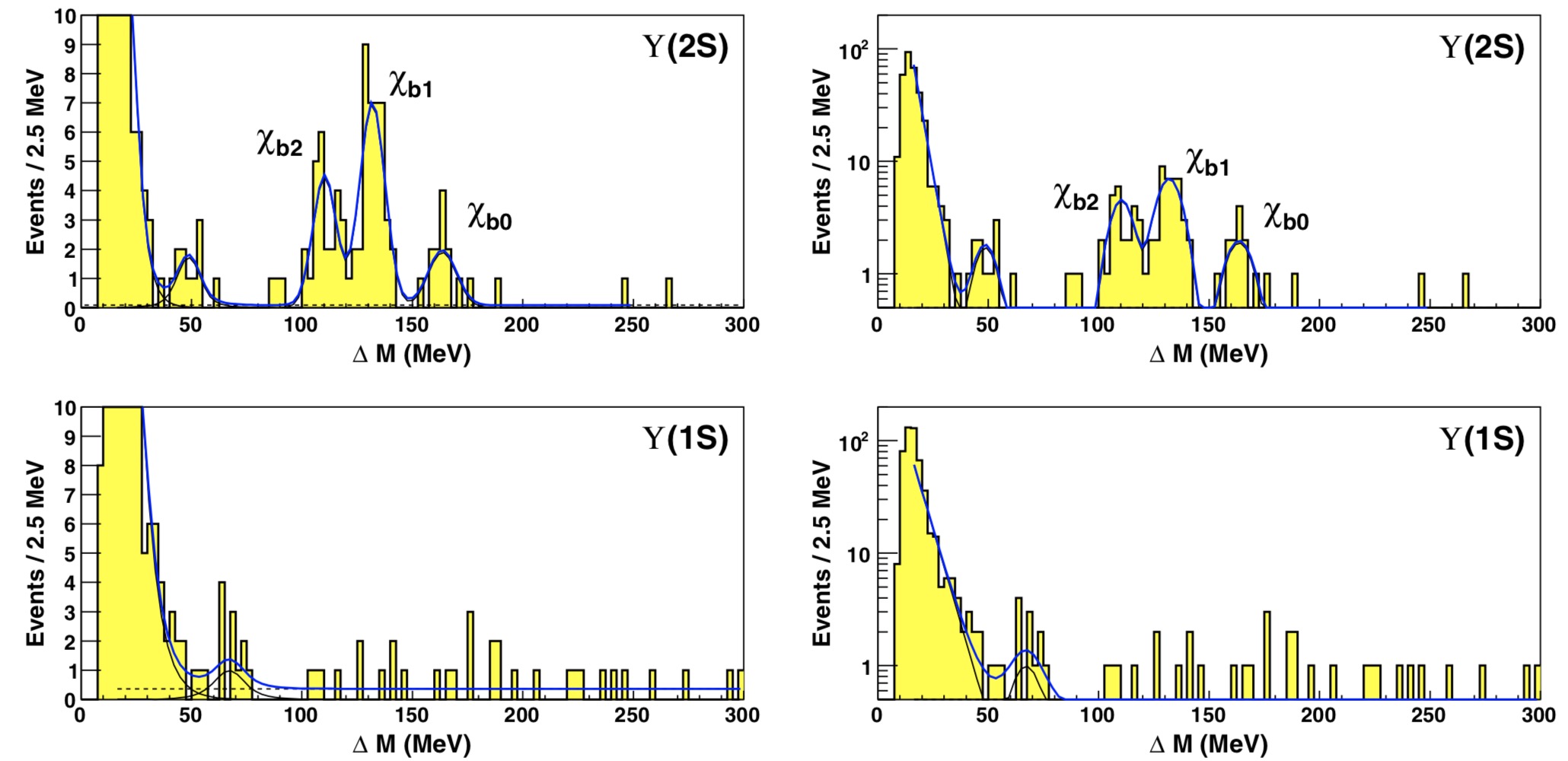}
\caption{Distributions of $\Delta M\equiv m_{\Upsilon(1S,2S)}-M(X_{b{\bar b}})$ in $\Upsilon(2S)$ data (top row) and $\Upsilon(1S)$ data (bottom row) are shown with both linear (left column) and logarithmic (right column) scales using CLEO data~\cite{Dobbs109}.}\label{SIIA36}
\end{figure*}

In addition to the searches for the $\eta_b(1S)$ state in the inclusive radiative transitions of $\Upsilon(2S,3S)$~\cite{Aubert101,Aubert161801,Fulsom121}, the exclusive decays of $\eta_b(1S)$ and $\eta_b(2S)$ to 26 different hadronic final states were also studied~\cite{Dobbs109}. The cascade decays $\Upsilon(1S,2S)\to \gamma\eta_b(nS)$, $\eta_b(nS)\to X_{b{\bar b}}~(X_{b{\bar b}}=4,6,8,10~\pi^{\pm},K^{\pm},p/\bar p)$ were studied in Ref.~\cite{Dobbs109} using CLEO data samples. Figure~\ref{SIIA36} shows the related distributions of $\Delta M\equiv M(X_{b{\bar b}}\gamma)-M(X_{b{\bar b}})$. The enhancement around 70 MeV/$c^2$ in the $\Upsilon(1S)$ data corresponds to the $\eta_b(1S)$ state with a mass of $(9393.2\pm4.1)$ MeV/$c^2$, and that around 50 MeV/$c^2$ in the $\Upsilon(2S)$ data corresponds to the $\eta_b(2S)$ state with a mass of $(9974.6\pm3.1)$ MeV/$c^2$. The evidence for $\eta_b(2S)$ and the transition of $\Upsilon(1S)\to \gamma\eta_b(1S)$ were reported for the first time. The products of the branching fractions $\BR(\Upsilon(1S)\to \gamma\eta_b(1S))\BR(\eta_b(1S)\to X_{b{\bar b}})$ and $\BR(\Upsilon(2S)\to \gamma\eta_b(2S))\BR(\eta_b(2S)\to X_{b{\bar b}})$ are measured to be $(30.1^{+33.5}_{-7.4}\pm7.5)\times10^{-6}$ and $(46.2^{+29.7}_{-14.2}\pm10.6)\times10^{-6}$, respectively.

\begin{figure}[htbp]
\centering
\includegraphics[width=4.8cm]{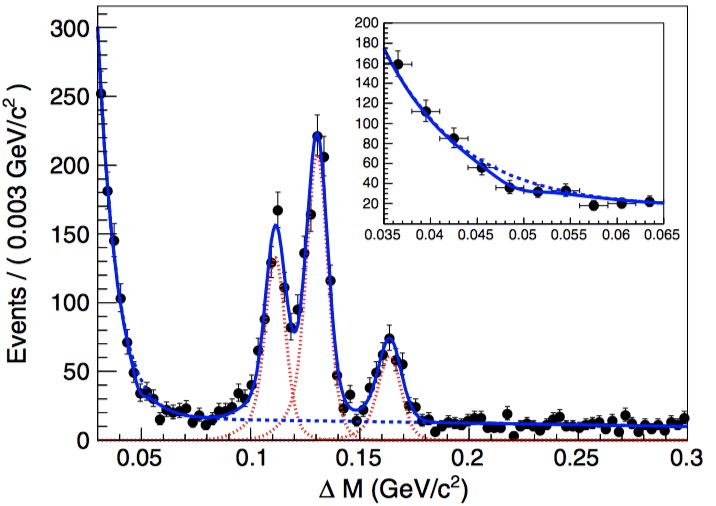}
\caption{The distributions of $\Delta M$ in $\Upsilon(2S)$ data from Belle~\cite{Sandilya111}. Dots with error bars are the data, the blue solid curve is the result of the fit for the signal-plus-background hypothesis, and blue dashed curve is background component, respectively. The three $\chi_{bJ}(1P)$ components indicated by the red dotted curves are here considered as part of the signal. The inset shows an expanded view of the $\Delta M$ distribution in the region of [0.035, 0.065] GeV/$c^2$.}\label{SIIA3dd2}
\end{figure}

With a much larger $\Upsilon(2S)$ data sample, Belle also searched for $\eta_b(2S)$ in the 26 exclusive hadronic final states as above~\cite{Sandilya111}. Figure~\ref{SIIA3dd2} shows the corresponding $\Delta M$ distribution~\cite{Sandilya111}. No evidence for $\eta_b(2S)$ was found, and a 90\% C.L. upper limit on the branching fraction $\BR(\Upsilon(2S)\to \gamma\eta_b(2S))\BR(\eta_b(2S)\to X_{b{\bar b}})$ $<$ 4.9$\times10^{-6}$ was obtained, which is an order of magnitude smaller than the result in Ref.~\cite{Dobbs109}. This result disfavors the observation of $\eta_b(2S)$ in $\Upsilon(2S)$ exclusive decays in Ref.~\cite{Dobbs109}. Previously, Belle reported the evidence for $\eta_b(2S)$ in the $h_b(2P)\to \eta_b(2S)\gamma$ transition using a 133.4 fb$^{-1}$ data sample collected at energies near the $\Upsilon(5S)$ resonance~\cite{Mizuk232002}. The measured mass of $\eta_b(2S)$ is $(9999.0\pm3.5^{+2.8}_{-1.9})$ MeV/$c^2$, which is far away from the mass reported in Ref.~\cite{Dobbs109}.~Therefore, the enhancement observed around 9974.6  MeV/$c^2$ in Ref.~\cite{Dobbs109} is likely attributed to statistical fluctuation.
	
\subsubsection{Search for exotic states in $\allu$ decays}

\vspace{0.5cm}
$\bullet$ Search for $XYZ$ states in $\Upsilon(1S,2S)$ radiative and inclusive decays
\vspace{0.1cm}

The radiative decays of the $\Upsilon$ states below open-bottom threshold are used to search for charge-parity-even charmonium-like states. Searches for the $X(3872)$, $X(3915)$, and $Y(4140)$ states in the $\Upsilon(1S,2S)$ radiative decays were carried out by Belle~\cite{Shen82,Wang84}.~The $X(3872)$ is reconstructed in the $\pi^+\pi^-J/\psi$ and $\pi^+\pi^-\pi^0J/\psi$ final states, and $X(3915)$ and $Y(4140)$ are reconstructed in $\omega J/\psi$ and $\phi J/\psi$. No $X(3872)$, $X(3915)$, or $Y(4140)$ signals are observed, and the corresponding production rates of the $\pi^+\pi^-J/\psi$ and $\pi^+\pi^-\pi^0J/\psi$, $\omega J/\psi$, or $\phi J/\psi$ modes are found to be less than a few times $10^{-6}$ at 90\% C.L..

Belle reported a search for some $XYZ$ states in $\Upsilon(1S)$ inclusive decays using the world's largest $\Upsilon(1S)$ data sample~\cite{Shen93}. The $XYZ$ states include the $X(3872)$, $X(4350)$, $Y(4140)$, $Y(4260)$, $Y(4360)$, $Y(4660)$, $Z_c(3900)^{\pm}$, $Z_c(4050)^{\pm}$, $Z_c(4200)^{\pm}$, $Z_c(4430)^{\pm}$, and $Z^{\pm}_{cs}$~\cite{PDG,Lee55,Dias88}. They are studied in the final states that contain a $J/\psi(\psi^{\prime})$ and up to two charged light hadrons ($K^{\pm}/\pi^{\pm}$). No significant signal is found in any of the studied modes and 90\% C.L. upper limits on the product branching fractions, i.e., $\BR(\Upsilon(1S)\to XYZ+anything)\BR(XYZ\to J/\psi(\psi^{\prime})+hadrons)$, are set, which are at or below the level of $10^{-5}$.

Considerable efforts in theory have been devoted to interpreting the charged charmonium-like states ($Z_c$) as tetraquarks, molecules, hybrids, or hadrocharmonia~\cite{Brambilla1534,Brambilla2981,A30,639,90,1907}. To distinguish among these explanations, experimental input is needed, especially that on the double $Z^{\pm}_c$ production in $e^+e^-$ annihilation. For $e^+e^-\to Z^+_cZ^-_c$, the dependence on the squared $e^+e^-$ C.M. energy, $s$, of the electromagnetic form factor, $F_{Z^+_cZ^-_c}$, is $1/s^3$ for a $Z_c$ state with tetraquark structure or $1/s$ for a $Z_c$ system of two tightly bound diquarks~\cite{Brodsky764,Brodsky91}. However, it remains unclear from which values of $s$ onwards this scaling is applicable. Belle searched for doubly charged charmonium-like state productions  using the largest data samples of $\Upsilon(1S)$ and $\Upsilon(2S)$~\cite{Jia97}. No significant signals are observed in any of the studied modes, and the 90\% C.L. upper limits on $\BR(\Upsilon(1S,2S)\to Z^+_cZ^{(\prime)-}_c)\times\BR(Z^+_c\to \pi^+ + c{\bar c})$ ($c\bar c=J/\psi,\chi_{c1}(1P),\psi^{\prime}$) are in the range of $(1-50)\times10^{-6}$. Here, $Z_c$ refers to $Z_c(3900)$ and $Z_c(4200)$ observed in the $\pi J/\psi$ final state, $Z_{c1}(4050)$ and $Z_{c2}(4250)$ in the $\pi \chi_{c1}$ final state, and $Z_c(4050)$ and $Z_c(4430)$ in the $\pi\psi^{\prime}$ final state.

\vspace{0.5cm}
$\bullet$ Search for glueballs and light tetraquraks in $\Upsilon(1S,2S)$ decays
\vspace{0.1cm}

The existence of glueballs, with a rich spectroscopy and a complex phenomenology, is one of the early predictions of QCD~\cite{Jaffe60}. However, despite many years of experimental efforts, none of these gluonic states has been established unambiguously. In the experimental aspect, the most outstanding obstacle is the isolation of glueballs from ordinary hadrons.

\begin{figure*}[htbp]
\centering
\includegraphics[width=13cm]{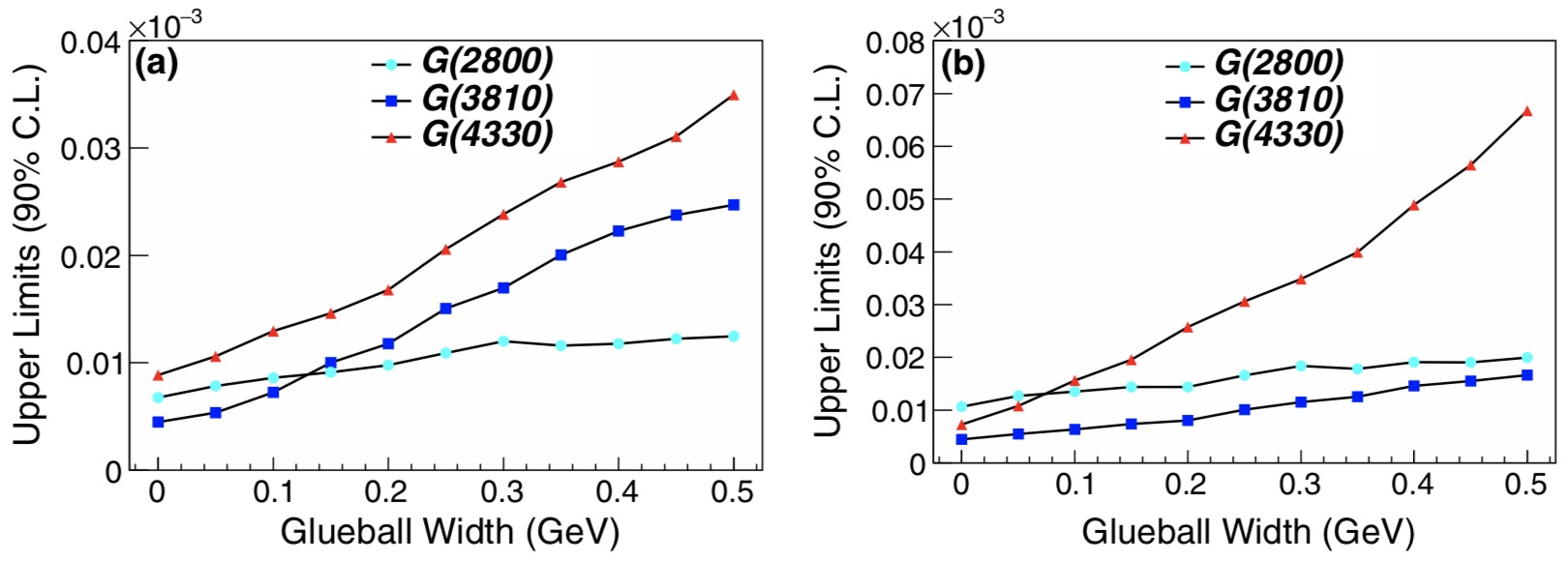}
\caption{The upper limits on the branching fractions for (a) $\Upsilon(1S )\to\chi_{c1} + G_{0^{--}}$ and (b) $\Upsilon(2S)\to\chi_{c1} + G_{0^{--}}$ as a function of the assumed $G_{0^{--}}$ decay width from Belle~\cite{Jia95}.}\label{SIIA4}
\end{figure*}

Recently Belle utilized the data samples of 102M $\Upsilon(1S)$ and 158M $\Upsilon(2S)$ events to search for $0^{--}$ glueballs ($G_{0^{--}}$) with quantum numbers incompatible with quark-antiquark bound states~\cite{Jia95}. Two $0^{--}$ glueballs are predicted using QCD sum rules~\cite{Qiao113} with masses of $(3.81\pm0.12)$ GeV/$c^2$ and $(4.33\pm0.13)$ GeV/$c^2$, and a $0^{--}$ glueball calculated using the dynamical holographic QCD model has a mass of 3.817 GeV/$c^2$~\cite{MDSM}. Belle searched for such $G_{0^{--}}$ in $\Upsilon(1S,2S)\to\chi_{c1}/f_1(1285)+G_{0^{--}}$ and $\chi_{b1}\to J/\psi/\omega+G_{0^{--}}$ processes~\cite{Jia95}. No evident signal is found at the predicted masses in all the studied processes, and 90\% C.L. upper limits are set on their branching fractions. As an example, Fig.~\ref{SIIA4} shows the 90\% C.L. upper limits on the branching fractions of $\Upsilon(1S)/\Upsilon(2S)\to\chi_{c1} + G_{0^{--}}$ as a function of the $G_{0^{--}}$ width. Interestingly, a signal at 3.92 GeV/$c^2$ is evident in $\Upsilon(1S)\to f_1(1285)+G_{0^{--}}$ with a significance of 3.7$\sigma$, which is consistent with the prediction in Ref.~\cite{Qiao113} within uncertainty.

Following the above searches for $G_{0^{--}}$, Belle utilized the low side of the recoil mass spectra of $\chi_{c1}$, $f_1(1285)$, $J/\psi$, and $\omega$ from Ref.~\cite{Jia95} to search for $X_{\rm tetra}$ in $\Upsilon(1S,2S)\to\chi_{c1}/f_1(1285)+X_{\rm tetra}$ and $\chi_{b1}\to J/\psi/\omega+X_{\rm tetra}$~\cite{Jia112002}. Here, the $X_{\rm tetra}$ is a four-quark state with $J^{PC}$ = $0^{--}$ or $1^{+-}$ calculated by Laplace sum rules and finite-energy sum rules using tetraquark-like currents~\cite{Huang95}. No evident signal is found below 3 GeV/$c^2$ in these processes, and 90\% C.L. upper limits on the branching fractions are set with $X_{\rm tetra}$ masses from 1.16 to 2.46 GeV/$c^2$ and widths from 0.0 and 0.3 GeV/$c^2$; those are found to a few times $10^{-4}$ or $10^{-5}$.

\vspace{0.5cm}
$\bullet$ Search for six-quark states in $\Upsilon(1S,2S,3S)$ decays
\vspace{0.1cm}

While the vast majority of known hadrons can be described as either quark-antiquark, three-quark, or three-antiquark combinations, other possibilities are allowed by QCD.  Among those, the six-quark configuration $uuddss$ is of particular interest, as its spatial wave function is completely symmetric. Generic arguments imply that it should be the most tightly bound six-quark state~\cite{Preskill177}. This property was already noticed by Jaffe in 1977~\cite{Jaffe38}. He predicted the existence of a loosely bound $uuddss$ state with a mass below the $2m_{\Lambda}$ threshold by about 80 MeV/$c^2$, dubbed the $H$ dibaryon. Recently, the motivation to search for $H$ has been strengthened by lattice QCD~\cite{Beane106,Beane26,Inoue106} and chiral constituent model calculations~\cite{Carames85}. With its mass approaching the $2m_{\Lambda}$ threshold from below (above), $H$ would behave more and more like a $\Lambda\Lambda$ analog of deuteron (dineutron), independent of its dynamical origin~\cite{Braaten428}. If its mass is below $2m_{\Lambda}$, $H$ would predominantly decay via $\Delta S=+1$ weak transitions to the $\Lambda n$, $\Sigma^- p$, $\Sigma^0 n$ or $\Lambda p \pi^-$ final states. If its mass is above $2m_{\Lambda}$, but below $m_{\Xi^0}+m_n~(2m_{\Lambda}+23.1~{\rm  MeV}/c^2)$, it would decay via strong interactions to $\Lambda\Lambda$ completely.

\begin{figure}[htbp]
\centering
\includegraphics[width=5.21cm]{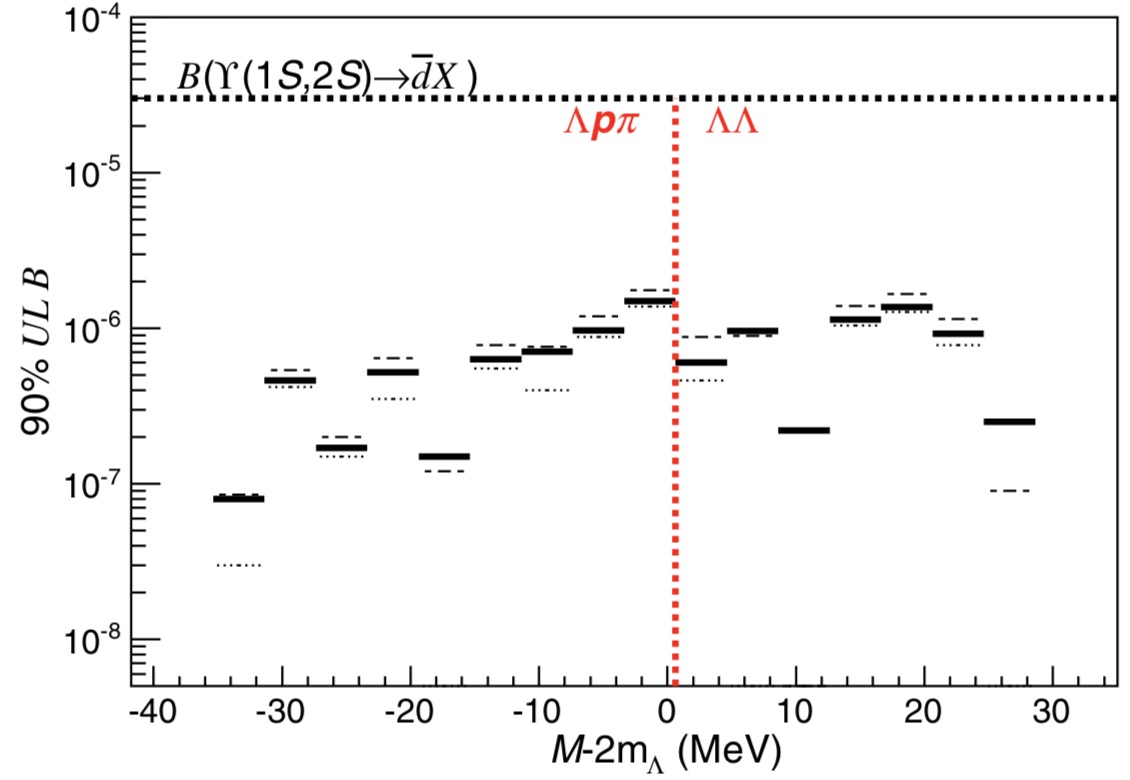}
\caption{Upper limits at 90\% C.L. for $\BR(\Upsilon(1S,2S)\to H+anything)$ for a narrow ($\Gamma=0$) $H$ dibaryon versus $M_H-2m_{\Lambda}$ are shown as solid horizontal bars from Belle~\cite{Kim110}. The +1$\sigma$ (+2$\sigma$) values from the fitted signal yields are shown as the dotted (dashed) bars. (For some mass bins, these are negative and not shown.) The vertical dotted line indicates the $M_H=2m_{\Lambda}$ threshold. The horizontal dotted line indicates the average PDG value for $\BR(\Upsilon(1S,2S)\to \bar d + anything)$.}\label{SIIA42}
\end{figure}

Belle reported a search for $H$ in the inclusive processes $\Upsilon(1S,2S)\to H +anything$, $H\to \Lambda p \pi^-$ and $\Lambda\Lambda$~\cite{Kim110}. The signals are extracted by a sequence of binned minimum fits to the distributions of $M(\Lambda p \pi^-)-2m_{\Lambda}$ and $M(\Lambda\Lambda)-2m_{\Lambda}$. In the fits, the signal peak position is confined to a 4 MeV/$c^2$ window that is scanned in 4 MeV/$c^2$ steps across the ranges $(m_{\Lambda}+m_p+m_{\pi^-})\le M(\Lambda p \pi^-)\le 2m_{\Lambda}$ and $2m_{\Lambda}\le M(\Lambda\Lambda)\le (2m_{\Lambda}+28$ MeV/$c^2$). None of the fits exhibits a positive signal with the significance greater than 3$\sigma$. The fit results are translated into 90\% C.L. upper limits on the signal yield, then used to determine upper limits on the inclusive product branching fractions ($\BR(\Upsilon(1S,2S)\to H+anything)\BR(H\to \Lambda p \pi^-/\Lambda\Lambda)$), as shown in Fig.~\ref{SIIA42}. The reported results  are some of the most stringent constraints to date on the existence of $H$.

\begin{figure}[htbp]
\centering
\includegraphics[width=6.5cm]{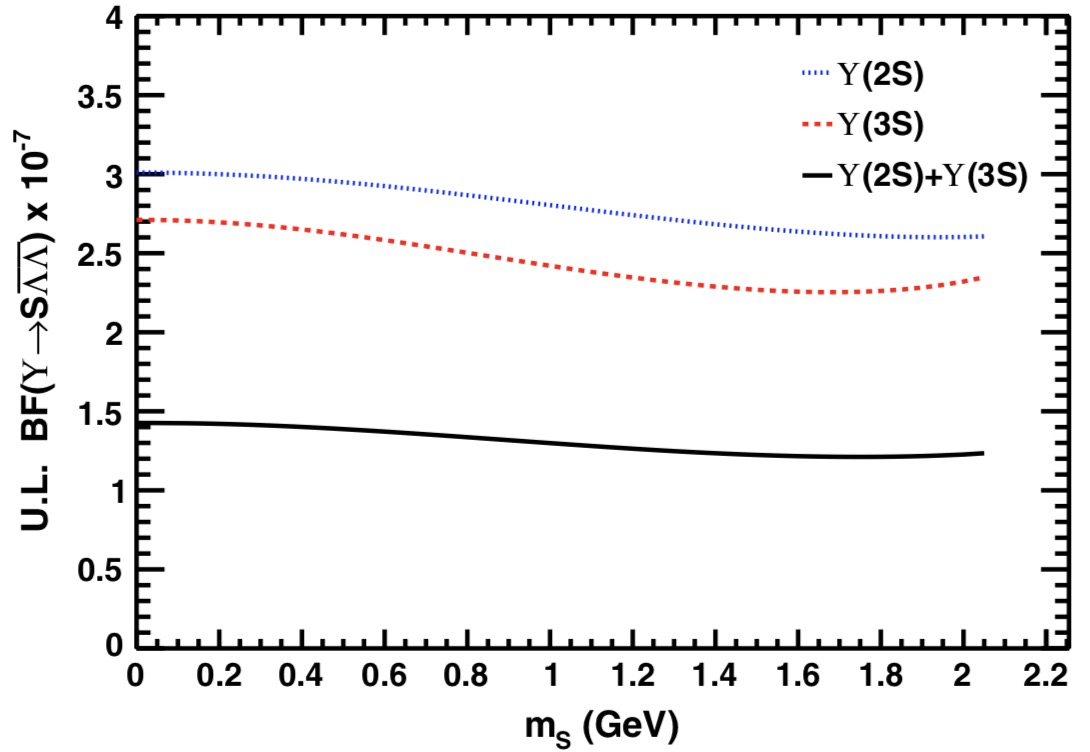}
\caption{The 90\% C.L. upper limits on the $\Upsilon(2S,3S)\to S{\bar \Lambda}{\bar \Lambda}$ branching fractions for the $\Upsilon(2S)$ and $\Upsilon(3S)$ datasets, as well as the combined sample assuming the same partial width from BaBar~\cite{Lees122}.}\label{SIIA43}
\end{figure}

A stable sexaquark, denoted as $S$, with the six-quark $uuddss$ component was proposed to be discovered in $\Upsilon\to S{\bar \Lambda}{\bar \Lambda}$ decays~\cite{Farrar}. If $S$ really exists, then $S$ is a good DM candidate. BaBar investigated such a state in $\Upsilon(2S)$ and $\Upsilon(3S)$ decays~\cite{Lees122}. Considering the $S$ is undetected, the total energy of clusters in the electromagnetic calorimeter not associated with charged particles, $E_{\rm extra}$, is required to be less than 0.5 GeV to remove hadronic events containing several strange baryons and additional charged and neutral particles. The mass of the unseen $S$ can be identified with the square of the mass recoiling against the $\Lambda\Lambda$ system. As a result, no significant signal is observed, and 90\% C.L. upper limits on the $\Upsilon(2S,3S)\to S{\bar \Lambda}{\bar \Lambda}$ branching fractions, scanning $S$ masses in the range 0 $<$ $m_S$ $<$ 2.05 GeV/$c^2$ in steps of 50 MeV/$c^2$ (approximately half the signal resolution), are derived, which are shown in Fig.~\ref{SIIA43} for the $\Upsilon(2S,3S)$ datasets, as well as the combined sample assuming the same partial width. These results set stringent bounds on the existence of the $S$ state.

\subsubsection{Search for new physics in $\allu$ decays}

\vspace{0.5cm}
$\bullet$ Invisible decay of $\Upsilon(1S)$
\vspace{0.1cm}

In the invisible decay modes, the final state particles interact so weakly with the detectors that they are not observable. In the SM, the invisible decay of the $\Upsilon(1S)$ meson into a $\nu {\bar \nu}$ pair is predicted to have a branching fraction of $(9.9\pm0.5)\times10^{-6}$~\cite{Chang441}. If the observed branching fraction is significantly larger than the SM prediction, physics beyond the SM will be implied. For instance, the low-mass DM  candidates could couple weakly to the SM particles to enhance the branching fraction to be about $6.0\times 10^{-3}$~\cite{McElrath72}.

\begin{figure}[htbp]
\centering
\includegraphics[width=7cm]{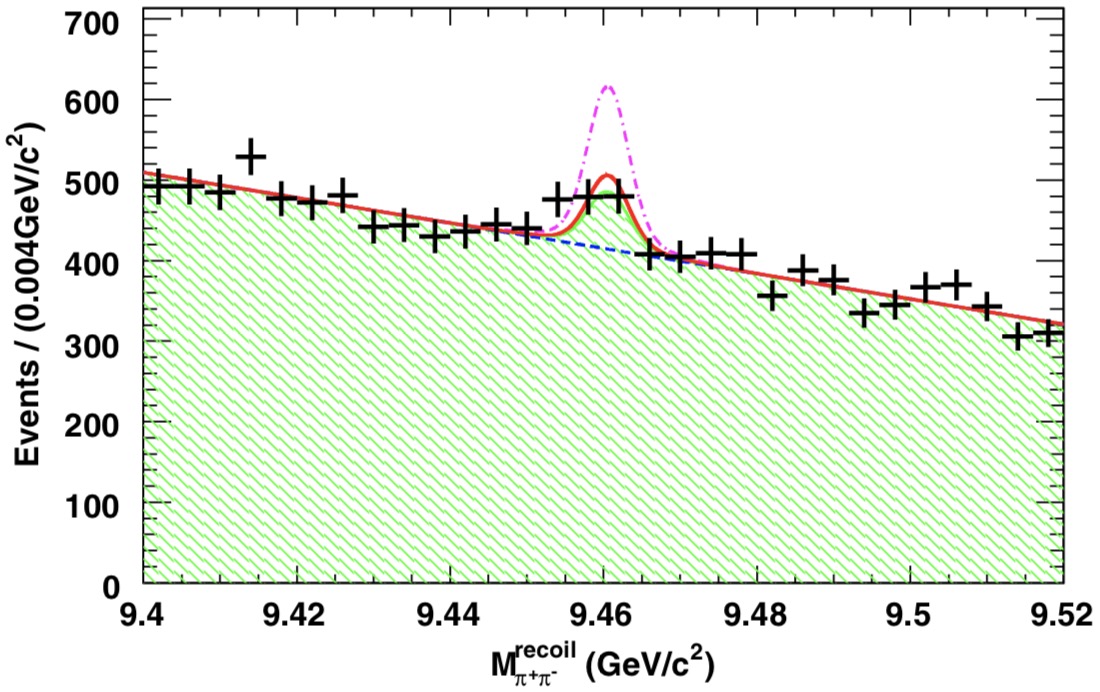}
\caption{Recoil mass distribution against two pions, $M^{\rm recoil}_{\pi^+\pi^-}$, from Belle~\cite{Tajima98}. The solid curve shows the result of the fit to the data, the shaded area shows the total background contribution, the dashed line shows the combinatorial background contribution, and the dot-dashed line shows the expected signal for $\BR(\Upsilon(1S)\to \chi\chi)=6.0\times 10^{-3}$~\cite{McElrath72}.}\label{SIIA51}
\end{figure}

Belle reported a search for the $\Upsilon(1S)$ invisible decay via the $\Upsilon(3S)\to \pi^+\pi^-\Upsilon(1S)$ transition using the $\Upsilon(3S)$ data sample~\cite{Tajima98}. For the selection of the invisible decay candidates, two oppositely charged tracks $\pi^+\pi^-$ are required in the event. Considering the products from $\Upsilon(1S)$ decay may go outside of the detector acceptance, the recoil mass distribution against two
pions in $\Upsilon(3S)$ data, $M^{\rm recoil}_{\pi^+\pi^-}$, still peaks at the $\Upsilon(1S)$ mass and becomes a peaking background to the invisible decay signal. Expected numbers of peaking background events based on MC calculations for $\Upsilon(1S)\to \mu^+\mu^-$, $e^+e^-$, $\tau^+\tau^-$, and $\nu {\bar \nu}$ are $77.3\pm12.0$, $50.3\pm8.2$, $5.2\pm1.0$, and $0.4\pm0.1$ assuming the world-average values of $\BR(\Upsilon(1S)\to \mu^+\mu^-/e^+e^-/\tau^+\tau^-)$ from PDG~\cite{PDG}, and $\BR(\Upsilon(1S)\to \nu {\bar \nu})$ = $(9.9\pm0.5)\times10^{-6}$~\cite{Chang441}. Figure~\ref{SIIA51} shows the $M^{\rm recoil}_{\pi^+\pi^-}$ distribution. The extracted signal yield, $38\pm39$ events, is consistent with zero. The dot-dashed line shows the expected signal for $\BR(\Upsilon(1S)\to \chi\chi)=6.0\times 10^{-3}$~\cite{McElrath72}. The resulting $\BR(\Upsilon(1S)\to invisible)$ $<$ $2.5\times 10^{-3}$ was obtained at 90\% C.L. This result disfavors the prediction in Ref.~\cite{McElrath72} for the $\Upsilon(1S)$ decay to a pair of DM particles lighter than the $b$ quark.

CLEO~\cite{Rubin75} and BaBar~\cite{Aubert251801} also measured the branching fraction of the invisible $\Upsilon(1S)$ decay. To provide a clean sample of $\Upsilon(1S)$ decays, the events in the transitions of $\Upsilon(2S)\to\pi^+\pi^-$ and $\Upsilon(3S)\to\pi^+\pi^-$ are selected by CLEO~\cite{Rubin75} and BaBar~\cite{Aubert251801}. The invisible $\Upsilon(1S)$ decay is also searched for in the mass recoiling against the dipion system. The resulting limits at 90\% C.L. are $\BR(\Upsilon(1S)\to invisible)$ $<$ $3.9\times 10^{-3}$ and $\BR(\Upsilon(1S)\to invisible)$ $<$ $3.0\times 10^{-4}$ from CLEO~\cite{Rubin75} and BaBar~\cite{Aubert251801}, respectively. Notably, the limit from BaBar is almost an order of magnitude closer to the SM prediction than the limits from Belle and CLEO.

\vspace{0.5cm}
$\bullet$ $\Upsilon(1S,2S,3S)\to \gamma + A^0$
\vspace{0.1cm}

A low mass DM particle, $\chi$, is predicted to be produced in interactions of SM particles via the exchange of a $CP$-odd Higgs boson $A^0$~\cite{Gunion73,Petraki28}, which is part of the NMSSM~\cite{Ellwanger491}. Transitions $\Upsilon(2S)\to\pi^+\pi^-\Upsilon(1S)$ and $\Upsilon(3S)\to\pi^+\pi^-\Upsilon(1S)$ offer a way to cleanly detect the production of $\Upsilon(1S)$ mesons, and enable searches for $A^0$ in the radiative decay of $\Upsilon(1S)$. The $\BR(\Upsilon(1S)\to \gamma A^0)$ is predicted to be as large as $5\times10^{-4}$, depending on $m_{A^0}$ and couplings~\cite{Dermisek76,Dermisek81}.
For the multi-body $\Upsilon(1S)\to \gamma\chi\chi$ decay, the branching fraction is suppressed by $\cal {O}(\alpha)$ in comparison with $\BR(\Upsilon(1S)\to \chi\chi)$, and the range $10^{-5}-10^{-4}$ is expected~\cite{Yeghiyan80}. The SM process $\Upsilon(1S)\to \gamma\nu {\bar \nu}$ has the same final state as the signal, but is predicted to have a branching fraction of the order of $10^{-9}$~\cite{Yeghiyan80}, which is three orders of magnitude below our experimental sensitivity.

\begin{figure*}[htbp]
\centering
\includegraphics[width=12cm]{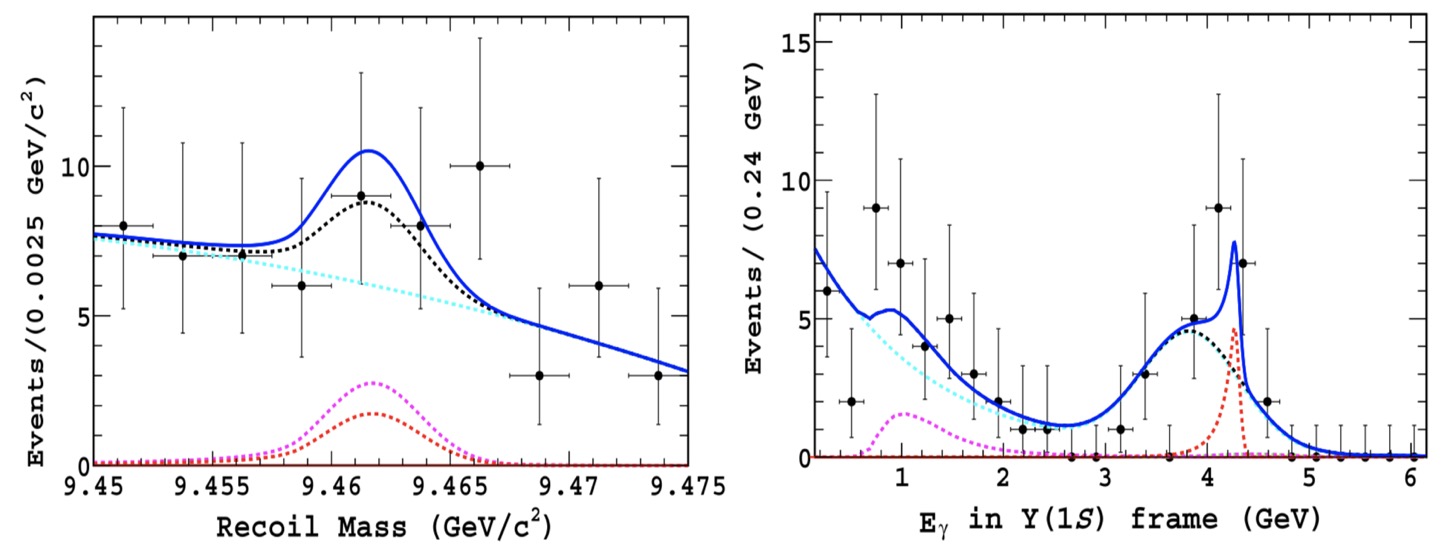}
\caption{Two-dimensional fit result for the on-shell process mass scan point with $M_{A^0}$ = 2.946 GeV/$c^2$, which has the highest local signal significance of 2.1$\sigma$ from Belle~\cite{Seong122}. Left: $M_{\rm recoil}$ distribution. Right: $E^{*}_{\gamma}$ distribution. The fitted components are tau-pair and continuum backgrounds (cyan dashed curve), $\Upsilon(1S)$ leptonic and hadronic decay backgrounds (magenta dashed curve), and the on-shell signal (red dashed curve). The blue solid curve shows the sum of all fitted components, and the black dashed curve shows the sum of all fitted background components.}\label{SIIA52}
\end{figure*}

Very recently, Belle searched for $A^0$ in the final states with a single photon and missing energy~\cite{Seong122}. Since the mass of $A^0$ is unknown, two processes are considered: the on-shell process $\Upsilon(1S)\to \gamma A^0$ with $A^0\to \chi\chi$; and the off-shell process $\Upsilon(1S)\to\gamma\chi\chi$. The $\Upsilon(1S)$ is tagged via the dipion transition $\Upsilon(2S)\to \Upsilon(1S)\pi^+\pi^-$. Both signal processes produce only three detectable particles: two charged pions, which have low transverse momenta, and a photon, which has the highest energy in the C.M. frame. Two observables are used to extract the signals: the invariant recoil mass of the dipion system ($M_{\rm recoil}$) and the energy of photon in the $\Upsilon(1S)$ frame ($E^{*}_{\gamma}$). Distributions of $M_{\rm recoil}$ and $E^{*}_{\gamma}$ are shown in Fig.~\ref{SIIA52}. Irreducible backgrounds can be categorized into four event types: tau-pair productions $\Upsilon(2S)\to \tau^+\tau^-$, continuum, leptonic decays $\Upsilon(1S)\to \LL$, and hadronic decays $\Upsilon(1S)\to \gamma hh$. Tau-pair productions from the $\Upsilon(2S)$ and continuum backgrounds do not peak either in the recoil mass distribution nor in the photon energy spectrum; therefore, they are combined, as shown by the cyan dashed curve in Fig.~\ref{SIIA52}. The background contributions from leptonic decays and hadronic decays of $\Upsilon(1S)$ are predicted to be $20.0\pm2.8$ and $1.2\pm0.7$ events, which are combined as shown by the magenta dashed curves in Fig.~\ref{SIIA52}.

\begin{figure*}[htbp]
\centering
\includegraphics[width=12cm]{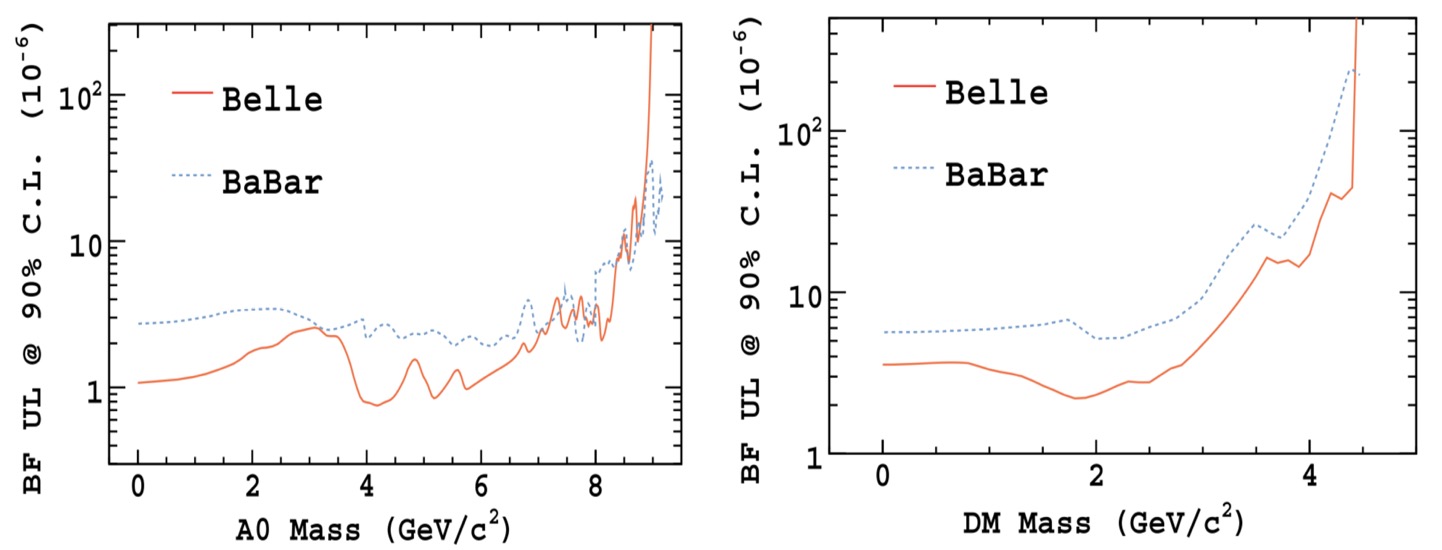}
\caption{90\% C.L. upper limits on the branching fractions of the on-shell process $\Upsilon(1S)\to\gamma A^0$ with $A^0\to \chi\chi$ (left) and the off-shell process $\Upsilon(1S)\to\gamma\chi\chi$ (right). The orange solid curves are the Belle limits~\cite{Seong122} and the blue dashed curves are the BaBar limits~\cite{Sanchez107}.}\label{SIIA53}
\end{figure*}

An unbinned extended likelihood fit in the two-dimensional ($M_{\rm recoil}$, $E^{*}_{\gamma}$) space is performed to estimate the yields of different event types. The fit is repeated for each value of $M_{A^0}$ or $M_{\chi}$ in the mass ranges 0 $<$ $M_{A^0}$ $<$ 8.97 GeV/$c^2$ (on-shell process) or 0 $<$ $M_{\chi}$ $<$ 4.44 GeV/$c^2$ (off-shell process). For the on-shell case, the photon energy is scaned in 353 steps that correspond to half the photon energy resolution, and step size in the range from 25 MeV/$c^2$ to 4 MeV/$c^2$. For the off-shell case, 45 scan points with a fixed step size of 100 MeV/$c^2$ are used.~When $M_{A^0}$ = 2.946 GeV/$c^2$, the local significance of 2.1$\sigma$ is largest; see Fig.~\ref{SIIA52}. No statistically significant signal is observed, and the upper limits (90\% C.L.) on the branching fractions of the on-shell and the off-shell signals are given by $N^{\rm UL}/[N_{\Upsilon(2S)}\times \BR(\Upsilon(2S)\to \Upsilon(1S)\pi^+\pi^-)\times\varepsilon]$, where $N^{\rm UL}$ an upper limit at 90\% C.L. on the signal yield and $\varepsilon$ is the signal efficiency, as shown in Fig.~\ref{SIIA53}.

BaBar also searched for the single-photon decays of $\Upsilon(1S)$: $\Upsilon(1S)\to \gamma A^0$ and $\Upsilon(1S)\to \gamma\chi\chi$~\cite{Sanchez107}. Two kinematic variables are used to extract the signals: the recoil mass of $\pi^+\pi^-$ and the missing mass squared of $\pi^+\pi^-\gamma$. The yields of signal events are extracted as a function of $m_{A^0}$ ($m_{\chi}$) in the interval 0 $\leq$ $m_{A^0}$ $\leq$ 9.2 GeV/$c^2$ (0 $\leq$ $m_{\chi}$ $\leq$ 4.5 GeV/$c^2$) by performing a series of unbinned extended maximum likelihood scans. No evidence for the single-photon decay $\Upsilon(1S)\to\gamma+invisible$ was found, and the 90\% C.L. upper limits on $\BR(\Upsilon(1S)\to\gamma A^0)\BR(A^0\to invisible)$ and $\BR(\Upsilon(1S)\to \gamma\chi\chi)$ were obtained; see Fig.~\ref{SIIA53}. The results from Belle~\cite{Seong122} and BaBar~\cite{Sanchez107} improve the existing limits by an order of magnitude or more, and significantly constrain light Higgs boson~\cite{Dermisek76,Dermisek81} and light DM~\cite{Yeghiyan80} models. The limits from Belle are more stringent than BaBar, especially for the off-shell case. Instead of $A^0\to invisible$, BaBar and CLEO searched for $A^0$ via the $\mu^+\mu^-$~\cite{Love101,Aubert081803,Lees87}, $\tau^+\tau^-$~\cite{Love101,Aubert181801,Lees88}, and hadronic final states~\cite{Lees107,Lees88a}, but did not see any significant signal in these final states. The most conservative upper limits at 90\% C.L. on $\BR(\Upsilon(1S)/\Upsilon(2S)/\Upsilon(3S)\to \gamma A^0)\BR(A^0\to \mu^+\mu^-/\tau^+\tau^-/hadrons)$ are summarized in Table~\ref{NP}~\cite{Love101,Aubert081803,Lees87,Aubert181801,Lees88,Lees88a}.
These measurements improve the constraints on the parameters of the NMSSM and similar theories with low-mass scalar degrees of freedom.

\linespread{1.2}
\begin{table*}[htbp]
\small
\caption{The most conservative upper limits at 90\% C.L. on the $\BR(\Upsilon(1S)/\Upsilon(2S)/\Upsilon(3S)\to \gamma A^0)\BR(A^0\to \mu^+\mu^-/\tau^+\tau^-/hadrons)$ from CLEO~\cite{Love101} and BaBar~\cite{Aubert081803,Lees87,Aubert181801,Lees88,Lees88a}.}
\vspace{0.2cm}
\label{NP}
\centering
\begin{tabular}{c  c  c  c}
\hline\hline
Mode & $\Upsilon(1S)\to \gamma A^0, A^0\to \mu^+\mu^-$ & $\Upsilon(2S)\to \gamma A^0, A^0\to \mu^+\mu^-$ & $\Upsilon(3S)\to \gamma A^0, A^0\to \mu^+\mu^-$ \\
Branching fraction & $9\times10^{-6}$~\cite{Love101,Lees87} & $8.3\times10^{-6}$~\cite{Aubert081803} & $5.5\times10^{-6}$~\cite{Aubert081803} \\\hline
Mode & $\Upsilon(1S)\to \gamma A^0, A^0\to \tau^+\tau^-$ & $\Upsilon(2S)\to \gamma A^0, A^0\to \tau^+\tau^-$ & $\Upsilon(3S)\to \gamma A^0, A^0\to \tau^+\tau^-$ \\
Branching fraction &  $1.3\times10^{-4}$~\cite{Love101,Lees88} & - &$1.6\times10^{-4}$~\cite{Aubert181801} \\\hline
Mode & $\Upsilon(1S)\to \gamma A^0, A^0\to hadrons$ & $\Upsilon(2S)\to \gamma A^0, A^0\to hadrons$ & $\Upsilon(3S)\to \gamma A^0, A^0\to hadrons$ \\
Branching fraction & - & $8\times10^{-5}$~\cite{Lees88a} & $8\times10^{-5}$~\cite{Lees88a} \\\hline\hline
\end{tabular}
\end{table*}

\vspace{0.5cm}
$\bullet$ Tests of lepton flavor universality in $\Upsilon(1S,2S,3S)$ decays
\vspace{0.1cm}

The measurements of the ratios $R_{\tau\mu}(\Upsilon(nS))$ = $\BR(\Upsilon(nS)\to\tau^+\tau^-)/\BR(\Upsilon(nS)\to\mu^+\mu^-)$ ($n$ = 1, 2, 3) are motivated as tests of the lepton flavor universality (LFU). In the SM these ratios are expected to be close to 1. Any significant deviations would violate LFU and could be introduced by the coupling to a light pseudoscalar Higgs boson in supersymmetry, or leptoquarks and compositeness models.~The values of $R_{\tau\mu}(\Upsilon(1S))$, $R_{\tau\mu}(\Upsilon(2S))$, and $R_{\tau\mu}(\Upsilon(3S))$ obtained by CLEO are $1.02\pm0.02\pm0.05$, $1.04\pm0.04\pm0.05$, and $1.05\pm0.08\pm0.05$~\cite{Besson98}, which are consistent with the expectations from the SM. Recently, BaBar reported a value of $R_{\tau\mu}(\Upsilon(3S))$ to be ($0.996\pm0.008\pm0.014$)~\cite{2005.01230}. The uncertainty in this result is almost an order of magnitude smaller than that in the CLEO result. BaBar also tested LFU in the decays $\Upsilon(3S)\to \Upsilon(1S)\pi^+\pi^-$, $\Upsilon(1S)\to \LL$, where $\ell$ = $\mu$, $\tau$~\cite{Sanchez104}. The resulting $R_{\tau\mu}(\Upsilon(1S))$ is found to be $1.005\pm0.013\pm0.022$ with improved precision. No significant deviation of $R_{\tau\mu}(\Upsilon(1S))$ from the SM expectation is observed.

The lepton flavor violating (LFV) processes were searched for by CLEO~\cite{Love201601} and BaBar~\cite{Lees104}. No significant signal was observed.~The 95\% C.L. upper limits on the branching fractions for $\Upsilon(1S)\to \mu^{\pm}\tau^{\mp}$, $\Upsilon(2S)\to \mu^{\pm}\tau^{\mp}$, and $\Upsilon(3S)\to \mu^{\pm}\tau^{\mp}$ are $6.0\times10^{-6}$, $14.4\times10^{-6}$, and $20.3\times10^{-6}$ from CLEO~\cite{Love201601}. The more stringent constraints were given by BaBar~\cite{Lees104}. The 90\% C.L. upper limits on the branching fractions for $\Upsilon(2S)\to e^{\pm}\tau^{\mp}$, $\Upsilon(2S)\to \mu^{\pm}\tau^{\mp}$, $\Upsilon(3S)\to e^{\pm}\tau^{\mp}$, and $\Upsilon(3S)\to \mu^{\pm}\tau^{\mp}$ are $3.2\times10^{-6}$, $3.3\times10^{-6}$, $4.2\times10^{-6}$, and $3.1\times10^{-6}$. Effective field theory allows one to relate the above fractions to the mass scale $\Lambda_{\ell\tau}$ of LFV beyond the SM physics using $\alpha^2_{\ell\tau}/\Lambda^4_{\ell\tau}$ = $(\BR(\Upsilon(nS)\to \ell^{\pm}\tau^{\mp})/\BR(\Upsilon(nS)\to \ell^{+}\ell^{-}))$($2q^2_b \alpha^2/(m_{\Upsilon(nS)})^4)$ ($n$ = 1, 2, 3)~\cite{Silagadze64,Black66}, where $\ell$ denotes $e$ or $\mu$, $\alpha_{l\tau}$ is the NP coupling constant, $q_b$ = $-1/3$ is the charge of $b$ quark, and $\alpha$ $\equiv$ $\alpha(m_{\Upsilon(nS)})$ is the fine structure constant at the $\Upsilon(nS)$ mass.
Assuming $\alpha_{e\tau}=\alpha_{\mu\tau}=1$, the branching fractions above can be translated to the 90\% C.L. lower limits $\Lambda_{e\tau}$ $>$ 1.6 TeV and $\Lambda_{\mu\tau}$ $>$ 1.7 TeV from BaBar~\cite{Lees104}, and the 95\% C.L. lower limit $\Lambda_{\mu\tau}$ $>$ 1.34 TeV from CLEO~\cite{Love201601}.

\subsection{Transitions between $\Upsilon(mS)$ and $\Upsilon(nS)$ ($m>n$)}

\vspace{0.5cm}
$\bullet$ $\Upsilon(2S,3S)\to \Upsilon(1S)+\eta/\pi^+\pi^-$
\vspace{0.1cm}

The QCD multipole expansion (QCDME) model can be used to describe the hadronic transitions between heavy
quarkonia~\cite{Kuang1}. It has succeeded in explaining the hadronic transitions between charmonia, e.g., the relative rate of $\psi^{\prime}\to\eta J/\psi$ to $\psi^{\prime}\to\pi^+\pi^- J/\psi$~\cite{Mendez78}. It also predicts that the transitions between bottomonium states involving an $\eta$ or $\pi^0$ meson are highly suppressed, since they require a spin flip of the heavy quark~\cite{Kuang1,Voloshin61,Simonov673}.

The first process involving a $b$-quark spin flip, $\Upsilon(2S)\to \eta\Upsilon(1S)$, was observed by CLEO with a statistical significance of 5.3$\sigma$~\cite{He192001}. Its branching fraction was measured to be $(2.1^{+0.7}_{-0.6}\pm0.3)\times10^{-4}$~\cite{He192001}. The transition was later confirmed  by BaBar with a more precise branching fraction of $(2.39\pm0.31\pm0.14)\times10^{-4}$~\cite{Lees84}. Meanwhile, the ratio $\BR(\Upsilon(2S)\to \eta\Upsilon(1S))$/$\BR(\Upsilon(2S)\to \pi^+\pi^-\Upsilon(1S))$ was measured to be $(1.35\pm0.17\pm0.08)\times 10^{-3}$~\cite{Lees84}. With a larger $\Upsilon(2S)$ data sample, Belle reported a larger branching fraction $\BR(\Upsilon(2S)\to \eta\Upsilon(1S))$ = $(3.57\pm0.25\pm0.21)\times10^{-4}$ and a larger ratio $\BR(\Upsilon(2S)\to \eta\Upsilon(1S))$/$\BR(\Upsilon(2S)\to \pi^+\pi^-\Upsilon(1S))$ = $(1.99\pm0.14\pm0.11)\times 10^{-3}$~\cite{Tamponi011104}. These  branching fractions obtained from CLEO, BaBar, and Belle are all slightly below the values predicted by QCDME in Ref.~\cite{Kuang1}. Up to now, the transition of $\Upsilon(3S)\to \eta\Upsilon(1S)$ has not yet been observed. The upper limits on $\BR(\Upsilon(3S)\to \eta\Upsilon(1S))$ at 90\% C.L. in units of $10^{-4}$ are 1.8 from CLEO~\cite{He192001} and 1.0 from BaBar~\cite{Lees84}.

\vspace{0.5cm}
$\bullet$ $\Upsilon(4S)\to \Upsilon(1S)+\pi^+\pi^-/\eta$
\vspace{0.1cm}

\begin{figure*}[htbp]
\centering
\includegraphics[width=12cm]{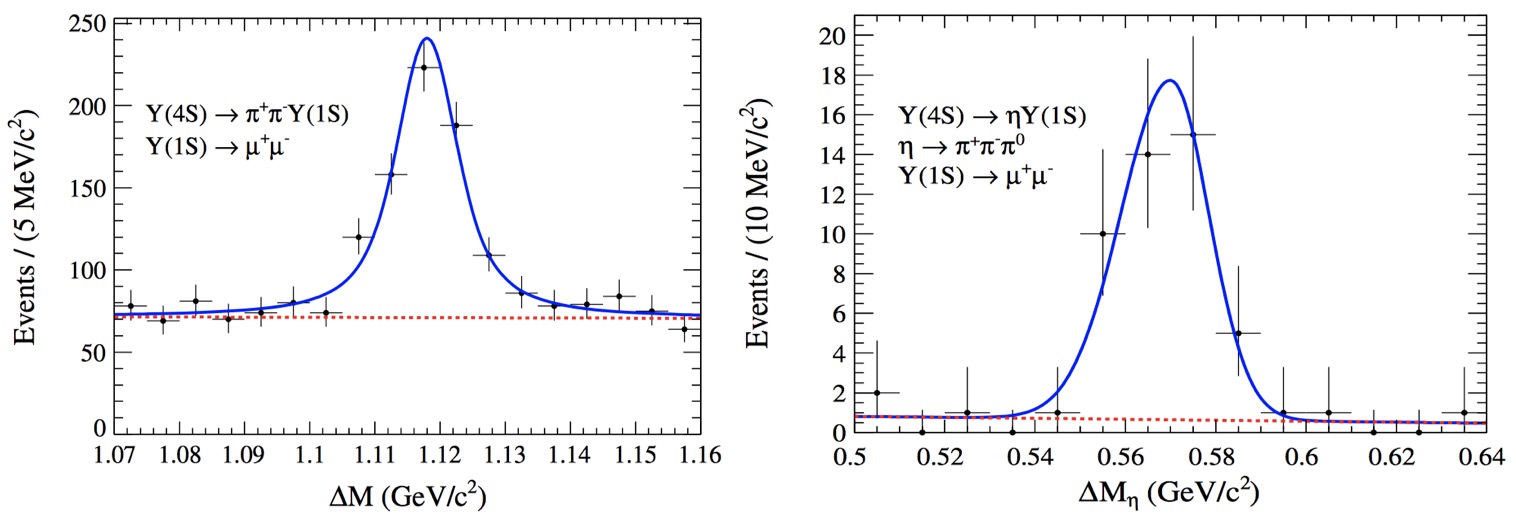}
\caption{The $\Delta M$ (left plot) and $\Delta M_{\eta}$ (right plot) distributions for $\Upsilon(4S)\to\pi^+\pi^-\Upsilon(1S)$ and $\Upsilon(4S)\to\eta\Upsilon(1S)$ from Belle~\cite{Guido96}. The data, best fits to them, and background contributions are shown by the dots with error bars, solid blue lines, and dashed red lines, respectively.}\label{SIIB1}
\end{figure*}

The hadronic transitions $\Upsilon(4S)\to\Upsilon(1S)+\pi^+\pi^-/\eta$ were studied by BaBar~\cite{Aubert78} and Belle~\cite{Guido96}.
The signature of $\Upsilon(4S)\to\pi^+\pi^-\Upsilon(1S)$ events is an invariant mass difference $\Delta M=M_{\pi\pi l l}-M_{ll}$ compatible with the difference between the masses of $\Upsilon(4S)$ and $\Upsilon(1S)$, where $M_{\pi\pi l l}$ is the $\pi^+\pi^- \LL$ invariant mass, and $M_{l l}$ is the $\LL$ invariant mass. For $\Upsilon(4S)\to\eta\Upsilon(1S)$, the signal events are identified by the invariant mass difference, $\Delta M_{\eta}=M_{3\pi \ell \ell}-M_{\ell\ell}-M_{3\pi}$ compatible with $M(\Upsilon(4S))-M(\Upsilon(1S))-M(\eta)$, where $M_{3\pi \ell\ell}$ is the $\pi^+\pi^-\pi^0 \LL$ invariant mass, and $M_{3\pi}$ is $\pi^+\pi^-\pi^0$ invariant mass. The $\Delta M$ and $\Delta M_{\eta}$ distributions from Belle are shown in Fig.~\ref{SIIB1}. Clear signals are observed in both $\Upsilon(4S)\to\pi^+\pi^-\Upsilon(1S)$ and $\Upsilon(4S)\to\eta\Upsilon(1S)$. The ratio of their branching fractions is determined to be $2.07\pm0.30\pm0.11$. From BaBar, the corresponding ratio is $2.41\pm0.40\pm0.12$. These results strongly disfavor the prediction by QCDME in Ref.~\cite{Kuang1}. This implies additional implementations for QCDME.

\vspace{0.5cm}
$\bullet$ $\Upsilon(4S)\to \Upsilon(1S)+\eta^{\prime}$
\vspace{0.1cm}

The kinematically allowed transition $\Upsilon(4S)\to \Upsilon(1S)+\eta^{\prime}$ is expected to be as strong as $\Upsilon(4S)\to \Upsilon(1S)+\eta$~\cite{Voloshin26}. The relative strength of the $\eta^{\prime}$ to $\eta$ transitions depends on the relative $u\bar u + d\bar d$ content of the mesons, and is predicted to range between 20\% and 60\%. Belle reported the observation of $\Upsilon(4S)\to \Upsilon(1S)+\eta^{\prime}$ with a significance of 5.7$\sigma$ ~\cite{Guido121}. The $\Upsilon(1S)$ meson is reconstructed via its decay to two muons, which is considerably cleaner than the dielectron mode. The $\eta^{\prime}$ meson is reconstructed via the $\rho^0(\to \pi^+\pi^-)\gamma$ and $\eta(\to\gamma\gamma)\pi^+\pi^-$ final states.

\begin{figure*}[htbp]
\centering
\includegraphics[width=12cm]{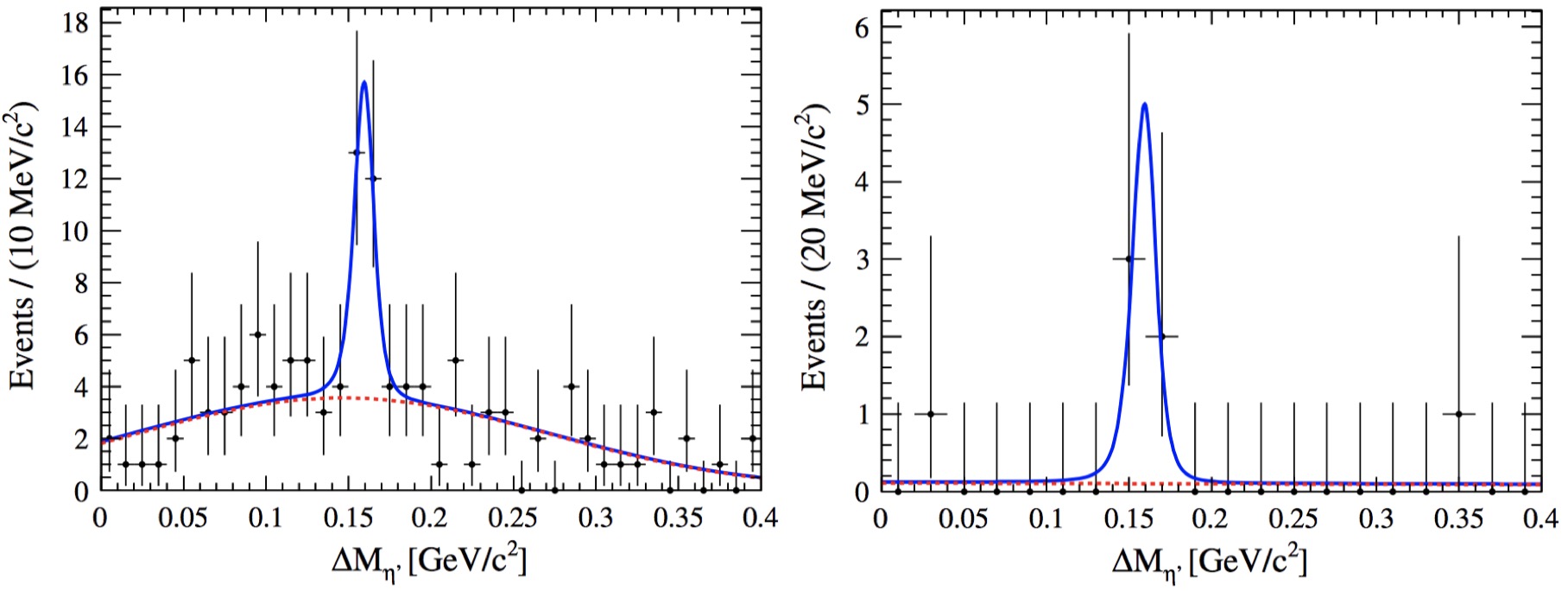}
\caption{A simultaneous fit to the $\Delta M_{\eta^{\prime}}$ distributions for (a) $\Upsilon(4S)\to \Upsilon(1S)+\eta^{\prime}(\to \rho^0\gamma)$ and (b) $\Upsilon(4S)\to \Upsilon(1S)+\eta^{\prime}(\to \eta\pi^+\pi^-)$ from Belle~\cite{Guido121}. The data, best fit to them, and background contributions are shown by the dots with error bars, solid blue lines, and dashed red lines, respectively.}\label{SIIB2}
\end{figure*}

The signal events are identified by the variable $\Delta M_{\eta^{\prime}} = M(\Upsilon(4S))-M(\Upsilon(1S))-M(\eta^{\prime})$, where $M(\eta^{\prime})=M(\pi^+\pi^-\gamma)$ or $M(\gamma\gamma\pi^+\pi^-)$, $M(\Upsilon(1S))=M(\mu^+\mu^-)$, and $M(\Upsilon(4S))$ = $M(\mu^+\mu^-\pi^+\pi^-\gamma)$ or $M(\mu^+\mu^-\gamma\gamma\pi^+\pi^-)$. The signal and background yields are determined by an unbinned maximum likelihood fit to the $\Delta M_{\eta^{\prime}}$ distributions, as shown in Fig.~\ref{SIIB2}. With a simultaneous fit to these $\Delta M_{\eta^{\prime}}$ distributions, the statistical significance is estimated to be 5.8$\sigma$. It is reduced to 5.7$\sigma$ with the systematic uncertainty included. The resulting branching fraction $\BR(\Upsilon(4S)\to \Upsilon(1S)+\eta^{\prime})$ is $(3.43\pm0.88\pm0.21)\times10^{-5}$. In addition, the ratios $\BR(\Upsilon(4S)\to \Upsilon(1S)+\eta^{\prime})/\BR(\Upsilon(4S)\to \Upsilon(1S)+\eta)$ and $\BR(\Upsilon(4S)\to \Upsilon(1S)+\eta^{\prime})/\BR(\Upsilon(4S)\to \Upsilon(1S)+\pi^+\pi^-)$ are obtained to be $0.20\pm0.06$ and $0.42\pm0.11$, where the uncertainties include both systematic and statistical uncertainties. The former ratio, in particular, is in agreement with the expected value in the picture of $\Upsilon(4S)$ as an admixture of a state containing light quarks in addition to the $b\bar b$ pair~\cite{Voloshin26}. The measurements of the $\eta^{\prime}$ transition may shed light on the puzzle of hadronic transitions between heavy quarkonia.

\vspace{0.5cm}
$\bullet$ $\Upsilon(5S)\to \Upsilon(1S,2S,3S)+\pi^+\pi^-/K^+K^-$
\vspace{0.1cm}

\begin{figure*}[htbp]
\centering
\includegraphics[width=12cm]{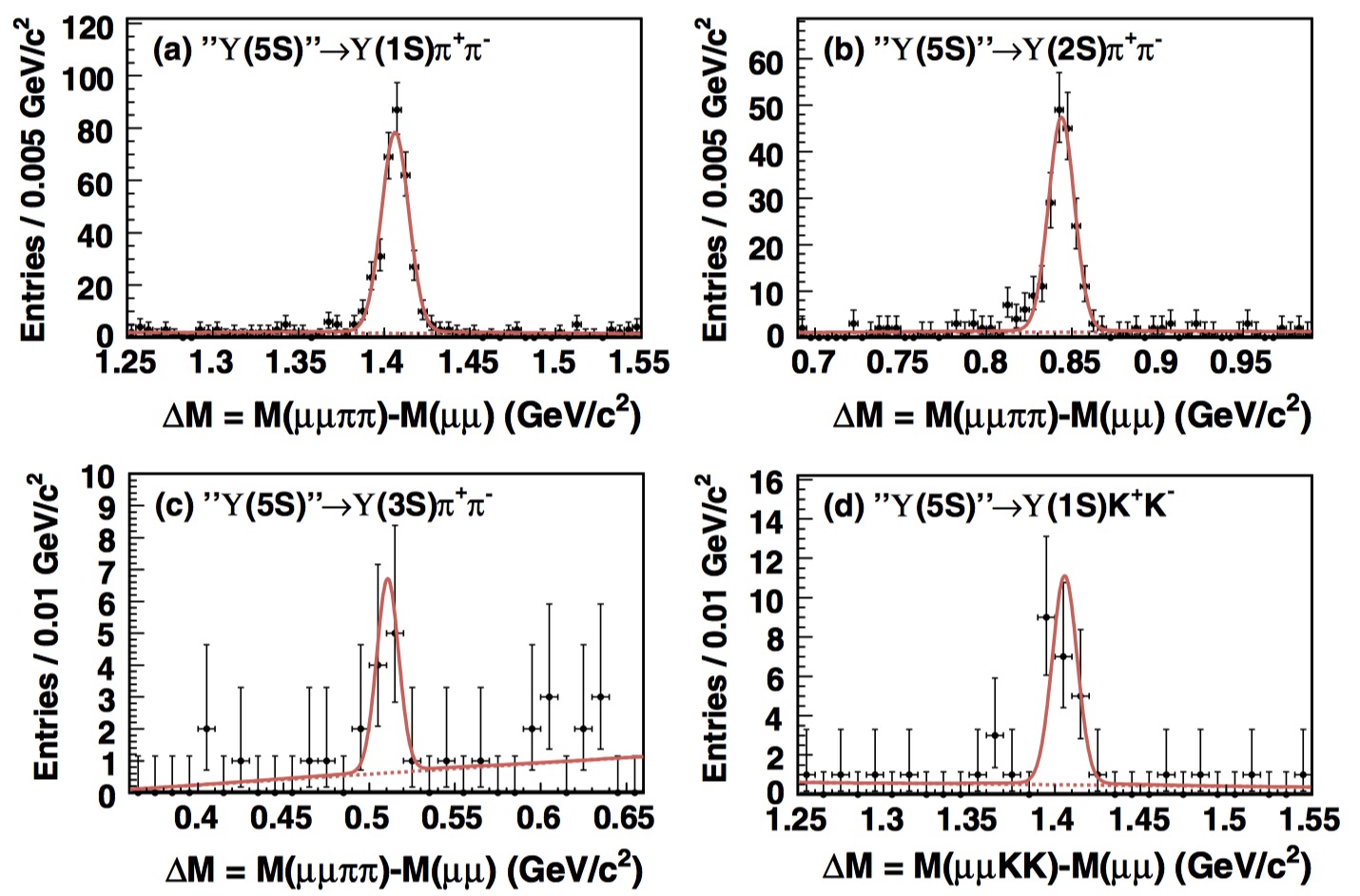}
\caption{The $\Delta M$ distributions for (a) $\Upsilon(5S)\to \Upsilon(1S)\pi^+\pi^-$, (b) $\Upsilon(5S)\to \Upsilon(2S)\pi^+\pi^-$, (c) $\Upsilon(5S)\to \Upsilon(3S)\pi^+\pi^-$, and (d) $\Upsilon(5S)\to \Upsilon(1S)K^+K^-$ from Belle~\cite{Chen100}. The solid red lines show the best fits to the data, while the dashed red lines show the backgrounds.}\label{SIIB3}
\end{figure*}

Besides the hadronic transitions between lower $\Upsilon$ resonances, the $\pi^+\pi^-$ and $K^+K^-$ transitions from $\Upsilon(5S)$ to $\Upsilon(1S,2S,3S)$ were also studied by Belle~\cite{Chen100}. The $\Delta M$ = $M(\mu\mu\pi\pi/\mu\mu KK)-M(\mu\mu)$ distributions for $\Upsilon(5S)\to \Upsilon(1S)\pi^+\pi^-$, $\Upsilon(5S)\to \Upsilon(2S)\pi^+\pi^-$, $\Upsilon(5S)\to \Upsilon(3S)\pi^+\pi^-$, and $\Upsilon(5S)\to \Upsilon(1S)K^+K^-$ are shown in Fig.~\ref{SIIB3}.
The branching fractions of these four decays are measured to be $(0.53\pm0.03\pm0.05)\%$, $(0.78\pm0.06\pm0.11)\%$, $(0.48^{+0.18}_{-0.15}\pm0.07)\%$, and $(0.061^{+0.016}_{-0.014}\pm0.010)\%$, respectively. Notably, these values exceed their counterparts between lower $\Upsilon$ resonances by more than 2 orders of magnitude~\cite{Aubert78,Guido96}. The unexpectedly large branching fractions disagree with the expectation of a pure $b\bar b$ state for the $\Upsilon(5S)$ resonance, unless there is a new mechanism to enhance the decay rate. The later observations of $Z_b(10610)$ and $Z_b(10650)$~\cite{Bondar108} decaying into $\pi\Upsilon(1S)$, $\pi\Upsilon(2S)$ and $\pi\Upsilon(3S)$ are supposed to contribute to the enhancements in the dipion transitions of $\Upsilon(5S)$.

\section{$\allu$ measurements at the LHC}

\subsection{The $pp$, $pPb$, and $PbPb$ datasets at the LHC}

\begin{figure*}[htbp]
\centering
\includegraphics[width=10cm]{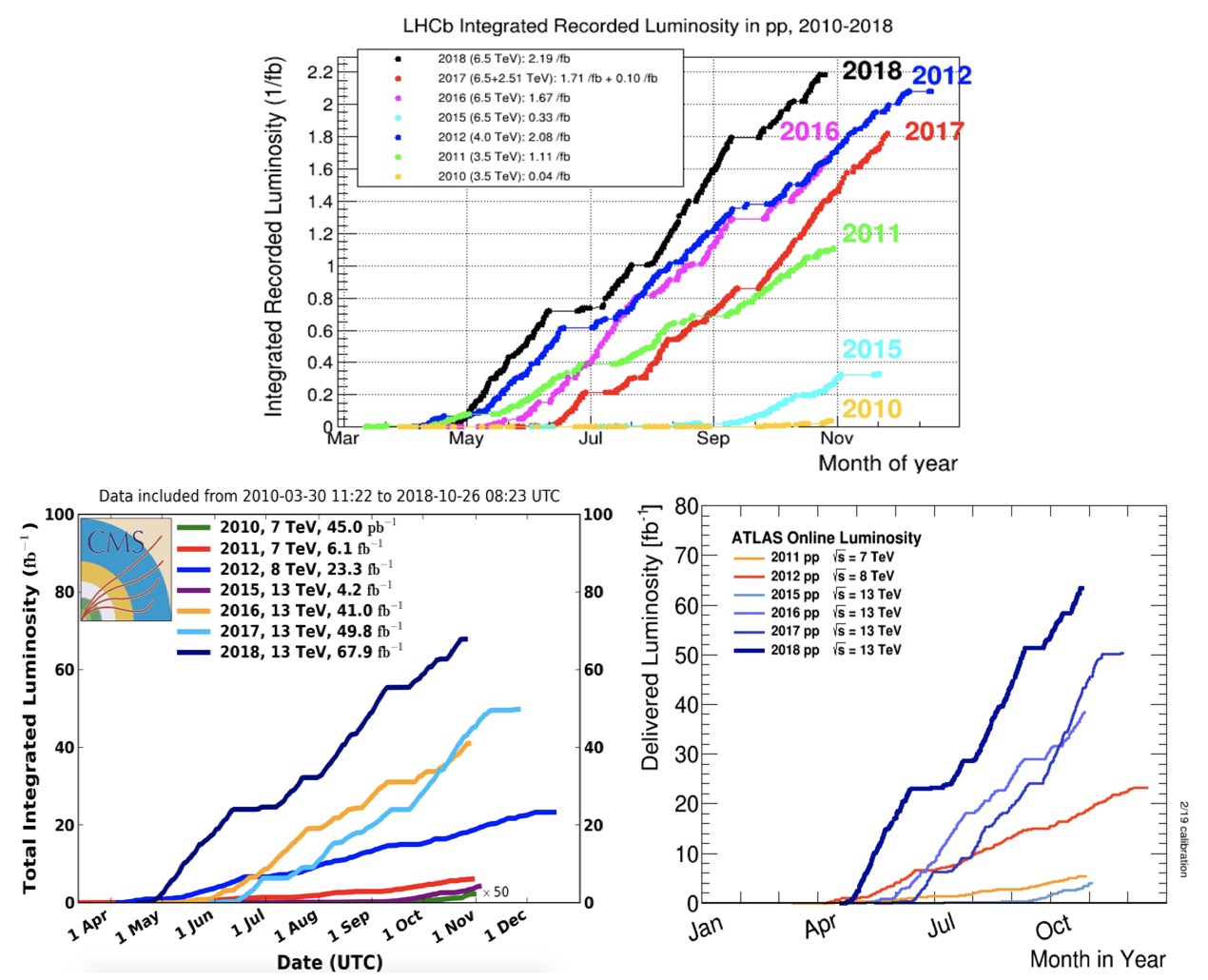}
\caption{The integrated luminosities of the $pp$ data recorded by LHCb, CMS, and ATLAS in the past decade~\cite{LHCpublic}.}\label{pp-lum}
\end{figure*}

The integrated luminosities of the $pp$ colliding data recorded by LHCb, CMS, and ATLAS in the past decade are displayed in Fig.~\ref{pp-lum}~\cite{LHCpublic}. The total integrated luminosities are up to hundreds of fb$^{-1}$ in CMS and ATLAS, and only a few fb$^{-1}$ in LHCb. The data samples were collected in a wide C.M. energy range from a few to 13 TeV. With these samples, LHCb, CMS, and ATLAS are able to measure the $\allu$ cross sections and polarizations precisely. In addition to the $pp$ colliding data, ALICE and CMS collected both $pPb$ and $PbPb$ colliding data, and LHCb collected $pPb$ colliding data, to probe the QGP mechanism by comparing the $\allu$ yields from $PbPb$ and $pPb$ collisions with those from $pp$ collisions. The total luminosities recorded in $pp$, $pPb$, and $PbPb$ collisions at the LHC are summarized in Table~\ref{features}.

The experiments at the LHC have different features in detecting muons. At LHCb, muons are identified by a system composed of alternating layers of iron and multiwire proportional chambers~\cite{S08005,P02022}. The trigger system of LHCb consists of a hardware stage based on information from the muon systems, followed by a software stage, which employs a full event reconstruction~\cite{P04022}. With high trigger efficiencies and good resolutions, the measurements consisting muon final states at LHCb are very promising. ATLAS~\cite{S08003} and CMS~\cite{S08004} own the muon spectrometers equipped with drift tubes, cathode strip chambers, and resistive plate chambers (CMS only). Owing to a larger magnetic field intensity (2 T at ATLAS and 3.8 T at CMS) and extra resistive plate chambers, the performances on the efficiency and resolution for muons at CMS, especially in the low momentum region, are better than those at ATLAS. The muon identification at ALICE utilizes a front absorber and an iron wall located between the tracking and trigger systems~\cite{S08002}. The performances on the efficiency and resolution for muons at ALICE are not very good compared to LHCb, ATLAS, and CMS. The features of the performances in $\Upsilon\to\mu^+\mu^-$ measurements at the LHC are shown in Table~\ref{features}, which are summarized from the related public webs at CERN~\cite{LHCpublic}, technical design reports~\cite{S08005,P02022,P04022,S08003,S08004,S08002}, and published papers~\cite{Khachatryan83,Aad052004,Aaij2025,
Acharya361}. We note that, the rapidity, transverse momentum, and resolution denote the related parameters of $\Upsilon$, while the trigger efficiency applies to a single muon.

\linespread{1.2}
\begin{table*}[htbp]
\small
\caption{The features of the performances in $\Upsilon\to\mu^+\mu^-$ measurements at the LHC~\cite{LHCpublic,S08005,P02022,P04022,S08003,S08004,S08002,Khachatryan83,Aad052004,Aaij2025,
Acharya361}.}
\vspace{0.2cm}
\label{features}
\centering
\begin{tabular}{c c c c c}
\hline\hline
Experiments & LHCb & CMS & ATLAS & ALICE \\\hline
\multirow{2}{*}{Total luminosity} & 9 fb$^{-1}$ ($pp$) & 192 fb$^{-1}$ ($pp$) & \multirow{2}{*}{182 fb$^{-1}$ ($pp$)} &  24.8 nb$^{-1}$ ($pPb$) \\
& 33.4 nb$^{-1}$ ($pPb$) & 216.2 nb$^{-1}$ ($pPb$), 2.5 nb$^{-1}$ ($PbPb$) & & 1.6 nb$^{-1}$ ($PbPb$) \\
Rapidity & $1.5<|y|<5$ & $|y|<2.4$ & $|y|<2.4$ & $2<|y|<4.5$ \\
Transverse momentum & $<$30 GeV/c & $<$100 GeV/c & $<$70 GeV/c & $<$15 GeV/c \\
Trigger efficiency & $>$95\% & 70\%$-$100\% & 40\%$-$90\% & $\sim$50\% \\
$\Upsilon$ mass resolution & $\sim$50 MeV/$c^2$ & 50$-$100 MeV/$c^2$ & $>$100 MeV/$c^2$ & $>$100 MeV/$c^2$ \\
\hline\hline
\end{tabular}
\end{table*}

\subsection{$\allu$ productions in $pp$ collisions}

\subsubsection{$\Upsilon(1S,2S,3S)$ cross sections}

\begin{figure*}[htbp]
\centering
\includegraphics[width=16cm]{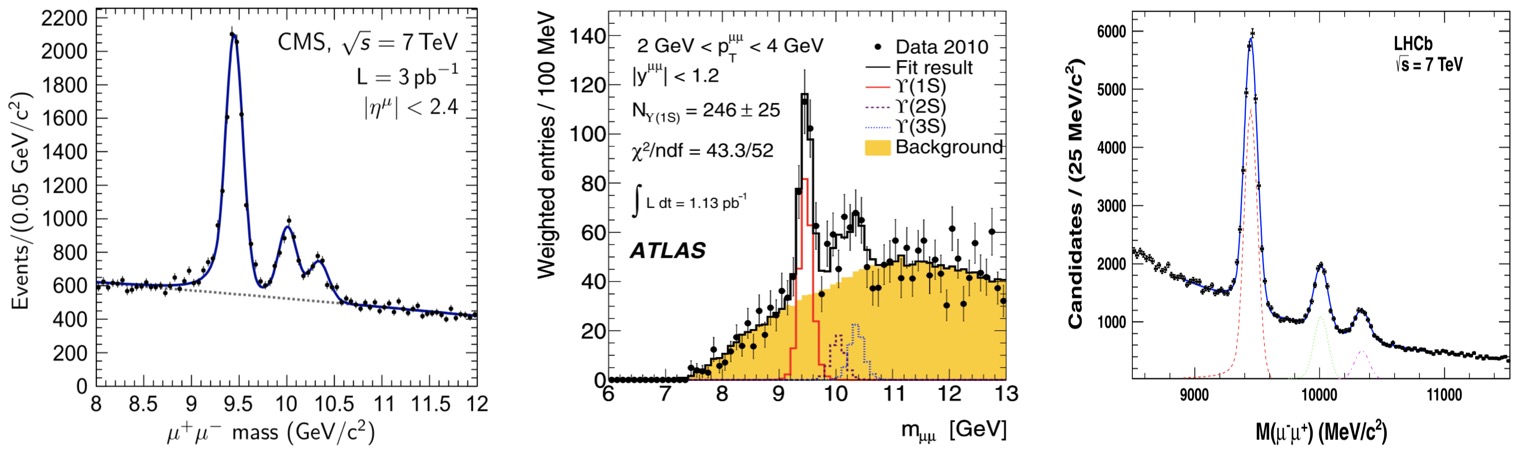}
\caption{The left: The dimuon invariant mass distribution with $p_{\rm T}$ $<$ 30 GeV/$c$ and $|{\rm y}|$ $<$ 2 from CMS~\cite{Khachatryan83}; the solid line shows the result of a fit to the invariant mass distribution, with the dashed line denoting the background component. The middle: The dimuon invariant mass distribution with 2 $<$ $p_{\rm T}$ $<$ 4 GeV/$c$ and $|{\rm y}|$ $<$ 1.2 from ATLAS~\cite{Aad705}; the shaded histogram shows the background contribution, and three other histograms show the three $\Upsilon$ states components. The right: The dimuon invariant mass distribution with $p_{\rm T}$ $<$ 15 GeV/$c$ and 2 $<$ $|{\rm y}|$ $<$ 4.5 from LHCb~\cite{Aaij2025}; the blue curve is the total fitted result, and the dashed curves show the $\Upsilon(1S)$, $\Upsilon(2S)$, and $\Upsilon(3S)$ components.}\label{plot1}
\end{figure*}

Using data samples of 3.1 pb$^{-1}$, 1.13 pb$^{-1}$, and 25 pb$^{-1}$, the $\Upsilon(1S,2S,3S)$ production cross sections in $pp$ collisions at $\sqrt{s}$ = 7 TeV were measured by CMS~\cite{Khachatryan83}, ATLAS~\cite{Aad705}, and LHCb~\cite{Aaij2025}, respectively. The $\Upsilon(1S,2S,3S)$ states are typically reconstructed via the dimuon decay channel: $\Upsilon(1S,2S,3S)\to \mu^+\mu^-$. To ensure that the trigger and muon reconstruction efficiencies are high and not rapidly changing within the acceptance window, muon candidates are required to satisfy: $p^{\mu}_{\rm T}$ $>$ 3.5 GeV/$c$ for $|\eta^{\mu}|$ $<$ 1.6 and $p^{\mu}_{\rm T}$ $>$ 2.5 GeV/$c$ for 1.6 $<$ $|\eta^{\mu}|$ $<$ 2.4 at CMS~\cite{Khachatryan83}, and $p^{\mu}_{\rm T}$ $>$ 4 GeV/$c$ for $|\eta^{\mu}|$ $<$ 2.5 at ATLAS~\cite{Aad705}, where $p^{\mu}_{\rm T}$ and $\eta^{\mu}$ are transverse momentum and pseudorapidity of the muon. The dimuon invariant mass distributions in ranges of $p_{\rm T}$ $<$ 30 GeV/$c$ and $|{\rm y}|$ $<$ 2 at CMS, 2 $<$ $p_{\rm T}$ $<$ 4 GeV/$c$ and $|{\rm y}|$ $<$ 1.2 at ATLAS, and $p_{\rm T}$ $<$ 15 GeV/$c$ and 2 $<$ $|{\rm y}|$ $<$ 4.5 at LHCb are shown in Fig.~\ref{plot1}. Here, $p_{\rm T}$ is the dimuon transverse-momentum, and ${\rm y} = \frac{1}{2}{\rm ln}(\frac{E+p_{||}}{E-p_{||}})$ is the rapidity, where $E$ is the energy and $p_{||}$ is the momentum parallel to the beam axis of the muon pair. The $\Upsilon(1S,2S,3S)$ resonances are clearly visible and fully resolved at CMS and LHCb. Assuming unpolarized $\Upsilon(1S,2S,3S)$ production, the products of the $\Upsilon(1S,2S,3S)$ cross sections and dimuon branching fractions in $pp$ collisions from CMS~\cite{Khachatryan83} and LHCb~\cite{Aaij2025} are summarized in Table~\ref{ppcs}.

\linespread{1.2}
\begin{table}[htbp]
\tiny
\caption{The products of the $\Upsilon(1S,2S,3S)$ cross sections and dimuon branching fractions in $pp$ collisions at 7 TeV from CMS~\cite{Khachatryan83} and LHCb~\cite{Aaij2025}. For the results from CMS~\cite{Khachatryan83}, the first uncertainty is statistical, the second is systematic, and the third is associated with the estimation of the integrated luminosity of the data sample. For the results from LHCb~\cite{Aaij2025}, the first uncertainty is statistical, the second is systematic, and the third is due to the unknown polarizations of the three $\Upsilon$ states.}
\vspace{0.2cm}
\label{ppcs}
\centering
\begin{tabular}{c c}
\hline\hline
\multicolumn{2}{c}{$|$y$|$ $<$ 2 \& $p_{\rm T}$ $<$ 30 GeV/$c$~\cite{Khachatryan83}} \\
$\sigma(pp\to \Upsilon(1S)X)\BR(\Upsilon(1S)\to \mu^+\mu^-)$ & $(7.37\pm 0.13^{+0.61}_{-0.42}\pm 0.81)~{\rm nb}$ \\
$\sigma(pp\to \Upsilon(2S)X)\BR(\Upsilon(1S)\to \mu^+\mu^-)$ & $(1.90\pm 0.08^{+0.20}_{-0.14}\pm 0.21)~{\rm nb}$ \\
$\sigma(pp\to \Upsilon(3S)X)\BR(\Upsilon(1S)\to \mu^+\mu^-)$ & $(1.02\pm 0.07^{+0.11}_{-0.08}\pm 0.11)~{\rm nb}$ \\\hline
\multicolumn{2}{c}{2.0 $<$ y $<$ 4.5 \& $p_{\rm T}$ $<$ 15 GeV/$c$~\cite{Aaij2025}} \\
$\sigma(pp\to \Upsilon(1S)X)\BR(\Upsilon(1S)\to \mu^+\mu^-)$ & $(2.29\pm 0.01\pm 0.01^{+0.19}_{-0.37})~{\rm nb}$\\
$\sigma(pp\to \Upsilon(2S)X)\BR(\Upsilon(1S)\to \mu^+\mu^-)$ & $(0.562\pm 0.007\pm 0.023^{+0.048}_{-0.092})~{\rm nb}$\\
$\sigma(pp\to \Upsilon(3S)X)\BR(\Upsilon(1S)\to \mu^+\mu^-)$ & $(0.283\pm 0.005\pm 0.012^{+0.025}_{-0.048})~{\rm nb}$\\\hline\hline
\end{tabular}
\end{table}

\begin{figure*}[htbp]
\centering
\includegraphics[width=11cm]{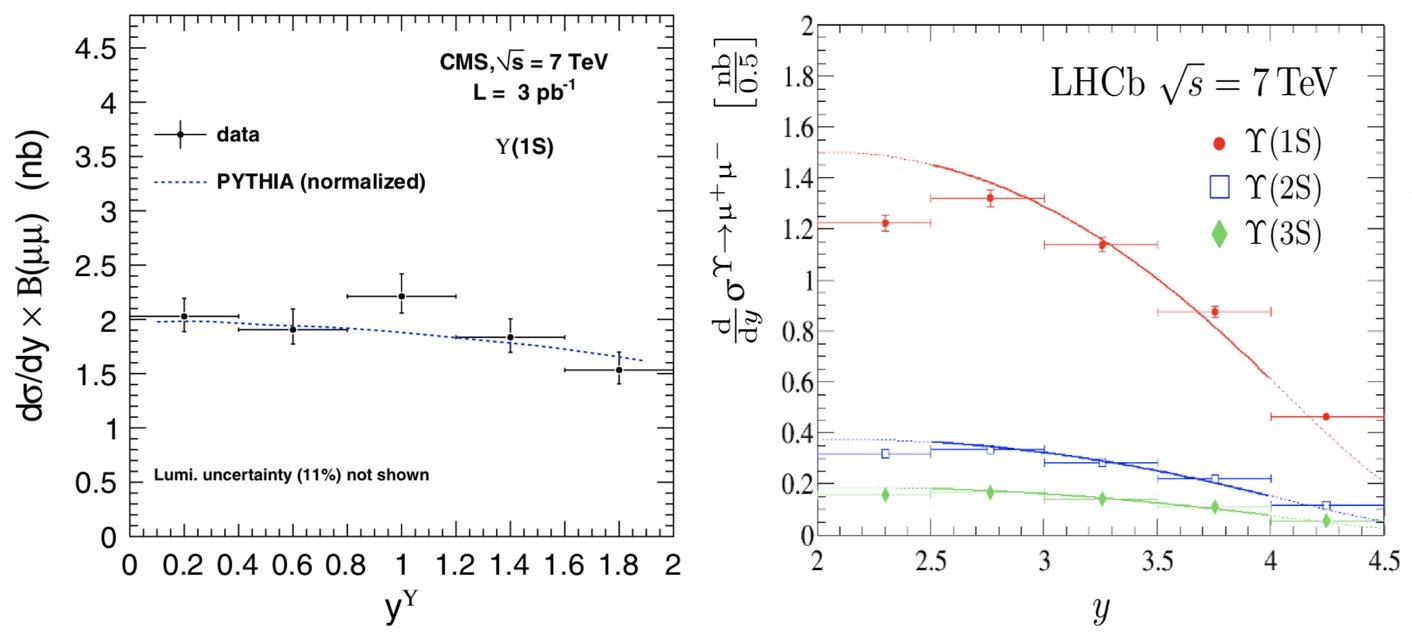}
\caption{The differential $\Upsilon$ cross sections as a function of rapidity y varying from 0.0 to 2.0 from CMS~\cite{Khachatryan83} (left) and 2.0 to 4.5 from LHCb~\cite{Aaij103} (right) in the transverse momentum range of $p_{\rm T}$ $<$ 30 GeV/$c$. In the left plot, the blue dashed line shows the normalized {\sc pythia} prediction~\cite{pythia}. In the right plot, the thick lines show the fit results with NRQCD color-octet model predictions from Refs.~\cite{Kisslinger039902,Kisslinger1350120} in the region 2.0 $<$ y $<$ 4.0, and dashed lines show the extrapolations to the full region 2.0 $<$ y $<$ 4.5.}\label{plot3}
\end{figure*}

The $\Upsilon(nS)$ ($n$ = 1, 2, 3) differential cross section is computed as~\cite{Khachatryan83}
\begin{equation}\nonumber
\begin{aligned}
\frac{d^2\sigma(pp\to\Upsilon(nS)X)}{d_{p_{\rm T}}d{\rm y}}\BR({\Upsilon(nS)\to\mu^+\mu^-}) = \\
\frac{N_{\Upsilon(nS)}(\mathcal{A},\varepsilon)}{\mathcal{L}\Delta_{p_{\rm T}}\Delta_{\rm y}},~~~~~~~~~~~~~~~~
\end{aligned}
\end{equation}
where $N_{\Upsilon(nS)}$ is the signal yield corrected by the dimuon event weights given by the inverse product of the detector acceptance $\mathcal{A}$ and the reconstruction efficiency $\varepsilon$, $\mathcal{L}$ is the integrated luminosity, and $\Delta_{p_{\rm T}}$ and $\Delta_{\rm y}$ are the widths of the bins in transverse momentum ($p_{\rm T}$) and rapidity (y). The symbol $X$ is used to indicate that the measurements include the feed-down contributions originating from higher-mass states, such as the $\chi_b$ family and $\Upsilon(3S)$.
The differential $\Upsilon(1S)$ cross sections as a function of rapidity y within y $<$ 2 were given by CMS, as illustrated in the left plot of Fig.~\ref{plot3}~\cite{Khachatryan83}. As a complement to the phase-space coverage of CMS, LHCb studied the $\Upsilon(1S)$ cross sections over the range of 2.0 $<$ y $<$ 4.5 in proton-proton collisions at $\sqrt{s}$ = 7 TeV (see Fig.~\ref{plot3} (right))~\cite{Aaij103}. The thick lines in the right plot of Fig.~\ref{plot3} show the fit results with NRQCD color-octet model predictions from Refs.~\cite{Kisslinger039902,Kisslinger1350120} in the region of 2.0 $<$ y $<$ 4.0, and dashed lines show the extrapolations to the full region of 2.0 $<$ y $<$ 4.5. From Fig.~\ref{plot3}, the rapidity y dependence of the cross section shows a slight decline towards the higher side.

\begin{figure*}[htbp]
\centering
\includegraphics[width=12cm]{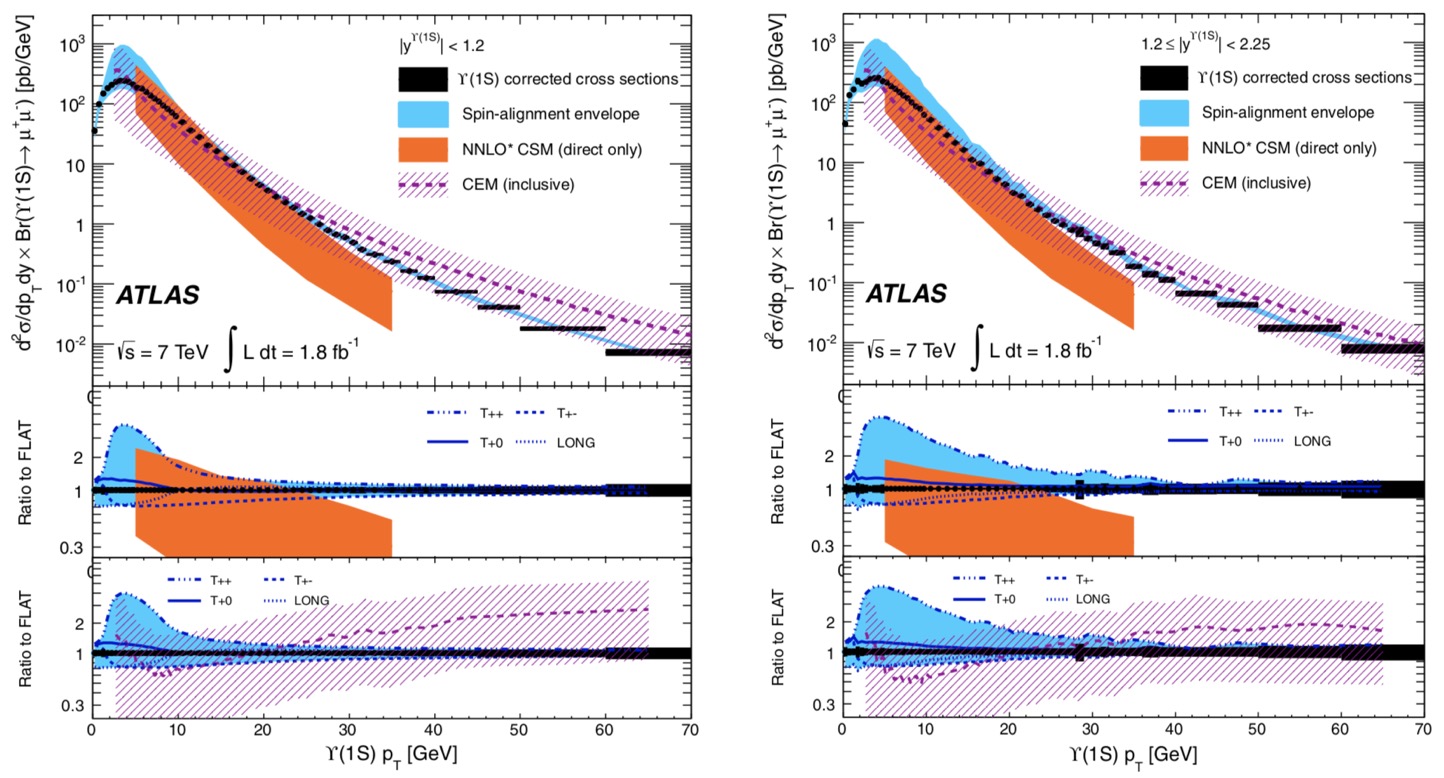}
\caption{Differential cross sections multiplied by the dimuon branching fractions for the $\Upsilon(1S)$ production in the ranges of $|{\rm y}|$ $<$ 1.2 (left) and 1.2 $<$ $|{\rm y}|$ $<$ 2.25 (right) from ATLAS~\cite{Aad052004}. Theoretical predictions are shown as ratios to the data in the lower panes for CEM (middle) and CSM (bottom), along with the variations of the cross section measurement under the spin-alignment scenarios.}\label{CMSPt}
\end{figure*}

With the increase of $p_{\rm T}$, several theoretical models with higher-order corrections become more sensitive. Therefore, the cross section measurement in high $p_{\rm T}$ is of great importance for distinguishing among the models. Using a $pp$ collision data sample of 1.8 fb$^{-1}$ at $\sqrt{s}=$ 7 TeV, ATLAS measured the production cross sections of $\Upsilon(1S,2S,3S)$ mesons in a larger momentum region of $p_{\rm T}$ $<$ 70 GeV/c and in the rapidity interval of y $<$ 2.25~\cite{Aad052004}. Figure~\ref{CMSPt} shows the measured differential cross sections of $\Upsilon(1S)$ as a function of $p_{\rm T}$, in comparison with theoretical calculations. By varying the spin-alignment assumption from the nominal isotropic assumption for muons, the derived difference is shown by the cyan bands in Fig~\ref{CMSPt}. The predictions from the next-to-next-to-leading-order (NNLO) QCD-based calculations using the color-singlet mechanism (CSM)~\cite{Owens1501,Kartvelishvili1315,Chang425,Berger1521,Baier1,Baier251} and the color evaporation model (CEM)~\cite{Fritzsch217,Halzen105,Gluck2324,Barger253,Amundson127,Amundson323} are indicated by the orange bands and cross hatched violet bands. Apparently, the discrepancy between the predictions from CSM and data points is large if one extends the CSM calculations to the higher $p_{\rm T}$ region. It is probably because CSM does not take into account the $\Upsilon$ productions that arise from higher $\Upsilon$ or $\chi_{bJ}(nP)$ decays, which mostly occur at modest rates.

\begin{figure}[htbp]
\centering
\includegraphics[width=7cm]{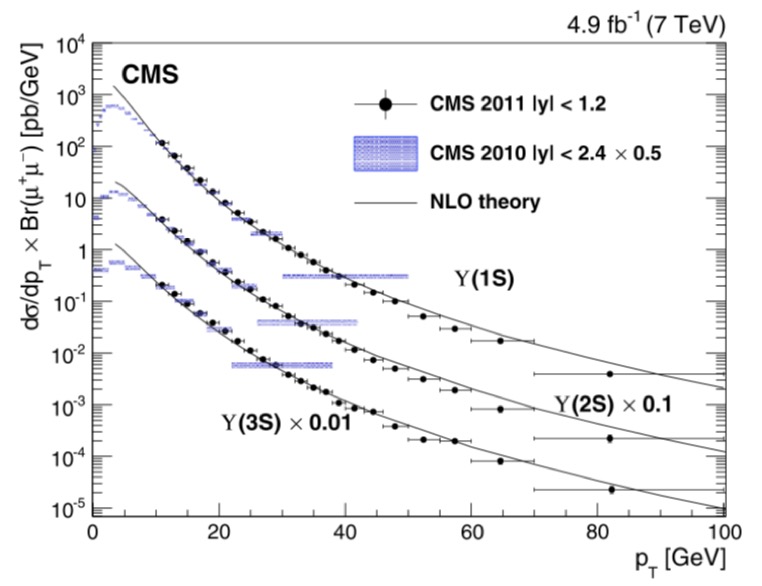}
\caption{The differential $\Upsilon(1S,2S,3S)$ cross sections times dimuon branching fractions as a function of $p_{\rm T}$ at $\sqrt{s}=$ 7 TeV in the rapidity range of $|$y$|$ $<$ 1.2 from CMS~\cite{Khachatryan14}. The solid lines are the NLO NRQCD color-octet calculations from Ref.~\cite{Gong032001} extended by the authors to cover the range of $p_{\rm T}$ $<$ 100 GeV/c.}\label{plot4}
\end{figure}

\begin{figure}[htbp]
\centering
\includegraphics[width=6cm]{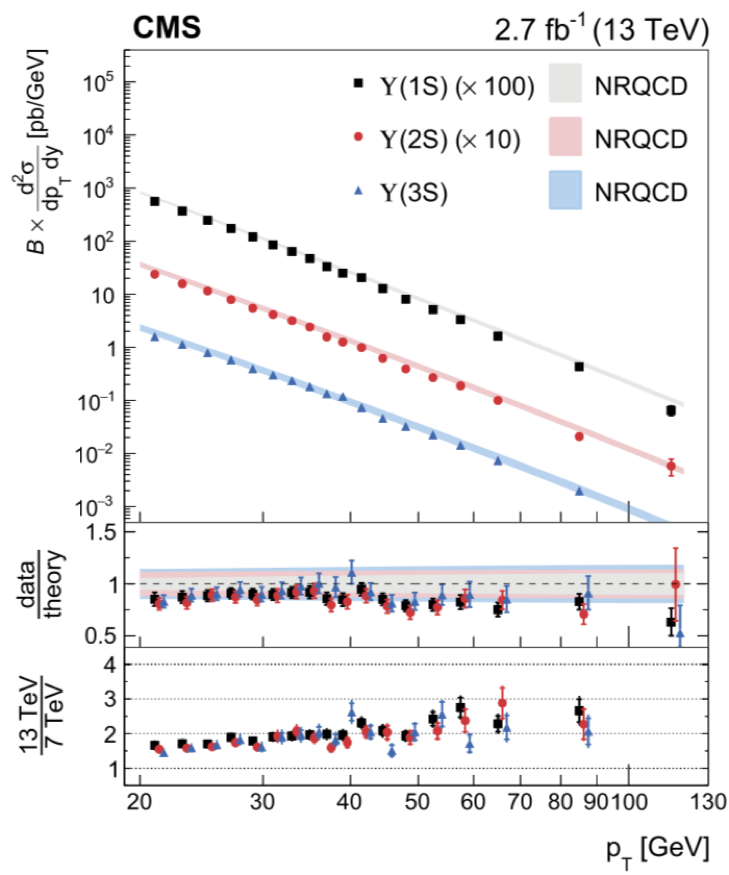}
\caption{The differential $\Upsilon(1S,2S,3S)$ cross sections times dimuon branching fractions as a function of $p_{\rm T}$ at $\sqrt{s}=$ 13 TeV in the rapidity range of $|$y$|$ $<$ 1.2 from CMS~\cite{Khachatryan2145411}. The bands show the predictions by NLO NRQCD~\cite{Ma042002,Han014028}. The middle panel shows the ratios of measurements to the theory, and the lower panel shows the ratios of cross sections measured at $\sqrt{s}$ = 13 TeV to those measured at 7 TeV~\cite{Khachatryan14}.}\label{plot4add}
\end{figure}

CMS extended the maximum of $p_{\rm T}$ to 100 GeV/c to study the $p_{\rm T}$ dependence using a 4.9 fb$^{-1}$ $pp$ collision data sample at $\sqrt{s}$ = 7 TeV~\cite{Khachatryan14}. The $\Upsilon(1S,2S,3S)$ differential cross sections times dimuon branching fractions for $|$y$|$ $<$ 1.2 are shown in Fig.~\ref{plot4}. The results are consistent with the measurements within $p_{\rm T}$ $<$ 70 GeV/c by ATLAS~\cite{Aad052004}. The solid lines are the NLO NRQCD color-octet calculations from Ref.~\cite{Gong032001} and extended by the authors to cover the range $p_{\rm T}$ $<$ 100 GeV/c. A similar measurement~\cite{Khachatryan2145411} was performed using a data set of 2.7 fb$^{-1}$, but at a higher energy of $\sqrt{s}$ = 13 TeV. In order to perform the measurement in a kinematical region where muon acceptance is high, the requirements are applied as $p^{\mu}_{\rm T}$ $>$ 4.5 GeV/$c$ for $|\eta^{\mu}|$ $<$ 0.3 and $p^{\mu}_{\rm T}$ $>$ 4.0 GeV/$c$ for 0.3 $<$ $|\eta^{\mu}|$ $<$ 1.4. Figure~\ref{plot4add}  shows the differential cross sections times dimuon branching fractions for 13 TeV CMS data, together with a comparison of the $\Upsilon(1S,2S,3S)$ differential cross sections for the 7 and 13 TeV data sets. The differential cross sections times dimuon branching fractions for 13 TeV are well described by NLO NRQCD~\cite{Ma042002,Han014028}. The cross sections of all the three $\Upsilon$ states at 13 TeV are factors of 2 to 3 larger than the corresponding cross sections at 7 TeV, changing slowly as a function of $p_{\rm T}$.

The measurements of the $\Upsilon(1S,2S,3S)$ cross sections in $pp$ collisions at a lower C.M. energy, i.e., $\sqrt{s}$ = 2.76 TeV were investigated by LHCb using a data sample of 3.3 pb$^{-1}$~\cite{Aaij2835}. The differential cross sections times dimuon branching fractions as functions of $p_{\rm T}$ and y were presented in Ref.~\cite{Aaij2835}. The theoretical predictions, based on the NLO NRQCD calculation~\cite{Wang85}, provide a good description of the data in the higher kinematic region. The products of the $\Upsilon(1S,2S,3S)$ cross sections and dimuon branching fractions in the regions of $p_{\rm T}$ $<$ 15 GeV/$c$ and 2.0 $<$ y $<$ 4.5, assuming unpolarised production, are listed in Table~\ref{ppcsLHCb}. The results at $\sqrt{s}$ = 2.76 TeV are approximately half of those at $\sqrt{s}$ = 7 TeV~\cite{Aaij2025}. These results, together with the comparison between $\sqrt{s}$ = 7 TeV and $\sqrt{s}$ =13 TeV from CMS~\cite{Khachatryan2145411}, provide useful information for characterizing the $s$-dependence of $\Upsilon(1S,2S,3S)$ cross sections in $pp$ collisions.

\linespread{1.2}
\begin{table}[htbp]
\tiny
\caption{The products of the $\Upsilon(1S,2S,3S)$ cross sections and dimuon branching fractions in $pp$ collisions at 2.76 TeV from LHCb~\cite{Aaij2835}.}
\vspace{0.2cm}
\label{ppcsLHCb}
\centering
\begin{tabular}{c c}
\hline\hline
$\sigma(pp\to \Upsilon(1S)X)\BR(\Upsilon(1S)\to \mu^+\mu^-)$ & $(1.111\pm 0.043\pm0.044)~{\rm nb}$\\
$\sigma(pp\to \Upsilon(2S)X)\BR(\Upsilon(1S)\to \mu^+\mu^-)$ & $(0.264\pm 0.023\pm0.011)~{\rm nb}$\\
$\sigma(pp\to \Upsilon(3S)X)\BR(\Upsilon(1S)\to \mu^+\mu^-)$ & $(0.159\pm 0.020\pm0.007)~{\rm nb}$\\
\hline\hline
\end{tabular}
\end{table}

\subsubsection{$\Upsilon(1S,2S,3S)$ polarizations}

Measurements of the $\Upsilon(1S,2S,3S)$ polarizations, complementing their cross section measurements, provide important information on their production mechanisms. CMS carried out these measurements by studying the angular distribution of the leptons produced in the $\Upsilon\to\mu^+\mu^-$ decay~\cite{Faccioli657,Faccioli151802}:
\begin{equation}\nonumber
\begin{aligned}
W(\theta,\phi) \propto~~~~~~~~~~~~~~~~~~~~~~~~~~~~~ \\
\frac{1}{3+\lambda_{\theta}} (1+\lambda_{\theta}{\rm cos}^2\theta+\lambda_{\phi}{\rm sin}^2\theta{\rm cos}2\phi+\lambda_{\theta\phi}{\rm sin}2\theta {\rm cos}\phi),
\end{aligned}
\end{equation}
where $\theta$ and $\phi$ are the polar and azimuthal angles of the outgoing leptons with respect to the quantization axis of the chosen polarization frame; $\lambda$ is the set of polarization parameters; the parameter $\lambda_{\theta}$ is 0 (1) for fully longitudinal (transverse) polarization. The polarization parameters depend on the reference frame in which they are measured. The three most commonly used reference frames are the helicity (HX)~\cite{Jacob404}, Collins-Soper (CS)~\cite{Collins2219}, and perpendicular helicity (PX)~\cite{Braaten014025} frames. In the HX frame, the $z$ axis is defined as the direction of the $\Upsilon$ momentum in the C.M. frame of the colliding protons. In the CS frame, the $z$ axis is defined such that it bisects the angle between ${\vec p}_1$ and $-{\vec p}_2$ in the rest frame of the $\Upsilon$ meson, where ${\vec p}_1$ and ${\vec p}_2$ are the three-momenta of the colliding protons in the rest frame of the $\Upsilon$ meson. The PX frame is orthogonal to the CS frame. The $y$ axis is always taken along the direction of the vector product of the two beam directions in the $\Upsilon$ rest frame. The $x$ axis is defined to complete a right-handed coordinate system.

\begin{figure*}[htbp]
\centering
\includegraphics[width=15cm]{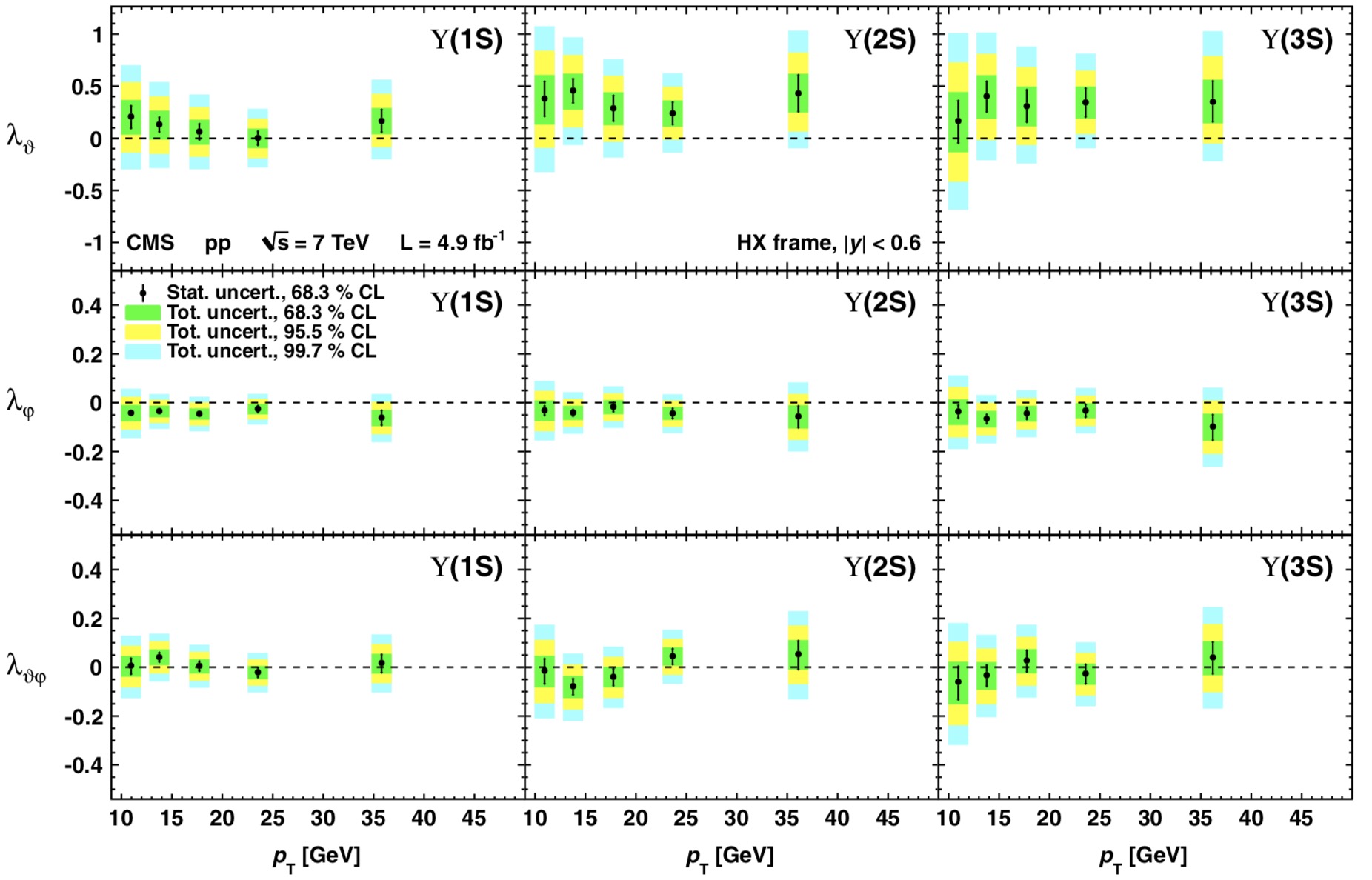}
\caption{Values of the $\lambda_{\theta}$ (top), $\lambda_{\phi}$ (middle), and $\lambda_{\theta\phi}$ (bottom) parameters for the $\Upsilon(1S)$ (left), $\Upsilon(2S)$ (middle), and $\Upsilon(3S)$ (right) resonances, in the HX frame, as a function of $p_{\rm T}$ in the rapidity range $|$y$|$ $<$ 0.6 from CMS~\cite{Chatrchyan081802}.}\label{plot6}
\end{figure*}

In the previous measurements from CDF~\cite{Aaltonen151802} and D0~\cite{Abazov182004}, only the $\lambda_{\theta}$ parameter in a single polarization frame was extracted. CMS measured all the polarization parameters $\lambda_{\theta}$, $\lambda_{\phi}$, and $\lambda_{\theta\phi}$, in all the three polarization frames mentioned above, plus the frame-invariant quantity $\tilde \lambda = (\lambda_\theta + 3\lambda_\phi)/(1-\lambda_\phi)$, using a $pp$ data sample of 4.9 fb$^{-1}$  at $\sqrt{s}$ = 7 TeV~\cite{Chatrchyan081802}. As an example, Fig.~\ref{plot6} shows the $\lambda_{\theta}$, $\lambda_{\phi}$, and $\lambda_{\theta\phi}$ measurements as functions of $p_{\rm T}$ in the rapidity range of $|$y$|$ $<$ 0.6 in the HX frame. All polarization parameters are compatible with zero or have small values in the three polarization frames.

\begin{figure*}[htbp]
\centering
\includegraphics[width=10cm]{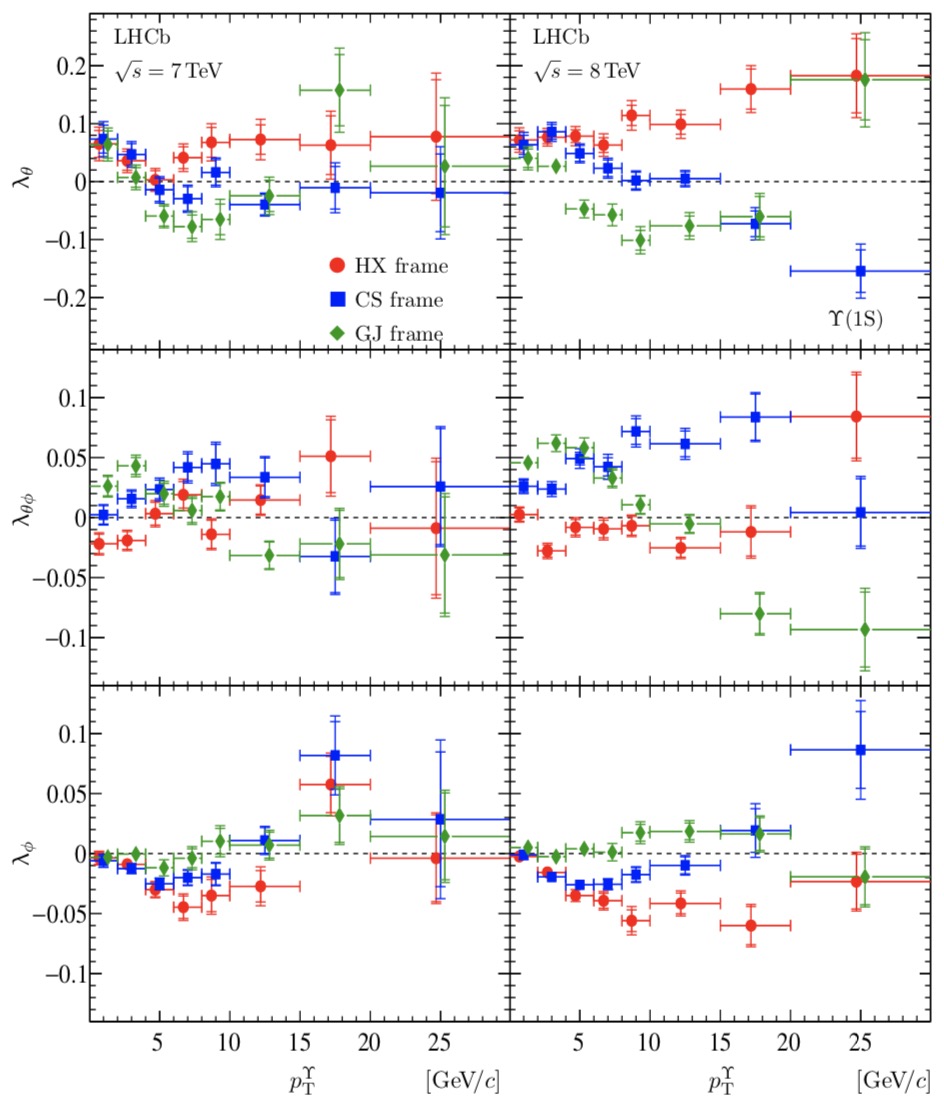}
\caption{The polarization parameters $\lambda_{\theta}$ (top), $\lambda_{\theta\phi}$ (middle), and $\lambda_{\phi}$ (bottom) measured in the HX, CS, and GJ frames for the $\Upsilon(1S)$ state in different bins of $p_{\rm T}$ and rapidity region of 2.2 $<$ y $<$ 4.5, for data collected at $\sqrt{s}$ = 7 (left) and 8 TeV (right) from LHCb~\cite{Aaij110}.}\label{plot7}
\end{figure*}

The polarizations of the $\Upsilon(1S,2S,3S)$ mesons produced in $pp$ collisions at $\sqrt{s}$ = 7 and 8 TeV in the complementary rapidity region of 2.2 $<$ y $<$ 4.5 were measured by LHCb using 1 and 2 fb$^{-1}$ data samples, respectively~\cite{Aaij110}. The measurements were performed in three polarization frames: HX, CS, and Gottfried-Jackson (GJ)~\cite{Gottfried309} frames. In the GJ frame, the $z$ axis is defined as the direction of ${\vec p}_1$ in the rest frame of the $\Upsilon$ meson. Figure~\ref{plot7} shows the polarization parameters $\lambda_{\theta}$, $\lambda_{\theta\phi}$, and $\lambda_{\phi}$ measured in the HX frame for the $\Upsilon(1S)$ state in different bins of $p_{\rm T}$ and rapidity region of 2.2 $<$ y $<$ 4.5, for data collected at $\sqrt{s}$ = 7 and 8 TeV. Similar cases are found for $\Upsilon(2S)$ and $\Upsilon(3S)$ states. No large polarization is observed.

The bottomonium states are heavier and approximately non-relativistic systems, allowing the application of theoretical tools that simplify and constrain the analyses of nonperturbative effects~\cite{Brambilla1534,Brambilla2981}. However, the corresponding predictions~\cite{Gong114021} of strong transverse polarizations are in stark contrast with the negligible polarizations observed by the experiments~\cite{Aaltonen151802,Abazov182004,Chatrchyan081802,Aaij110}. Accordingly, the measurements of $\Upsilon$ polarization have led to new theoretical interpretations for the quarkonium puzzle. For instance, in Ref.~\cite{Faccioli98}, global fits using both cross section and polarization measurements are performed to determine the nonperturbative parameters of bound-state formation. This study reveals unexpected hierarchies in the phenomenological long-distance parameters, which brings a new understanding of the bound-state formation mechanism in QCD.

\subsection{Sequential $\Upsilon$ suppression in heavy ion collisions}

\subsubsection{$\Upsilon$ suppression in $PbPb$ collisions}

\begin{figure*}[htbp]
\centering
\includegraphics[width=13cm]{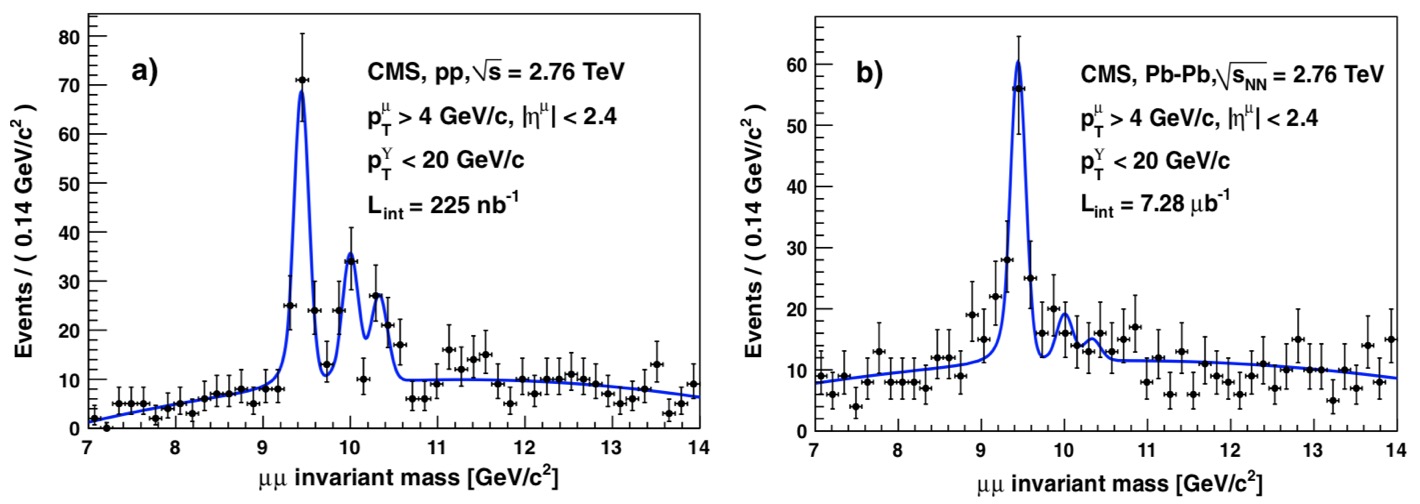}
\caption{Dimuon invariant mass distributions obtained with the (a) 225 nb$^{-1}$ $pp$ and (b) 7.28 $\mu$b$^{-1}$ $PbPb$ data samples at $\sqrt{s}$ = 2.76 TeV from CMS~\cite{Chatrchyan052302}. The solid lines show the best fits to the data.}\label{plot8}
\end{figure*}

The first indication of $\Upsilon$ suppression in heavy ion collisions was reported by CMS at $\sqrt{s}$ = 2.76 TeV~\cite{Chatrchyan052302}. The integrated luminosity used in this measurement corresponds to 7.28 $\mu$b$^{-1}$ for $PbPb$ and 225 nb$^{-1}$ for $pp$ collisions, the latter corresponding approximately to the equivalent nucleon-nucleon luminosity of the $PbPb$ run. Thanks to the good momentum resolution of the CMS detector, the three $\Upsilon$ resonances in the dimuon mass spectrum can be well resolved. The same reconstruction algorithm and analysis criteria are applied to both data sets.~The results are shown in Fig.~\ref{plot8}, where the three $\Upsilon$ peaks are clearly observed in the $pp$ case, but $\Upsilon(2S)$ and $\Upsilon(3S)$ are not significant in the $PbPb$ case.
The suppression effects for $\Upsilon(2S)$ and $\Upsilon(3S)$ in $PbPb$ collisions are more obvious than that for $\Upsilon(1S)$. The ratio on the production rates of $\Upsilon(2S+3S)$ and $\Upsilon(1S)$ in $PbPb$ and $pp$ collisions can be derived according to the $\Upsilon(2S+3S)$ and $\Upsilon(1S)$ signal yields directly since the acceptance and efficiency differences among the reconstructed resonances are cancelled. An extended unbinned maximum likelihood fit to the two invariant mass distributions of Fig.~\ref{plot8} gives the double ratio of the observed signal yields
\begin{equation}\nonumber
\begin{aligned}
\frac{N^{\rm obs}(\Upsilon(2S+3S))/N^{\rm obs}(\Upsilon(1S))|_{PbPb}}{N^{\rm obs}(\Upsilon(2S+3S))/N^{\rm obs}(\Upsilon(1S))|_{pp}} = \\
0.31^{+0.19}_{-0.15}\pm0.03,~~~~~~~~~~~~~~~
\end{aligned}
\end{equation}
in the kinematic region of $p^{\mu}_{\rm T}$ $>$ 4 GeV and $|\eta^{\mu}|$ $<$ 2.4.
The statistical significance of the effect was evaluated to be 2.4$\sigma$ using an ensemble of $1\times10^{6}$ pseudoexperiments.

\begin{figure*}[htbp]
\centering
\includegraphics[width=10cm]{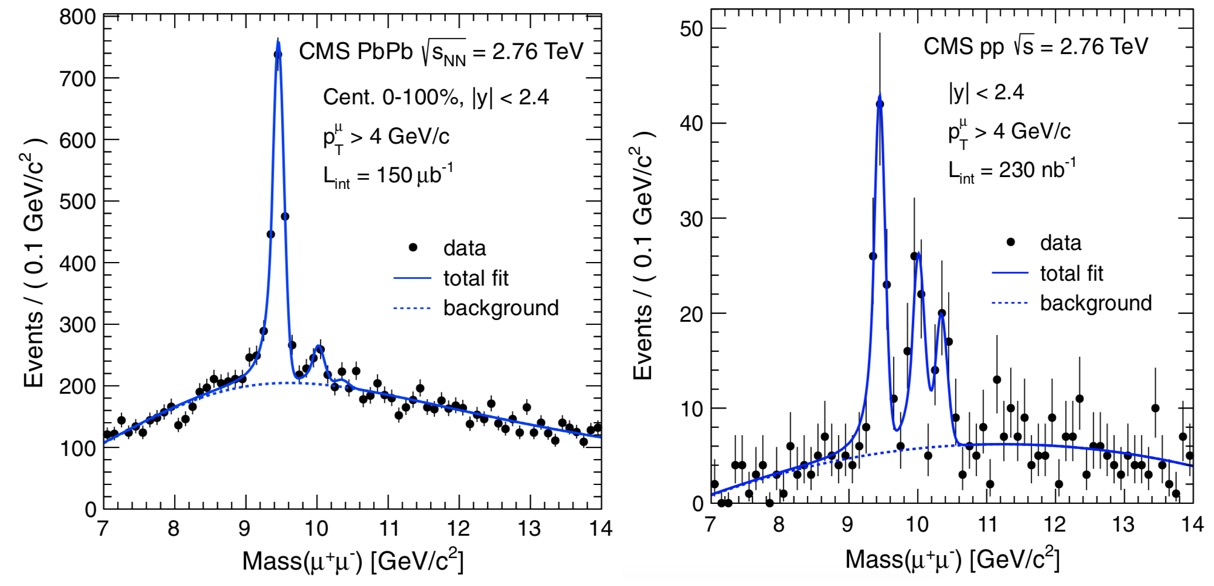}
\caption{Dimuon invariant mass distributions from the (a) 150 $\mu$b$^{-1}$ $PbPb$ and (b) 230 nb$^{-1}$ $pp$ data samples at $\sqrt{s}$ = 2.76 TeV from CMS~\cite{Chatrchyan222301}. The solid lines show the best fits to the data, while the dashed lines show the backgrounds.}\label{plot9}
\end{figure*}

In 2011, CMS collected a number of $pp$ and $PbPb$ data samples at $\sqrt{s}$ = 2.76 TeV to measure the sequential $\Upsilon$ suppression~\cite{Chatrchyan222301}. The total integrated luminosity of the data samples is 150 $\mu$b$^{-1}$, which is approximately 20 times larger than those used in Ref.~\cite{Chatrchyan052302}. The resultant dimuon invariant mass spectra are shown in Fig.~\ref{plot9} for the $pp$ and $PbPb$ datasets. The three $\Upsilon$ peaks are clearly observed in the $pp$ case; the $\Upsilon(3S)$ peak is not prominent in the $PbPb$ case. The simultaneous fit to the dimuon invariant mass spectra in the $PbPb$ and $pp$ datasets gives the double ratios of the observed signal yields
$$
\frac{N^{\rm obs}(\Upsilon(2S))/N^{\rm obs}(\Upsilon(1S))|_{PbPb}}{N^{\rm obs}(\Upsilon(2S))/N^{\rm obs}(\Upsilon(1S))|_{pp}} = 0.21\pm0.07\pm0.02
$$
and
\begin{equation}\nonumber
\begin{aligned}
\frac{N^{\rm obs}(\Upsilon(3S))/N^{\rm obs}(\Upsilon(1S))|_{PbPb}}{N^{\rm obs}(\Upsilon(3S))/N^{\rm obs}(\Upsilon(1S))|_{pp}} = \\
0.06\pm0.06\pm0.06~(< 0.17~{\rm at}~95\%~{\rm C.L.}).
\end{aligned}
\end{equation}
The measured values are considerably smaller than unity. The significance of the observed suppression exceeds 5$\sigma$.

In addition, the absolute suppressions of all three individual $\Upsilon$ states are also studied using the nuclear modification factor, $R_{AA}$, defined as the yield per nucleon-nucleon collision in $PbPb$ relative to that in $pp$. The $R_{AA}$ observable,
$$
R_{AA} = \frac{{\cal L}_{pp}}{T_{AA}N_{MB}} \frac{N^{\rm obs}(\Upsilon(nS))|_{PbPb}}{N^{\rm obs}(\Upsilon(nS))|_{pp}}\frac{\varepsilon_{pp}}{\varepsilon_{PbPb}},
$$
is evaluated from the ratio of total $\Upsilon(nS)$ yields in $PbPb$ and $pp$ collisions corrected for the difference in efficiencies $\varepsilon_{pp}$ and $\varepsilon_{PbPb}$, with the average nuclear overlap function $T_{AA}$, number of minimum-bias events sampled by the event selection $N_{MB}$, and integrated luminosity of the $pp$ dataset ${\cal L}_{pp}$ accounting for the normalization. The values of the $R_{AA}$ are determined to be
$$
R_{AA}(\Upsilon(1S)) = 0.56\pm0.08\pm0.07,
$$
$$
R_{AA}(\Upsilon(2S)) = 0.12\pm0.04\pm0.02,
$$
and
$$
R_{AA}(\Upsilon(3S)) < 0.10~(95\%~{\rm C.L.}).
$$

\begin{figure*}[htbp]
\centering
\includegraphics[width=12cm]{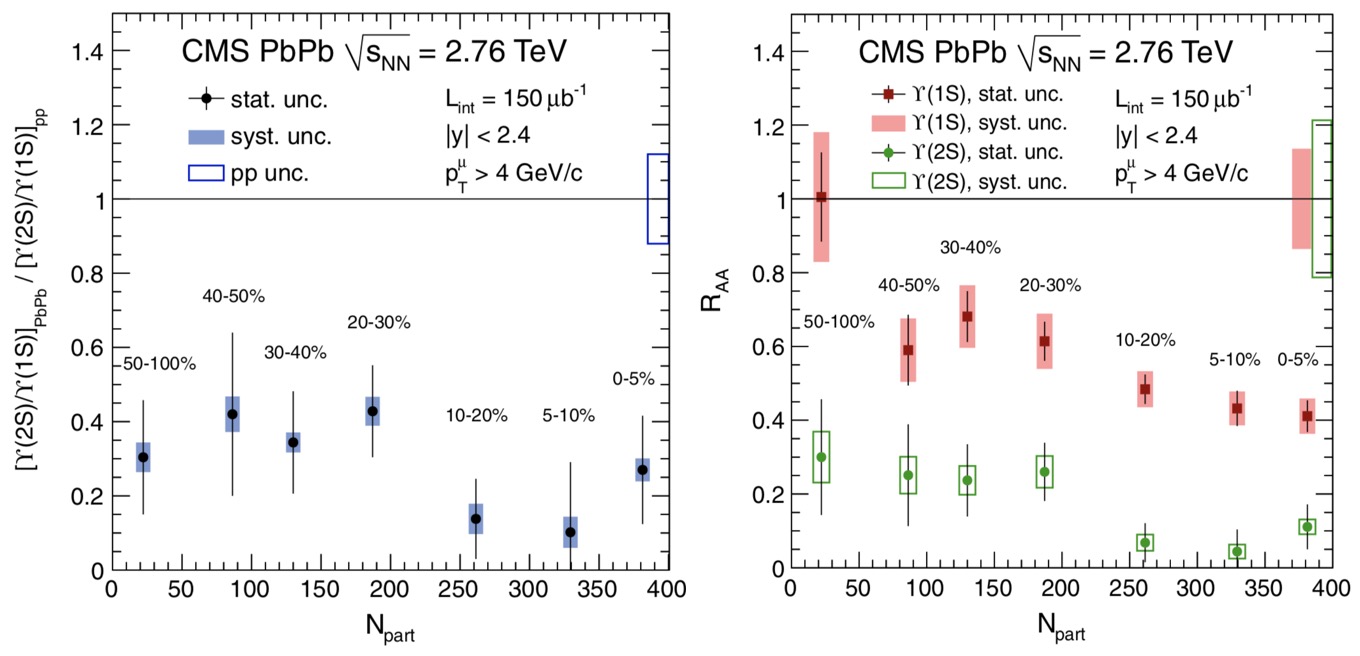}
\caption{Centrality dependence of the double ratio (left) and of $R_{AA}$ (right) for the $\Upsilon(1S)$ and $\Upsilon(2S)$ states from CMS~\cite{Chatrchyan222301}. }\label{plot10}
\end{figure*}

Centrality is an important parameter for the QGP formation since it is directly related to the overlap region of the colliding nuclei. The extensive forward calorimetry of the CMS apparatus provides a good chance for the centrality determination in $PbPb$ collisions. The event centrality observable corresponds to the fraction of the total inelastic cross section, starting at 0  for the most central collisions.
The centrality classes used in this study are $(50 - 100)$\%, $(40 - 50)$\%, $(30 - 40)$\%, $(20-30)$\%, $(10-20)$\%, $(5-10)$\%, and $(0-5)$\%, ordered from the lowest to the highest  energy deposit. Using a Glauber-model calculation as described in Ref.~\cite{Chatrchyan024906}, the centrality variable can be expressed by the average number of nucleons participating in the collisions (${\rm N}_{\rm part}$). The double ratio of the observed signal yields $\frac{N^{\rm obs}(\Upsilon(2S))/N^{\rm obs}(\Upsilon(1S))|_{PbPb}}{N^{\rm obs}(\Upsilon(2S))/N^{\rm obs}(\Upsilon(1S))|_{pp}}$ and $R_{AA}$ for $\Upsilon(1S)$ and $\Upsilon(2S)$ as a function of ${\rm N}_{\rm part}$ are shown in Fig.~\ref{plot10}~\cite{Chatrchyan222301}. From Fig.~\ref{plot10} (left), the double ratio dependence on centrality is not pronounced. While in Fig.~\ref{plot10} (right), the $\Upsilon(1S)$ and $\Upsilon(2S)$ suppressions are observed to increase with collision centrality. These results indicate a significant suppression of the $\Upsilon$ states in heavy-ion collisions compared to $pp$ collisions at the same per-nucleon-pair energy. The $\Upsilon(1S)$ is the least suppressed and the $\Upsilon(3S)$ is the most suppressed of the three states, which supports the hypothesis of increased suppression of less strongly bound states.

\begin{figure*}[htbp]
\centering
\includegraphics[width=15cm]{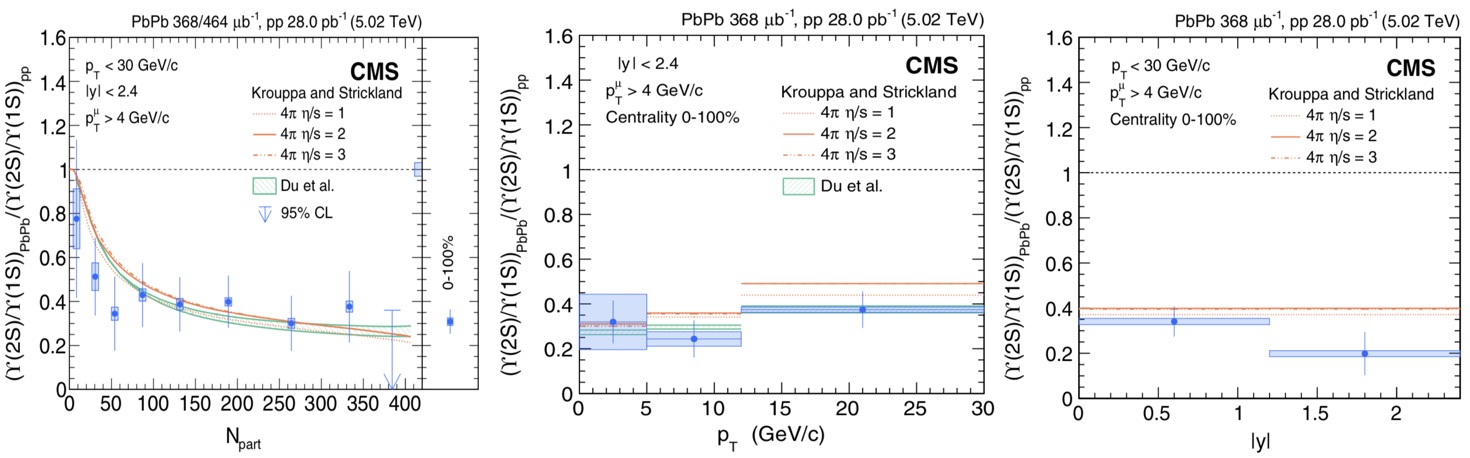}
\caption{The double ratios for $\Upsilon(2S)$ as a function of the collision centrality, transverse momentum, and rapidity from CMS~\cite{Sirunyan142301}.}\label{plot12}
\end{figure*}

The $\Upsilon$ suppression in $PbPb$ collisions at a higher C.M. energy, i.e.~$\sqrt{s}$ = 5.02 TeV, was studied by CMS~\cite{Sirunyan142301}. The double ratio $[\Upsilon(nS)/\Upsilon(1S)]_{PbPb}/[\Upsilon(nS)/\Upsilon(1S)]_{pp}$ was measured to be $0.308\pm0.055\pm0.019$ for $\Upsilon(2S)$ and less than 0.26 at 95\% C.L. for $\Upsilon(3S)$. The double ratio for $\Upsilon(2S)$ was studied as a function of collision centrality, as well as the $\Upsilon$ transverse momentum and rapidity, as shown in Fig.~\ref{plot12}. Predictions of $\Upsilon$ suppression from Krouppa and Strickland~\cite{Krouppa16}, incorporating color-screening effects on the bottomonium family and reflecting feed-down contributions from decays of heavy quarkonia, are in overall agreement with the $\Upsilon(2S)$ double ratio results presented in Fig.~\ref{plot12}. In this model, the dynamical evolution is treated using anisotropic hydrodynamics, where the relevant initial conditions are changed by varying the viscosity to entropy ratio, $\eta/s$, and the initial momentum-space anisotropy. Another theoretical curve from Du {\em et al.}\ in Ref.~\cite{Du054901}, based on a kinetic-rate equation approach and containing a small component of regenerated bottomonia, shows a similar level of agreement with the data. No significant variations with $p_{\rm T}$ (middle) or $|{\rm y}|$ (right) in Fig.~\ref{plot12} are observed. Predictions of $\Upsilon$ suppression as functions of $p_{\rm T}$~\cite{Krouppa16,Du054901} and $|{\rm y}|$~\cite{Krouppa16} are in overall agreement with the data.

\begin{figure*}[htbp]
\centering
\includegraphics[width=12cm]{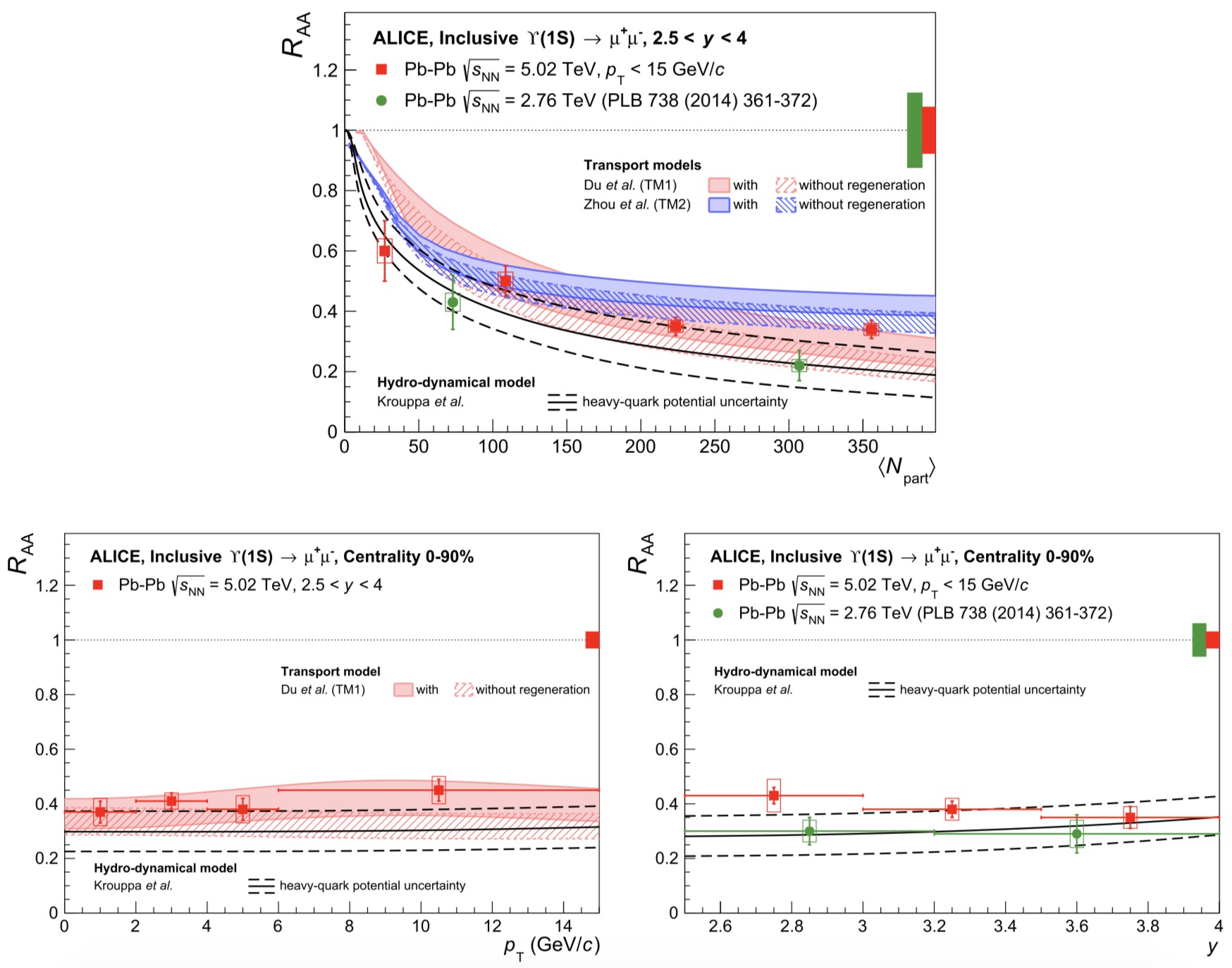}
\caption{The $R_{AA}$ factor of $\Upsilon(1S)$ as a function of centrality (top), $p_{\rm T}$ (left) and y (right) at $\sqrt{s}$ =  2.76 TeV~\cite{Acharya361} and $\sqrt{s}$ =  5.02 TeV~\cite{Acharya89} from ALICE.}\label{ALICEfig1}
\end{figure*}

The $\Upsilon$ suppressions at $\sqrt{s}$ = 2.76 TeV and $\sqrt{s}$ = 5.02 TeV were also explored by ALICE in a complementary rapidity range of 2.5 $<$ y $<$ 4~\cite{Acharya361,Acharya89}. The integrated luminosities of the data samples used for $\sqrt{s}$ = 2.76 TeV and $\sqrt{s}$ = 5.02 TeV are 68.8 $\mu$b$^{-1}$ and 225 $\mu$b$^{-1}$, respectively. The obtained values of $R_{AA}$ for $\Upsilon(1S)$ as a function of centrality, $p_{\rm T}$, and y are shown in Fig.~\ref{ALICEfig1}. ALICE has no ability to measure the cross sections of $\Upsilon$ in $pp$ collisions. Therefore, they applied an interpolation method utilizing the LHCb measurements in Ref.~\cite{Aaij2835}. From Fig.~\ref{ALICEfig1}, $R_{AA}$ decreases with increasing centrality, and no significant dependencies on $p_{\rm T}$ and y are seen. The suppressions of the $\Upsilon(1S)$ state on the centrality and y at $\sqrt{s}$ = 5.02 TeV are slightly weaker than those measured at $\sqrt{s}$ = 2.76 TeV, which is not consistent with our expectations since the bound state is more likely to dissolve at higher temperatures. Two transport models (TM1 and TM2)~\cite{Du054901,Zhou654} and one hydro-dynamical model~\cite{Krouppa016017} are compared with data points. Though TM1 and TM2 use different rate equations, both of them take account of the feed-down contributions from higher-mass bottomonia to $\Upsilon(1S)$. Deviation between TM1 (TM2) and data is 2$\sigma$ (1.4$\sigma$). The trends of $R_{AA}$ as a function of centrality, $p_{\rm T}$, and y can be described by the hydro-dynamical model to some extent.

\begin{figure}[htbp]
\centering
\includegraphics[width=7cm]{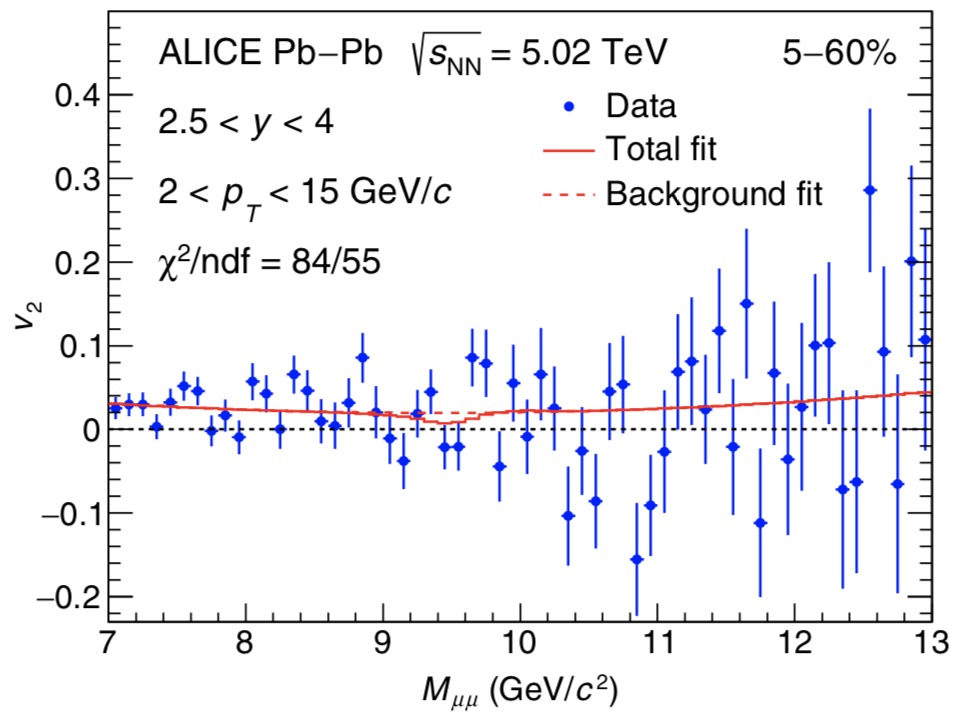}
\caption{The graph of $\nu_2$ with respect to $M_{\mu\mu}$ in the ranges of 2.5 $<$ y $<$ 4 and 2 $<$ $p_{\rm T}$ $<$ 15 GeV/c from ALICE~\cite{Acharya192301}. The red solid line shows the fitted result, and the red dashed line shows the fitted backgrounds.}\label{v21}
\end{figure}

\begin{figure}[htbp]
\centering
\includegraphics[width=7cm]{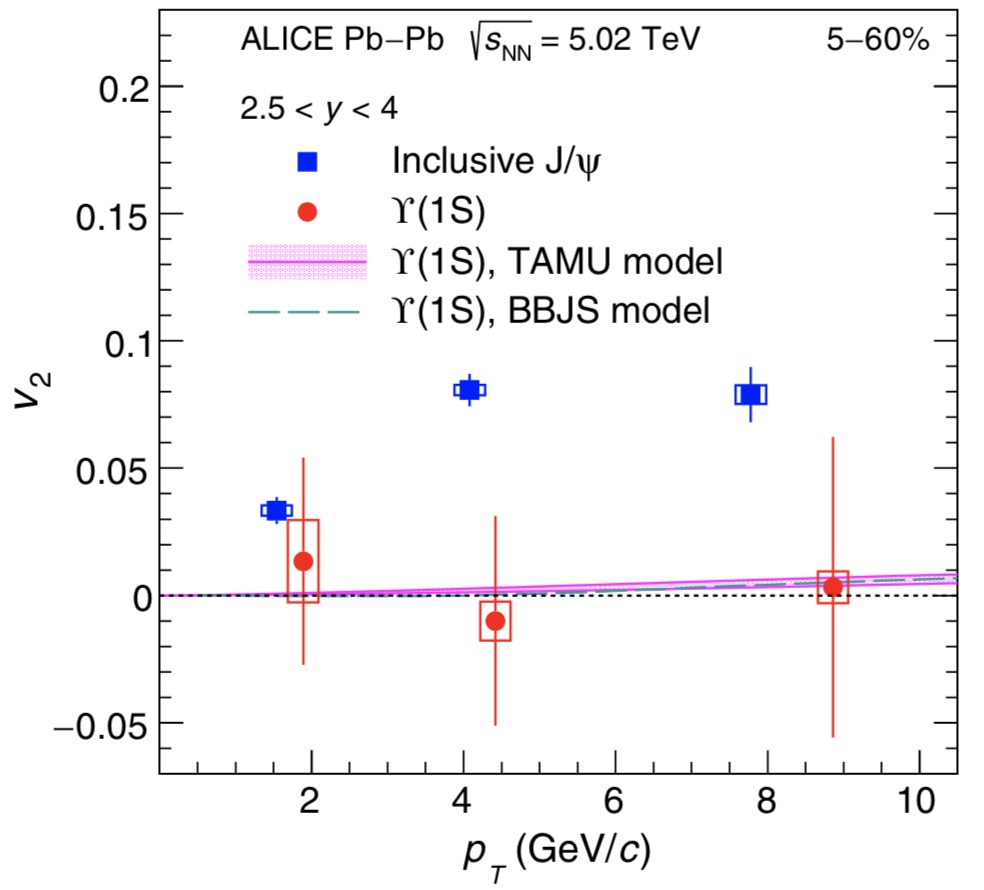}
\caption{The $\nu_2$ coefficient of $\Upsilon(1S )$ as a function of transverse momentum in the rapidity region of 2.5 $<$ y $<$ 4, compared to that of inclusive $J/\psi$ from ALICE~\cite{Acharya192301}.}\label{v22}
\end{figure}

The elliptic flow coefficient, denoted by $\nu_2$~\cite{Voloshin665}, reflects the response of the QGP medium to the initial anisotropy in non-central $PbPb$ collisions. Using a $PbPb$ collision data sample of 750 $\mu$b$^{-1}$ at $\sqrt{s}$ = 5.02 TeV, ALICE performed the first measurement of the $\nu_2$ coefficient of $\Upsilon(1S)$ in the forward rapidity region of 2.5 $<$ y $<$ 4~\cite{Acharya192301}. The $\Upsilon$ mesons are reconstructed via their $\mu^+\mu^-$ decay channel. The dimuon $\nu_2$ coefficient was determined using the scalar product method~\cite{Adler034904,Voloshin293}, correlating the reconstructed dimuons with the second-order harmonic event flow vector~\cite{Voloshin665,Barrette2532}. Figure~\ref{v21} shows the graph of $\nu_2(M_{\mu\mu})$ as well as the fitted result, where a negative yield around $\Upsilon(1S)$ mass is observed. Here, only 5\%$-$60\% centrality interval is considered since the eccentricity of the initial collision geometry is small for the central collision (0\%$-$5\%), and the signal yield is low for the peripheral collision (60\%$-$100\%). Figure~\ref{v22} shows the $\nu_2$ coefficient of $\Upsilon(1S)$ as a function of transverse momentum. The results are compatible with 0 and consistent with the small positive values predicted by the available theoretical models within uncertainties. The values of $\nu_2$ for $\Upsilon(1S)$ are lower than those for $J/\psi$ by 2.6 standard deviations, which indicates the limited dissociation for $\Upsilon(1S)$ in the early $PbPb$ collision stage, and an additional regeneration component needed for $J/\psi$.

\subsubsection{$\Upsilon$ suppression in $pPb$ collisions}

In addition to hot-nuclear-matter (HNM) effects, regarding the QGP formation, the differences in quarkonium production between $PbPb$ and $pp$ collisions can also arise from cold-nuclear-matter (CNM) effects~\cite{Vogt044903}. Therefore, it is necessary to study $\Upsilon(1S,2S,3S)$ productions in $pPb$ collisions in order to study the CNM effect separately. This knowledge can be extrapolated into $PbPb$ collisions so that the fraction of the suppression due to HNM in $PbPb$ collisions can be determined~\cite{Hu1730015}.

\begin{figure}[htbp]
\centering
\includegraphics[width=6cm]{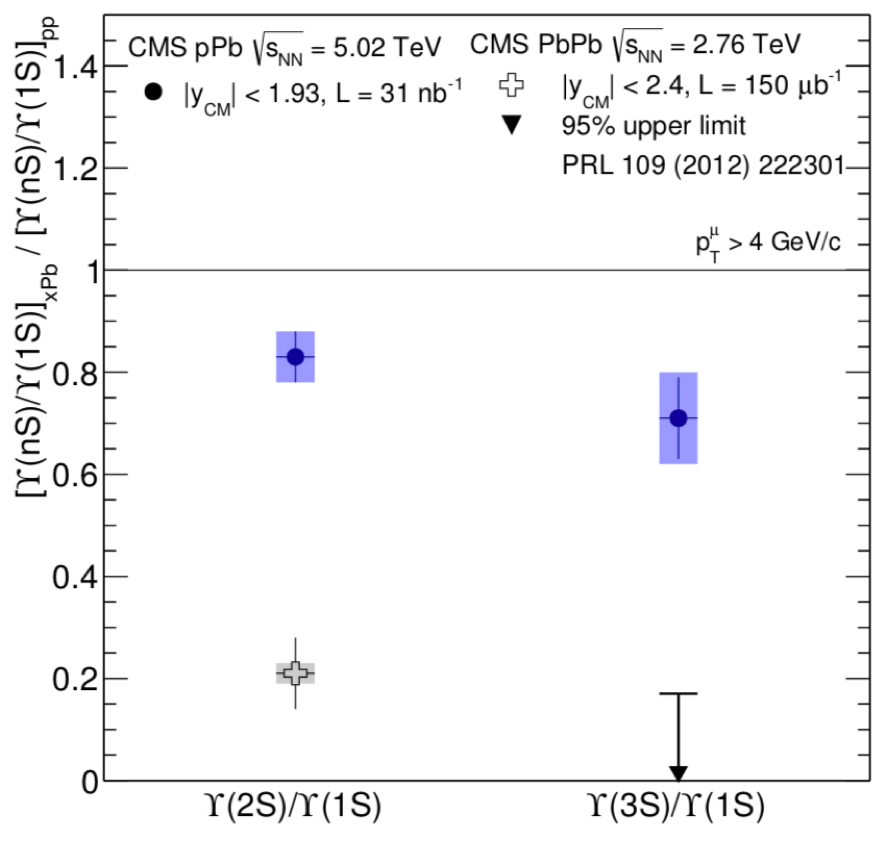}
\caption{Event activity integrated double ratios of the excited states, $\Upsilon(2S)$ and $\Upsilon(3S)$, to the ground state, $\Upsilon(1S)$, in $pPb$ collisions at $\sqrt{s}$ = 5.02 TeV with respect to $pp$ collisions at $\sqrt{s}$ = 2.76 TeV from CMS~\cite{Chatrchyan103}, compared to their counterparts for $PbPb$ collisions at $\sqrt{s}$ = 2.76 TeV from CMS~\cite{Chatrchyan222301}.}\label{plot13}
\end{figure}

\begin{figure*}[htbp]
\centering
\includegraphics[width=12cm]{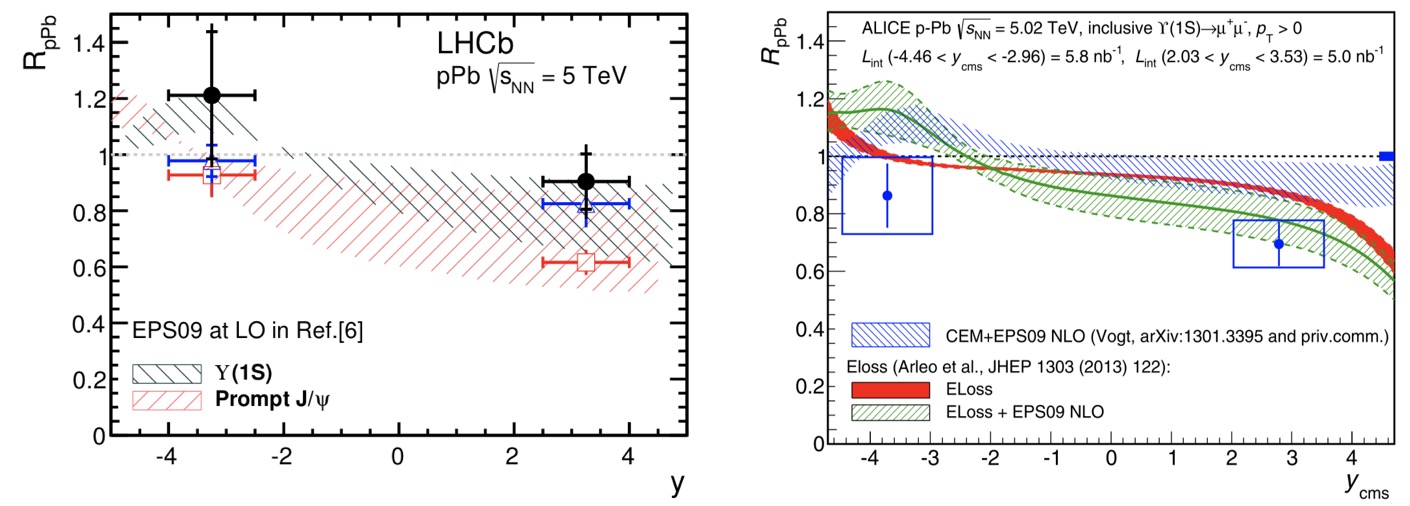}
\caption{The left: the nuclear modiﬁcation factors for $\Upsilon(1S)$ (black dots), prompt $J/\psi$ (red squares), and $J/\psi$ from $b$-hadron decays (blue triangles) in $pPb$ collisions at $\sqrt{s}$ = 5.02 TeV as a function of rapidity from LHCb~\cite{Aaij094}. The right: the nuclear modiﬁcation factors for $\Upsilon(1S)$ in $pPb$ collisions at $\sqrt{s}$ = 5.02 TeV as a function of rapidity from ALICE~\cite{Acharya105}.}\label{pPb1}
\end{figure*}

The $\Upsilon$ suppressions in $pPb$ collisions at $\sqrt{s}$ = 5.02 TeV were investigated by CMS~\cite{Chatrchyan103}, LHCb~\cite{Aaij094}, and ALICE~\cite{Acharya105}, using data samples of 31 nb$^{-1}$, 1.6 nb$^{-1}$, and 10.82 nb$^{-1}$, respectively. The studied rapidity intervals in CMS, LHCb, and ALICE are $|{\rm y}|$ $<$ 1.93, $-5$ $<$ y $<$ $-2.5$ (backward) \&\& $1.5$ $<$ y $<$ $4.0$ (forward), and $-4.46$ $<$ y $<$ $-2.96$ (backward) \&\& $2.03$ $<$ y $<$ $3.53$ (forward), respectively. By utilizing the results in $pp$ collisions at $\sqrt{s}$ = 2.76 TeV, CMS gave the $pPb$ double ratios, which are shown in Fig.~\ref{plot13}, together with their counterparts in $PbPb$ collisions at $\sqrt{s}$ = 2.76 TeV for comparison. Clearly, the $pPb$ ratios are much larger than the corresponding $PbPb$ ratios~\cite{Chatrchyan222301}. The nuclear modification $R_{AA}$ as a function of rapidity y was determined by LHCb~\cite{Aaij094} and ALICE~\cite{Acharya105} in both of their forward and backward regions, where the input cross sections of $\Upsilon$ in $pp$ collisions were obtained by a power-law interpolation from LHCb measurements~\cite{Aaij2025,Aaij2835,Aaij064}. The resulting $R_{AA}$ distributions are shown in Fig.~\ref{pPb1} with the calculations from several theoretical models. The values of $R_{AA}$ for $\Upsilon(1S)$ in the forward region from LHCb and forward and backward regions from ALICE are below 1, while the value of $R_{AA}$  for $\Upsilon(1S)$ in the backward region from LHCb is above 1. The blue band in Fig.~\ref{pPb1} (left) shows the predicted results based on CSM at LO with EPS09 corrected parameter~\cite{Ferreiro2427}. The calculation based on CEM at LO with EPS09~\cite{Albacete1330007} is shown by the blue band in Fig.~\ref{pPb1} (right). The parton energy loss calculations with and without EPS09 at NLO~\cite{Arleo122} are shown by green and red bands in Fig.~\ref{pPb1} (right), respectively. All of the above theoretical predictions~\cite{Ferreiro2427,Albacete1330007,Arleo122} favor the LHCb result that the value of $R_{AA}$ in the backward region is at least 1. The nuclear modification factors measured from LHCb also deliver information that the suppression for $\Upsilon(1S)$ is smaller than that for $J/\psi$.

\begin{figure*}[htbp]
\centering
\includegraphics[width=12cm]{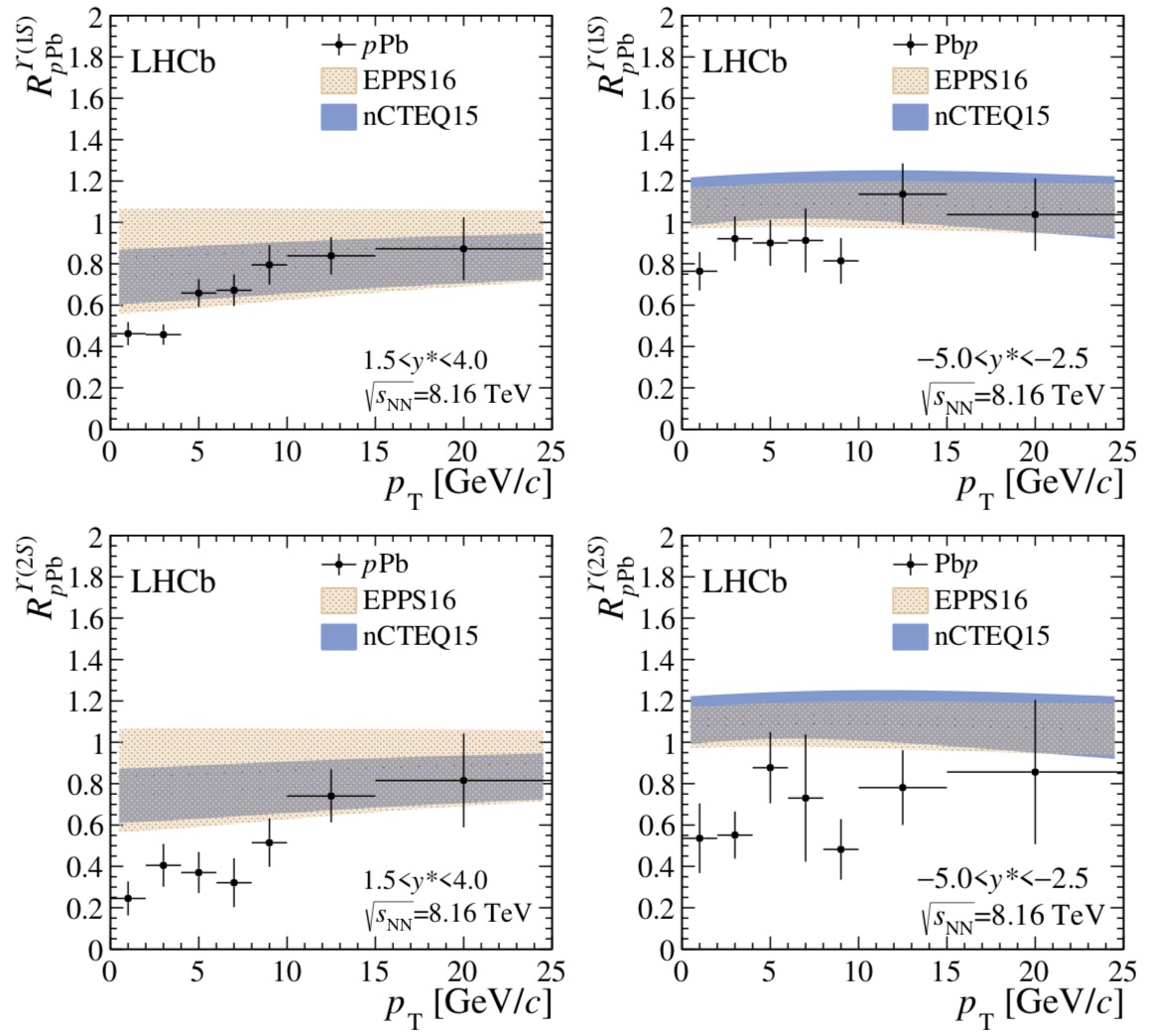}
\caption{The nuclear modification factors for $\Upsilon(1S)$ (above) and $\Upsilon(2S)$ (below) in $pPb$ collisions as a function of $p_{\rm T}$ in the forward (left) and backward (right) regions from LHCb~\cite{Aaij194}.}\label{LHCbPt}
\end{figure*}

\begin{figure*}[htbp]
\centering
\includegraphics[width=12cm]{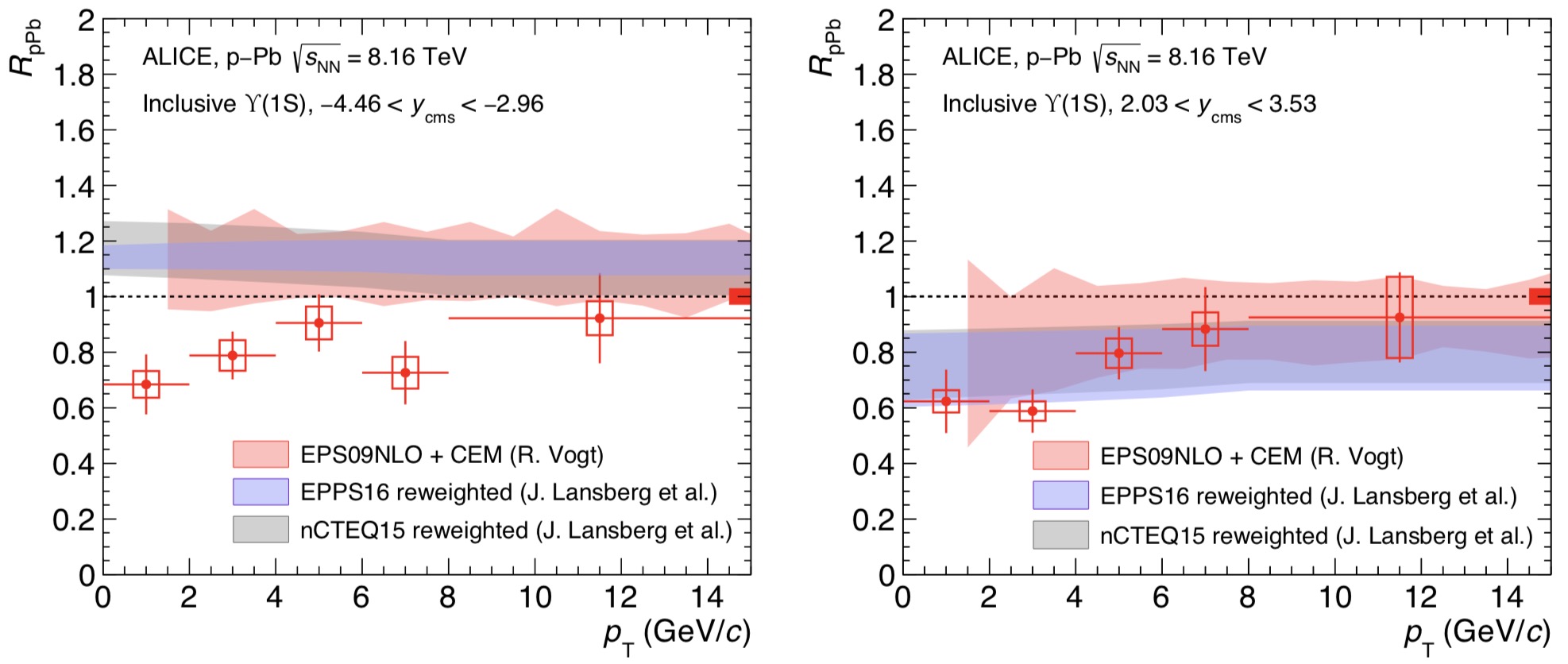}
\caption{The nuclear modification factors for $\Upsilon(1S)$ in $pPb$ collisions as a function of $p_{\rm T}$ in the backward (left) and forward (right) regions from ALICE~\cite{Acharya2019}.}\label{ALICEPt}
\end{figure*}

The $\Upsilon$ suppressions in $pPb$ collisions at a higher C.M. energy, i.e.\ $\sqrt{s}$ = 8.16 TeV were studied by LHCb~\cite{Aaij194} and ALICE~\cite{Acharya2019}. Particularly, the $\Upsilon$ suppression performances with increasing transverse momentum $p_{\rm T}$ were intensively investigated. Figure~\ref{LHCbPt} shows the nuclear modification factor $R_{AA}$ for $\Upsilon(1S)$ and $\Upsilon(2S)$ as a function of $p_{\rm T}$ in both the forward and backward regions. From the figure, the values of $R_{AA}$ increase with $p_{\rm T}$. The similar trends were also found in the $R_{AA}$ measurements from ALICE~\cite{Acharya2019}, as shown in Fig.~\ref{ALICEPt}. The predictions with EPPS16~\cite{Shao238,Shao2562,Eskola,Lansberg1} and nCTEQ15~\cite{Kovarik085037,Eskola163,Kusina052004}, as two different nuclear parton distribution functions, are used to compare the experimental results.
The calculations based on EPS09 NLO~\cite{Albacete18,Vogt034909} are also used to compare the ALICE result, as shown by the red bands in Fig.~\ref{ALICEPt}. A good agreement between experimental data and theoretical predictions is found in the forward region for both $\Upsilon(1S)$ and $\Upsilon(2S)$. While in the backward region, the values of $R_{AA}$ in data are mostly less than 1, different  from theoretical predictions. Overall, the suppression effects are seen in $pPb$ collisions, but much smaller than those in $PbPb$ collisions, which indicates the presence of the CNM effect. Such contributions can make small corrections to the complete HNM suppression effects in $PbPb$ collisions.

\section{Summary and prospects}

\subsection{Summary at $e^+e^-$ colliders and prospects at Belle II}

Although the $B$-factories stopped taking data more than ten years ago, the Belle II as a next-generation $B$-factory experiment came back in March 2019. Datasets of $\Upsilon(1S,2S,3S)$ collected at Belle II are expected to be about 200 fb$^{-1}$ for each in the schedule~\cite{Kou1808}, which offers a better chance to study $\Upsilon(1S,2S,3S)$ decays further.

The measurements of exclusive $\Upsilon(1S,2S)$ decays into light hadrons were performed by Belle and authors using CLEO data. From Belle, the evidences were found in $K^*(892)^0{\bar K}^*_2(1430)^0$, $\phi K^+ K^-$, $K^*(892)^0K^-\pi^+$, $\pi^+\pi^-\pi^0\pi^0$, and $K^0_SK^+\pi^-$ final states in both $\Upsilon(1S)$ and $\Upsilon(2S)$ decays; from the measurements in Ref.~\cite{Dobbs86}, significance strengths were found in 17 decay modes consisting of 4 -- 10 light hadrons, pions, kaons, and protons in both $\Upsilon(1S)$ and $\Upsilon(2S)$ decays. The $Q_{\Upsilon}$ values measured by Belle are close to the ``80\% rule" line. However, most of $Q_{\Upsilon}$ values from the measurements using CLEO data in Ref.~\cite{Dobbs86} are below the ``80\% rule" line.~The discrepancy is expected to be resolved by using a large number of $\Upsilon(1S,2S)$ events collected at Belle II.

Abundant events of $\Upsilon(1S)\to J/\psi+anything$ were seen by CLEO and Belle, and the distribution of their momenta favors the expectation of the color-singlet process. So far, no double charmonium process is evident except for $\Upsilon(1S)\to J/\psi+\chi_{c1}$ with a significance of 4.6$\sigma$. This needs to be studied further at Belle II, in order to confirm the related calculations with the NRQCD factorization approach.

The radiative decays of $\Upsilon(1S,2S,3S)$ to charmonia and bottomonia were systematically studied by CLEO, BaBar, and Belle. The first radiative decay of the $\Upsilon(1S)$ into a charmonium state, i.e., $\Upsilon(1S)\to\gamma \chi_{c1}$ was observed by Belle using a $\Upsilon(1S)$ tagging technique. The branching fraction of $\Upsilon(1S)\to\gamma \chi_{c1}$ was obtained with large uncertainties, and is slightly higher than the previous upper limit and much higher than the theoretical expectation from NRQCD. At Belle II, the precise measurement of $\Upsilon(1S)\to\gamma \chi_{c1}$ and the searches for more $\Upsilon(1S,2S,3S)$ radiative decays to a charmonium, including $\Upsilon(1S,2S,3S)\to\gamma \chi_{c0}/\chi_{c2}/\eta_c(1S)/\eta_c(2S)$, are expected to be done. Among the photon transitions of $\Upsilon(2S,3S)$ to $\chi_{bJ}$, the branching fractions of $\Upsilon(3S)\to\gamma \chi_{bJ}(1P)$ decrease by one order of magnitude compared to those of $\Upsilon(2S)\to\gamma \chi_{bJ}(1P)$ and $\Upsilon(3S)\to\gamma \chi_{bJ}(2P)$. The $\eta_b(1S)$ resonance was observed in the $\Upsilon(2S,3S)$ radiative decays by Belle and CLEO in the inclusive photon energy spectra. However, the observations of $\eta_b(1S)$ and $\eta_b(2S)$ in the exclusive hadronic final states of the $\Upsilon(1S)$ and $\Upsilon(2S)$ radiative decays are much debated.

The charmonium-like states, exotic glueballs, light tetraquarks, and a stable six-quark state were searched for in $\Upsilon(1S,2S,3S)$ inclusive or radiative decays. Unfortunately, no clear signals were observed in all the studied modes, and only the 90\% C.L.\ upper limits on the production rates were determined. Nonetheless, searching for exotic states is still an important topic in hadronic physics. At Belle II, special attention should be paid to $\Upsilon(1S)\to f_1(1285)+G_{0^{--}}$ to check whether there is a peak around 3.92 GeV/$c^2$. The other exotic glueballs with $J^{PC}$ = $0^{+-}$, $1^{-+}$, and $2^{+-}$ predicted in Ref.~\cite{Tang282} will also be searched for at Belle II. Related studies with the radiative decays of $\Upsilon(1S,2S,3S)$ will be conducted further at Belle II.

The CLEO, Belle, and BaBar experiments studied the $\Upsilon(1S) \to invisible$ decay to search for low mass DM. No significant signal was observed in all these experiments. The most stringent upper limit was set by BaBar; it is $\BR(\Upsilon(1S) \to invisible$) $<$ 3.0 $\times$ $10^{-4}$ at 90\% C.L. With a tagging method, Belle and BaBar also set the upper limits at 90\% C.L. on $\BR(\Upsilon(1S) \to \gamma A^0)\BR(A^0\to invisible$) and $\BR(\Upsilon(1S) \to \gamma \chi\chi$) to a few times $10^{-5}$ and a few times $10^{-4}$, respectively. At Belle II, one can expect a sensitivity of $1.3\times10^{-5}$ at 90\% C.L. for $\BR(\Upsilon(1S) \to invisible)$ which is comparable to the SM prediction $\BR(\Upsilon(1S)\to \nu {\bar \nu})$ = $1.0\times10^{-5}$~\cite{Chang441}. As to the search for $\Upsilon(1S) \to \gamma + invisible$, Belle II has the possibility to discover an excess of events at 90\% C.L. if $\BR(\Upsilon(1S) \to \gamma A^0)\BR(A^0\to invisible)$ $>$ 5 $\times$ 10$^{-7}$ and $\BR(\Upsilon(1S) \to \gamma \chi\chi)$ $>$ 5 $\times$ 10$^{-6}$.

Leptonic decays of $\Upsilon(1S,2S,3S)$ can be used to test LFU. To date, the measured values of $\BR(\Upsilon(1S,2S) \to \ell^+\ell^-)$ ($\ell$ = e, $\mu$, $\tau$) are consistent with each other, but with large uncertainties of a few percent. The LFV processes have not been observed, and the 95\% C.L. upper limits on the branching fractions are determined to be at a level of 10$^{-7}$ by CLEO and BaBar.
Belle II experiment is expected to achieve better control of statistical and systematic effects, allowing improved determinations of $\BR(\Upsilon(1S) \to \ell^+\ell^-)$ and $\BR(\Upsilon(1S) \to \ell^+\ell^{\prime-})$, where $\ell^+$ and $\ell^{\prime-}$ are different types of leptons.

The transitions between $\Upsilon(mS)$ and $\Upsilon(nS)$ ($m > n$) were intensively investigated in $e^+e^-$ collision experiments. The ratio $\BR(\Upsilon(2S)\to\eta\Upsilon(1S))/\BR(\Upsilon(2S)\to \pi^+\pi^-\Upsilon(1S))$ is slightly below the value predicted by QCDME,~while $\BR(\Upsilon(4S)\to\eta\Upsilon(1S))/\BR(\Upsilon(4S)\to\pi^+\pi^-\Upsilon(1S))$ strongly disfavors the prediction from QCDME, which may imply additional implementations for QCDME are needed.~The $\BR(\Upsilon(4S)\to \eta^{\prime}\Upsilon(1S))/\BR(\Upsilon(4S)\to \eta\Upsilon(1S))$ was measured to be $(0.20\pm0.06)$, which is in agreement with the expected value in the case of an admixture of a state containing light quarks in addition to the $b\bar b$ pair for $\Upsilon(4S)$. The production rates of $\Upsilon(5S)\to \pi^+\pi^-\Upsilon(1S,2S,3S)$ exceed those of the dipion transitions between lower resonances by more than two orders of magnitude, which indicates a new mechanism to enhance the decay rates. The latter observations of $Z_b(10610)$ and $Z_b(10650)$ decaying into $\pi\Upsilon(1S)$, $\pi\Upsilon(2S)$ and $\pi\Upsilon(3S)$ are supposed to contribute such enhancement in the dipion transitions of $\Upsilon(5S)$.

\subsection{Summary in $pp$, $pPb$, and $PbPb$ collisions and outlook at the LHC}

Measurements of the $\Upsilon(1S,2S,3S)$ cross sections in $pp$ collisions at the unprecedented C.M. energies of 2.76, 5.02, 7, 8, and 13 TeV have been undertaken, within the rapidity window of $-2.0$ $<$ y $<$ 4.5 and the dimuon momentum range of $p_{\rm T}$ $<$ 100 GeV/c. In addition, the angular distribution of the muons produced in the $\Upsilon(1S,2S,3S)$ decays has been analyzed in different reference frames to determine the polarization parameters. From the results of CMS and LHCb, $\Upsilon(1S,2S,3S)$ polarization parameters are all so close to zero that polarizations can be neglected. The measurements of these cross sections and polarizations have shed light on the $\Upsilon(1S,2S,3S)$ production mechanisms in $pp$ collisions. Analysis of new LHC data will extend the reach of the kinematics to test the NRQCD with higher-order corrections which becomes more sensitive with the increase of $p_{\rm T}$. Additional studies of other charmonia, bottomonia, and hadron jets are expected to be conducted at the LHC to gain a more complete understanding of the mechanisms of hadron productions.

The QGP formation is a hot topic in modern particle physics. One of its most striking characteristics is the suppression of quarkonium states. With the increased collision energy and detector capability, CMS, LHCb, and ALICE have established the pattern of sequential $\Upsilon(1S,2S,3S)$ suppression in collisions involving heavy ions, where the excited states are more suppressed than the ground state. Besides, they proved that the $\Upsilon(1S,2S,3S)$ suppression increases with the collision centrality and energy, and does not change along with the transverse momentum or rapidity. In addition to $PbPb$ collisions, the $\Upsilon(1S,2S,3S)$ suppression was also studied in $pPb$ collisions by CMS, LHCb, and ALICE to distinguish the non-HNM effects in the QGP formation.  The slight $\Upsilon(1S,2S,3S)$ suppression observed in $pPb$ collisions implies that both HNM and CNM can influence the QGP formation. In the future, the large datasets collected at the LHC will be used to explore the sequential suppression of $P$-wave bottomonia. New observables, such as azimuthal anisotropies and polarizations, will be also studied via angular analyses to figure out the properties of HNM in hadron collisions.

\begin{center}
{\bf Acknowledgements}
\end{center}

\vspace{0.0cm}
We would like to thank Prof. Zhen Hu for fruitful discussions. This work is supported by the National Natural Science Foundation of China under Grants No. 11575017, No. 11761141009, No. 11975076, and No. 11661141008; National Key R\&D Program of China under the contract No. 2018YFA0403902; and the CAS Center for Excellence in Particle Physics (CCEPP).

\end{document}